\documentclass[seceq]{ptptex}

\usepackage{graphicx}
\usepackage{wrapft}

\font\mbf=cmmib10  
\font\calg=cmsy10  
\def\RMNa{\uppercase\expandafter{\romannumeral 1}} 
\def\RMNb{\uppercase\expandafter{\romannumeral 2}}
\def\RMNc{\uppercase\expandafter{\romannumeral 3}}
\def\RMNd{\uppercase\expandafter{\romannumeral 4}}

\title{
Resonances in $^{28}$Si$+^{28}$Si. I
}

\subtitle{Dinuclear Molecular Model 
with Axial Asymmetry}    

\author{
Eiji \textsc{Uegaki}$^1$%
 and Yasuhisa \textsc{Abe}$^2$
} 

\inst{
$^1$Graduate School of Engineering and Resource Science, Akita University,\\
    Tegata, Akita City 010-8502, Japan\\
$^2$Research Center for Nuclear Physics, Osaka University,\\ 
	10-1 Mihogaoka, Ibaraki, Osaka 567-0047, Japan
}

\abst{
A molecular model developed for resonances observed 
in medium light heavy-ion collisions is described. 
At high spins in $^{28}\rm Si+{}^{28}Si$ (oblate-oblate system), 
a stable dinuclear configuration is found to be equator-equator 
touching one. 
The normal modes around the equilibrium are investigated. 
These modes are expected to be the origin of a large number of 
resonances observed. Furthermore, 
due to the axially asymmetric shape of the stable configuration 
of $^{28}\rm Si+{}^{28}Si$, the system rotates preferentially 
around the axis with the largest moment of inertia, 
which gives rise to wobbling motion ($K$-mixing). 
Energy spectra for the normal modes and for the extended model 
including the wobbling motion are given.  
}

\PTPindex{223}  

\begin{document}
\maketitle

\section{Introduction}

Intermediate resonances observed in heavy-ion scattering have offered 
intriguing subjects in nuclear physics.  
High-spin resonances well above the Coulomb barrier in the 
$^{24}\rm Mg+{}^{24}Mg$ and $^{28}\rm Si+{}^{28}Si$ systems 
exhibit very narrow widths,
which suggest rather long lived compound nuclear 
states.\cite{BettsC-a,BettsC-c} 

Betts et al. firstly observed a series of resonance-like enhancements 
at $\theta_{\rm cm}=90^\circ$ in elastic scattering of 
$^{28}\rm Si+{}^{28}Si$, in the energy range from 
$E_{\rm lab}=101$MeV to $128$MeV with broad bumps of about 2MeV width.
They gave spin assignments of $J=34 \sim 42$ by the Legendre-fits 
to the elastic angular distributions for each bump, which correspond to 
the grazing partial waves.\cite{BettsPRL1,BettsPL}

They further closely investigated angle-averaged excitation functions 
for the elastic and inelastic scattering 
in the energy region corresponding to $J=36 - 40$, 
and found, in each bump, several sharp peaks correlating among 
the elastic and inelastic channels.\cite{BettsPRL2,SainiBetts} 
The total widths of those resonances are about  $150{\rm keV}$, and 
the inelastic decay strengths are enhanced and stronger than the elastic one,
which suggests that they are special eigenstates of the compound system. 
Similar sharp resonance peaks are observed by Zurm\"uhle et al. in  
the $^{24}\rm Mg+{}^{24}Mg$ system.\cite{Zurm}
The level densities observed in those systems are over one per MeV, 
which suggests activation of internal degrees of freedom, 
in addition to the radial motion. 
The decay widths of the elastic and inelastic channels 
up to high spin members of the $^{24}\rm Mg$ or $^{28}\rm Si$ ground 
rotational band exhaust about $30\%$ of the total widths, whereas those 
into $\alpha$-transfer channels are much smaller.\cite{Saini,Beck2000}
These enhancements of symmetric-mass decays strongly suggest 
dinuclear molecular configurations for the resonance states. 
It is also noted that the widths of the elastic channel are rather small, 
for example, a few keV, being quite different 
from high spin resonances in lighter systems such as 
${}^{12}\rm {C} +{}^{12}\rm {C}$ and 
${}^{16}\rm {O} +{}^{16}\rm {O}$, 
which are well explained by the Band Crossing Model (BCM),\cite{BCMSuppl} 
i.e., by couplings between the relative motion of the incident 
ions and the low-lying collective excitations of the ions.

From viewpoints of nuclear structure studies, 
one immediately thinks of secondary minima in fission of heavy nuclei, 
or of superdeformations 
which have been intensively studied in medium mass nuclei.\cite{Faber}
Actually Bengtsson {\it et al.} made 
Nilsson-Strutinsky calculations for shape isomers of $^{56}$Ni  
and obtained an energy minimum at large deformation, 
which appears to correspond to a dinuclear configuration.\cite{Bengtsson}
Recently, a couple of microscopic calculations have been performed with 
expectation of existing shape-isomer bands.\cite{Gupta, Darai} 
For the $^{A}{\rm Mg} +{}^{A}{\rm Mg}$ system ($A=24$ or $26$) with 
very high spins, theoretical works were made to obtain 
stable configurations.\cite{MaassPL, Broglia} 
All those models, however, are not able to reproduce the level density 
of the sharp resonances as well as the decay properties observed.

Taking into account the difference from resonances in lighter systems 
and the level density of the sharp resonances observed, 
we have proposed a new dinucleus-molecular model 
for the high spin resonances in the $^{24}\rm Mg +{}^{24}Mg$ 
and $^{28}\rm Si +{}^{28}Si$ systems,\cite{Ue89,Ue93,Ue94} 
in which two incident ions are supposed to form a united composite system. 
It rotates as a whole in space with the internal degrees of freedom 
originating from interaction of the deformed constituent ions.  
This is in contrast with the viewpoint of BCM.

Actually, we have already applied the model 
to the $^{24}\rm Mg +{}^{24}Mg$ system to obtain a stable dinucleus 
configuration, using the folding potential.  
Normal modes of motion around the stable minimum were solved with 
harmonic approximations, and several characteristic modes were obtained, 
such as butterfly one, etc., which are expected to be responsible 
to the observed sharp resonances.
Decay properties of those resonance states were analyzed and strong 
enhancements to the mutual excitation channels are obtained in agreement 
with experiments for the $^{24}\rm Mg +{}^{24}Mg$ system.\cite{UeSuppl}

The same model has been applied to $^{28}\rm Si +{}^{28}Si$.\cite{Ue94}  
As is expected from the experience on $^{24}\rm Mg +{}^{24}Mg$, 
there are several intrinsic modes with excitation energy of a few MeV 
to several MeV.  They, thus, are expected to correspond to the sharp 
resonance peaks within each bump of the grazing $J$.  
Therefore, the present model appears promising also for the sharp 
high spin resonances observed in $^{28}\rm Si +{}^{28}Si$.

Recently, a new development has been obtained, 
giving attention to the remarkable difference between 
the $^{24}\rm Mg +{}^{24}Mg$ and $^{28}\rm Si +{}^{28}Si$ systems.  
In the former, the stable configuration is pole-to-pole one 
due to the prolate deformation of $^{24}\rm Mg$, while in the latter, 
it is the equator-to-equator configuration due to the oblate deformation 
of $^{28}\rm Si$.  Therefore, the former composite system is axially 
symmetric in the equilibrium, while the latter is triaxial.  Then, 
in the latter, strong $K$-mixing is kinematically induced, and results 
in a wobbling motion.\cite{Wobb, Nouicer99}   
Hence we have extended our molecular model so as to include couplings 
between states with different $K$-quantum numbers (projection of 
the total angular momentum on the molecular $z'$-axis).  As a result, 
we have obtained new low-lying states due to a triaxial shape of 
the equilibrium configuration.  In practice, 
we do not treat Coriolis terms in the hamiltonian explicitly, 
but we diagonalize the hamiltonian of the asymmetric rotator 
to obtain the rotational spectrum.  

Since the two different kind of models, i.e., the dinuclear 
molecular model and the asymmetric rotator are used to 
obtain the results, it is necessary to clarify the relation 
between them. 
We have studied simple examples of dinuclear systems by using 
the molecular model, and have found that the molecular model 
hamiltonian reduces to that of the asymmetric rotator 
in the sticking limit.\cite{Bass}

The present paper has the twofold aim. One is to describe 
the molecular normal-mode analyses\cite{Ue94, UeSuppl} 
as the full paper, 
and the other is to describe the development newly obtained. 
As for the former, a brief reminder of the molecular model 
is given in \S2, where we present the coordinate system 
and the model hamiltonian in the rotating molecular frame.  
In addition, in Appendix A, 
we take up simple examples of quantization, to compare the 
kinetic energy expression described by the angular momenta 
in the laboratory frame with that in the molecular frame, 
and to clarify the role of the Coriolis terms. 
There, the sticking-limit condition of sharing the total angular 
momentum between the orbital motion and the fragment spins 
is also derived. 
In \S3, structures of the $^{28}\rm Si+{}^{28}Si$ system are 
investigated. We begin with inspecting the multi-dimensional energy 
surface and look for the equilibrium configuration of the system. 
In \S3.2,  
harmonic approximation is adopted to solve normal modes around 
the equilibrium. Firstly, an energy spectrum with good $K$-quantum 
numbers will be given. 
The symmetries of the system and the practical expressions 
of the wave functions are described in Appendices B and C.

In order to present the new development,
section~4 is devoted to the analyses for the dinuclear system 
with axial asymmetry, which gives rise to wobbling motions 
($K$-mixing) in extremely high spins.
After $K$-mixing, the $K$-states are recombined into new states.  
The sequence of energy levels obtained by the diagonalization 
of the asymmetric rotator hamiltonian is given. 
A simple analytic solution is also discussed. 
In \S4.2, 
we take up simple examples of the molecular model hamiltonians 
and see how they reduce to the asymmetric rotator hamiltonians. 
As a summary, in \S5, we discuss on the structures 
of the $^{28}\rm Si-{}^{28}Si$ molecule theoretically explored.

Those molecular states are expected to be the origin of 
a large number of resonances observed, 
and hence theoretical analyses have been made. 
The results are in good agreements with 
the experimental data,\cite{Beck2000, Nouicer99} 
which will be given in the succeeding paper, 
no.~{\uppercase\expandafter{\romannumeral 2}}.\cite{UeNewII}

\section{Dinuclear molecular model of the oblate-oblate system
}

First, we briefly recapitulate the new molecular model for heavy-ion 
resonances. Definitions and derivations of the expressions are 
given in detail in Ref.~\citen{Ue93}, for the prolate-prolate system.  
We have already proposed a new description of interacting two 
oblate-deformed nuclei such as 
${}^{28}\rm Si+{}^{28}Si$.\cite{Ue94, UeSuppl}
In \S2.1, we extend our consideration to the coordinates of 
the system including axially-asymmetric deformed constituent nuclei, 
for later descriptions in \S4. 
Subsections~2.2 and 2.3 are devoted for the descriptions of the 
kinetic energy and the nucleus-nucleus potential, respectively, 
some expressions of which are already published 
in Ref.~\citen{Ue94} and in a part of Ref.~\citen{UeSuppl}.

The total system is described in terms of rotation of the whole system 
in space and of internal motions of all the other degrees of freedom 
such as the orientations of the deformation axes of two nuclei relative 
to the rotating molecular axes.  
We anticipate that there exists a stable geometrical configuration, i.e., 
a minimum in the potential energy of the internal degrees of the system.
Actually as explained later, each typical stable configuration appears 
by strong attractive nuclear interaction between tips of two deformed 
nuclei, for prolate or oblate deformations, respectively.
Accordingly, motions of their pole orientations should be treated 
as vibrational degrees of freedom around the geometrical equilibrium 
configuration, which is quite different from the usual description 
using "channels" in the weak coupling picture with the orbital angular 
momentum and the spins of the interacting nuclei.

\subsection{Coordinate systems}

\begin{figure}[t]
 \parbox[b]{\halftext}{
   \begin{center}
     \includegraphics[scale=0.63]{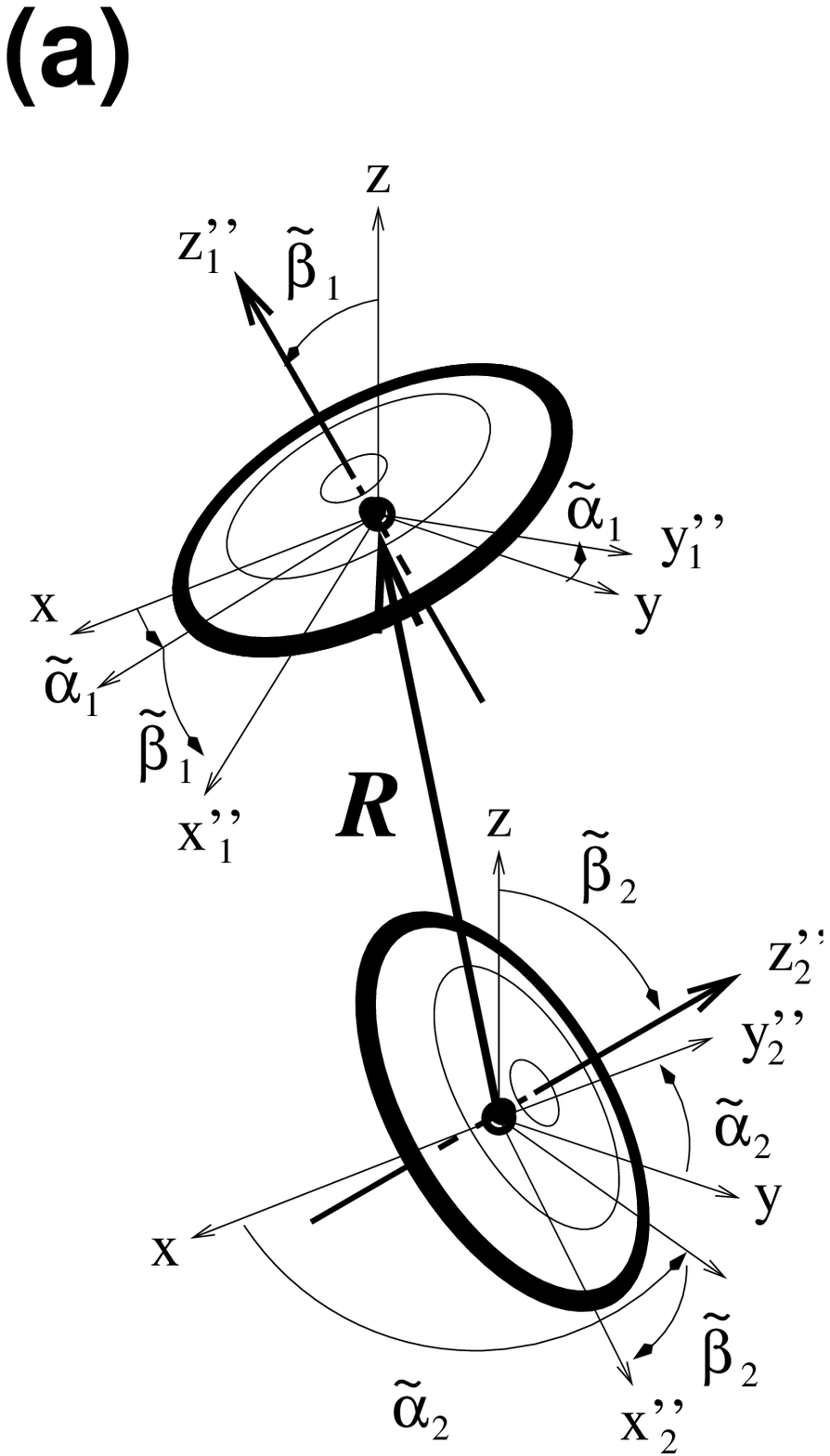}
   \end{center}
 }
 \parbox[b]{\halftext}{
   \begin{center}
     \includegraphics[scale=0.63]{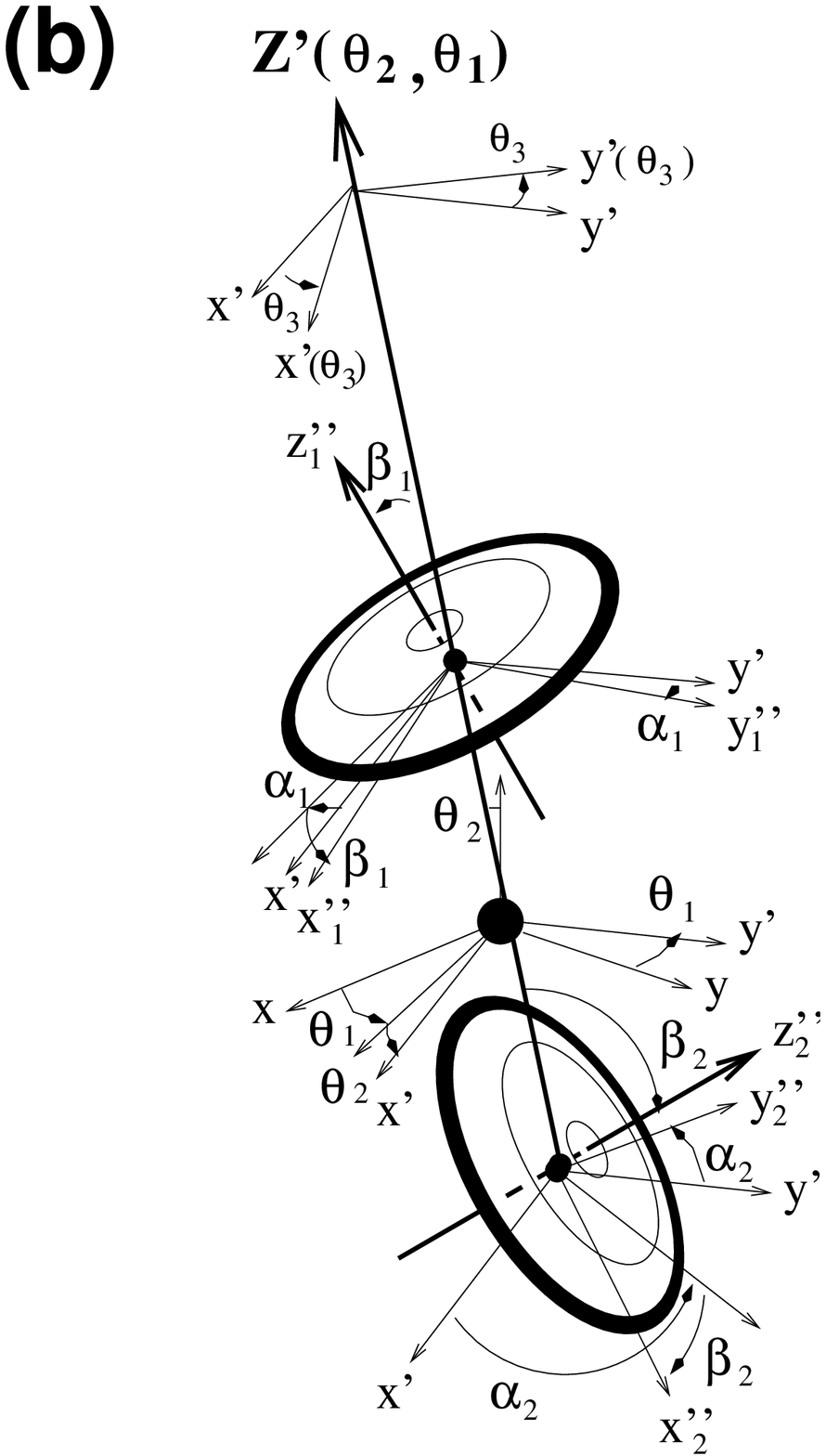}
   \end{center}
 }
\caption{The coordinates of an interacting dinuclear system. 
(a) shows the relative vector $\boldsymbol{ R} =(R,\theta_2,\theta_1)$ 
and usual Euler angles $({\tilde \alpha}_i, {\tilde \beta}_i)$ 
of the $i$-th nucleus referring to the laboratory frame. 
In (b), the molecular $z'$-axis and the seven degrees of freedom 
of the system are displayed, 
where the distance $R$ is not indicated explicitly. 
The third angle $\theta_3$ is defined by 
$\theta _{3} =(\alpha_1 + \alpha_2)/2$ 
to give the whole rotation around the $z'$-axis.
}
                 \label{fig:1}
\end{figure}

The total system to be solved 
consists of two deformed nuclei interacting with each other. 
We expect the axial symmetry of the constituent nuclei and their 
constant deformations, corresponding to the states of the $K=0$ 
ground rotational band. 
We thus start with seven degrees of freedom illustrated in Fig.~1(a), 
that is, the relative vector $\boldsymbol{ R} =(R,\theta_2,\theta_1)$ 
and the Euler angles of the interacting nuclei 
$({\tilde \alpha}_1, {\tilde \beta}_1)$ 
and $({\tilde \alpha}_2, {\tilde \beta}_2)$, 
where the deformations of the constituent nuclei are taken 
to be oblate for $^{28}$Si nuclei.\cite{SiShape} 
When the constituent nuclei contact and interact strongly with each other,
their deformations in the ground state may change, i.e., 
additional deformations may be induced, 
such as those associated with the surface $\gamma$-vibrations 
or the static asymmetric ones.\cite{Davydov}
In that case, we have additional degrees of freedom, 
${\tilde \gamma}_1$ and ${\tilde \gamma}_2$, by which the nuclei rotate 
around their intrinsic $z$-axes.
(Those degrees of freedom are not illustrated in Fig.~1, for simplicity.) 
We define the rotating molecular axis $z'$ of the whole system 
with the direction of the relative vector of two interacting nuclei, 
as is shown in Fig.~1(b). 
In the molecular model, the intrinsic axes of each deformed nucleus are 
referred to the molecular frame as usual.
We introduce new Euler angles of the interacting nuclei 
in the molecular frame  $(\alpha_1,\beta_1,\gamma_1)$ and 
$(\alpha_2,\beta_2,\gamma_2)$ as in Fig.~1(b), which are related to 
$({\tilde \alpha}_i, {\tilde \beta}_i, {\tilde \gamma}_i)$ by 
\begin{equation}
 \Omega_i (\alpha_i, \beta_i, \gamma_i) 
        =   \Omega_M^{-1}(\theta_1, \theta_2) 
       \Omega_i ({\tilde \alpha}_i, {\tilde \beta}_i, {\tilde \gamma}_i),  
                                                 \qquad   i=1,\,\,  2, 
\label{eq:1}
\end{equation}
where $\Omega$'s denote Euler rotations with respective angles, 
with indications of the rotations for the each constituent nucleus 
no.~1 or no.~2 by $i$. 
To obtain the configuration of Fig.~1, it may be more useful 
to describe in terms of successive rotations as
$ \Omega_i({\tilde \alpha}_i, {\tilde \beta}_i, {\tilde \gamma}_i)
  =\Omega'_i (\alpha_i, \beta_i, \gamma_i) \Omega_M(\theta_1, \theta_2) $,
where the second rotations $\Omega'_i$ refer to the molecular axes, 
i.e., to be operated on the intrinsic axes 
which are parallel to the rotated molecular axes $(x',y',z')$.
Correspondingly $\Omega_M(\theta_1, \theta_2)$ 
appears to be multiplied from the right-hand side in this case.
Later, we introduce $\theta_3 =(\alpha_1 + \alpha_2)/2$ 
as the third Euler angle for the rotation of the total system. 
The axes $x'$ and $y'$ in Fig.~1(b) indicate the axes after the rotation 
$\Omega_M(\theta_1, \theta_2)$, while the molecular axes 
$x'(\theta_3)$ and $y'(\theta_3)$ indicate after the whole rotation 
$\Omega_M(\theta_1, \theta_2, \theta_3)$. 
Note that for the constituent nuclei with the axial symmetry, 
${\tilde\gamma}_i$ are not necessary, and we put  ${\tilde\gamma}_i =0$.
Generally we obtain $\gamma_i$ not to be zero due to the
transformations between the coordinate systems, 
but these $\gamma_i$ are physically meaningless. 
They appear in the rotational matrices, but they practically disappear 
in the inertia tensor of the total system; see Appendix B 
of Ref.~\citen{Ue93}.

%
\begin{figure}[t]
\centerline{\includegraphics[height=4.5 cm]
                             {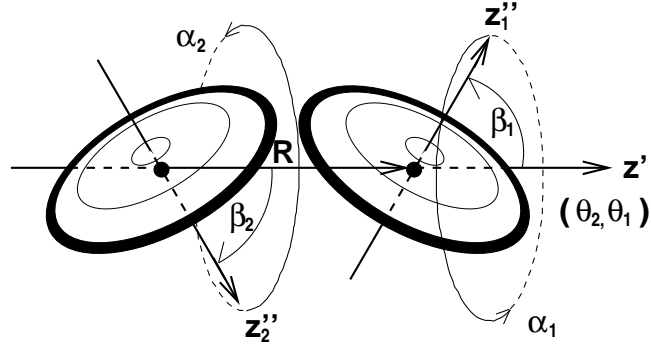}}
\caption{ The dinuclear configuration and the coordinates 
in the rotating molecular frame for an oblate-oblate system. 
The molecular $z'$-axis and the seven degrees of freedom of 
the system are displayed, where the 
$\alpha_1$- and $\alpha_2$-degrees are to be combined 
into $\theta_3 = (\alpha_1 + \alpha_2)/2$ 
and the degree of twisting $\alpha=(\alpha_1 - \alpha_2)/2$. 
The figure is the same as published in Refs.~\citen{Ue94} 
and \citen{UeSuppl}. 
%
}
                 \label{fig:2}
\end{figure}

Large deformations of the constituent nuclei may be induced 
in the deeply touching configurations of the resonances, 
then their axial symmetry of the deformations 
would be lost and the  $\gamma_i$-degrees of freedom may appear. 
In the scope of the present model, we are able to introduce 
those degrees of freedom. 
However without information on the extent of the induced 
deformations nor on the dynamical properties about the 
deformations in the touching configurations, 
such efforts would bring no fruitful result.
Later in \S4.1,  
we consider configurations with such large induced deformations, 
and investigate an example of the molecular model 
with the $\gamma$-degree of freedom in \S4.2,  
but in this section we restrict ourselves to descriptions 
without induced deformations.
This means that we assume the axial symmetry of the constituent nuclei 
with their moments of inertia $I_x = I_y$ and $I_z = 0$ 
in their principal axes, and that we start with the seven degrees 
of freedom as already mentioned. They are illustrated again in Fig.~2; 
the relative vector $(R,\theta_{2},\theta_{1})$ 
and Euler angles of the interacting nuclei in the molecular frame  
$(\alpha_1,\beta_1)$ and $(\alpha_2,\beta_2)$. 
The variables $\alpha _{1}$ and $\alpha _{2}$ are combined 
into variables $\theta _{3} =(\alpha _{1} + \alpha _{2})/2$ and 
$\alpha  =(\alpha _{1} - \alpha _{2})/2$. 
Then we have 
\begin{equation}
(q_i)= 
    (\theta _1,\theta _2,\theta _3, R, \alpha, \beta _1,\beta _2),
\label{eq:2}
\end{equation}
where $\theta_1, \theta_2$ and $\theta_3$ 
are the Euler angles of the rotating molecular frame 
with the other fours being internal variables.

\subsection{Kinetic energy of the dinuclear molecule}

Firstly we obtain an expression of the kinetic energy operator 
in terms of the above coordinates. 
We start with the classical kinetic energy of the system, which can be 
given in terms of the energies associated with the relative motion 
(the radial motion and the rotational motion of the two-ion centers) 
and the rotational motions of the two constituent nuclei,
\begin{equation}
T  =  {1 \over 2} \mu {\dot R}^2 
       + {1 \over 2} {}^t \mbox{\boldmath $\omega$}' 
                     \boldsymbol{ I}_\mu (R)  \mbox{\boldmath $\omega$}' 
       +  {1 \over 2} {}^t \mbox{\boldmath $\omega$}_1 
                     \boldsymbol{ I}_1 \mbox{\boldmath $\omega$}_1 
       +  {1 \over 2} {}^t \mbox{\boldmath $\omega$}_2 
                     \boldsymbol{I}_2 \mbox{\boldmath $\omega$}_2 , 
\label{eq:3}
\end{equation}
where  $R$ denotes the relative distance between the two-ion centers,  
$\mu$ being the reduced mass $m_1 m_2 / (m_1 + m_2)$ 
of the two nuclei with masses $m_1$ and $m_2$, 
and the c.m. energy of the total system is omitted.  
The second term of the r.h.s. of Eq.~(\ref{eq:3}) is the rotational 
energy of the two-ion centers given by the angular velocity of 
the molecular frame $\mbox{\boldmath $\omega$}'$ 
and the moment of inertia tensor  $\boldsymbol{I}_\mu (R)$.
The diagonal components $I_{11}$ and $I_{22}$ of the inertia tensor 
are $\mu R^2$,  the others being zero, which is associated with 
masses $m_1$ and $m_2$ at the relative distance $R$.
Then the expression of the rotational energy is equal to usual one, 
$ {1 \over 2} \mu R^2 ({\dot \theta_2}^2 
                     + {\dot \theta_1}^2 \sin^2 \theta_2 )  $.
The vectors  
$ \mbox{\boldmath $\omega$}_1$  and $ \mbox{\boldmath $\omega$}_2$ 
denote the angular velocities 
of the rotational motions of the two constituent nuclei, 
$ {}^t \mbox{\boldmath $\omega$}_i$   being the transpose 
of $\mbox{\boldmath $\omega$}_i$. 
The inertia tensors of the two nuclei 
$\boldsymbol{I}_1$ and $\boldsymbol{I}_2$ 
are defined in the coordinate frames of their principal axes. 
Then, they are diagonal, elements of which are determined 
by the excitation energies of the members of the ground rotational 
bands of the constituent nuclei.

At this stage, the angular velocities of the constituent nuclei 
$\mbox{\boldmath $\omega$}_i$ in Eq.~(\ref{eq:3}) 
are still those referred to the laboratory frame, 
so we have to express them in the molecular coordinate system, 
i.e., in terms of the angular velocity of the molecular frame 
$\mbox{\boldmath $\omega$}'$ and those $\mbox{\boldmath $\omega$}''_i$ 
referred to the molecular frame. 
Then we express the total kinetic energy  
as a sum of three parts, 
the total rotational energy $T_{\rm rot}$ associated with 
$\mbox{\boldmath $\omega$}'$, 
the internal kinetic energy $T_{\rm int}$ 
and the Coriolis coupling term $T_{\rm C}$, 
as follows; 
\begin{eqnarray}
    T            &=&  T_{\rm rot} + T_{\rm int} + T_{\rm C} ,  
\label{eq:4}
\\ 
\cr 
    T_{\rm rot} 
    &=& {1 \over 2} {}^t \mbox{\boldmath $\omega$}' 
                            \boldsymbol{I}_{\rm s} 
                                 \mbox{\boldmath $\omega$}' ,
\label{eq:5}
\\
    T_{\rm int}  &=&  {1 \over 2} \mu {\dot R}^2 
      +{1 \over 2} {}^t \mbox{\boldmath $\omega$}''_1 
                     \boldsymbol{I}_1 \mbox{\boldmath $\omega$}''_1 
      +{1 \over 2} {}^t \mbox{\boldmath $\omega$}''_2 
                     \boldsymbol{I}_2 \mbox{\boldmath $\omega$}''_2, 
\label{eq:6}
\\
    T_{\rm C}    &=&  {}^t \mbox{\boldmath $\omega$}' \,
 \bigl\{ {}^t R'(\alpha_1 \beta_1 \gamma_1) \, \boldsymbol{I}_1 
                                    \mbox{\boldmath $\omega$}''_1 
 + {}^t R'(\alpha_2 \beta_2 \gamma_2) \, \boldsymbol{I}_2 
                               \mbox{\boldmath $\omega$}''_2  \bigr\},
\label{eq:7}
\end{eqnarray}
where  $R'(\alpha_i \beta_i \gamma_i)$  denotes the transformation 
matrix (rotation matrix)
which connects the axes of the molecular frame 
and the principal axes of each constituent nucleus. 
The total rotational energy $T_{\rm rot}$ is the rotational energy 
of the interacting constituent nuclei as a whole system,
which rotates with the angular velocity $\mbox{\boldmath $\omega$}'$. 
The inertia tensor is given by 
\begin{equation}
    \boldsymbol{I}_{\rm s}  =     \boldsymbol{I}_\mu (R)  
   +{}^t R'(\alpha_1 \beta_1 \gamma_1) \, \boldsymbol{I}_1 
                                         R'(\alpha_1 \beta_1 \gamma_1)  
   +{}^t R'(\alpha_2 \beta_2 \gamma_2) \, \boldsymbol{I}_2 
                                         R'(\alpha_2 \beta_2 \gamma_2),   
\label{eq:8}
\end{equation}
where the first term of the r.h.s. denotes just the moments of inertia 
of two-ion centers, and the second and third terms are contributions 
from the constituent nuclei individually, though their ''rotations'' 
are already taken into account in Eq.~(\ref{eq:6}). 
The internal kinetic energy $T_{\rm int}$ is those associated 
with the orientation degrees of freedom of the constituent nuclei 
in addition to the radial motion between them. 
The last two terms of the r.h.s. of Eq.~(\ref{eq:6}) have a form of 
rotational energy, but their motions are not necessarily rotational. 
This is why the quotations are put on the word rotations above. 
Actually the nucleus-nucleus interaction favors cohesion of two 
constituent nuclei, which obstructs rotations of the constituent nuclei. 
Motions in the orientations are, therefore, not necessary to be 
rotational but would be rather confined, such as a sticking of the 
constituent nuclei and small fluctuations thereabout. 
In the sticking limit, the angular velocities 
$\mbox{\boldmath $\omega$}''_i$ are zero, 
while they are constant in free rotations. 
We, of course, anticipate intermediate states between the sticking limit 
and the rotation, i.e., fluctuations around the sticking configuration. 
For vibrational motions, for example, 
we consider fluctuations of the values of  $\mbox{\boldmath $\omega$}''_i$ 
around zero, average values of them being to be zero.

After expressing those angular velocities with time derivatives of 
the corresponding Euler angles, we obtain a classical kinetic energy 
expression ${1 \over 2} \sum g_{ij}{\dot q}_i{\dot q}_j$.
And then we quantize it by using the general formula 
for the curve-linear coordinate system, 
\begin{equation}
  {\hat T} = -{\hbar^2 \over 2}
              \sum_{ij} {1 \over \sqrt g}{\partial \over \partial q_i} 
                       \sqrt g(g^{-1})_{ij}{\partial \over \partial q_j}, 
\label{eq:9}
\end{equation}
where $g$ and $g^{-1}$ denote the determinant 
and the inverse matrix of $(g_{ij})$, respectively. 
As the classical kinetic energy consists of the three parts, 
i.e., the total rotation, the internal motions 
and their couplings, the quantum mechanical operator for the 
kinetic energy ${\hat T}$ is also given as a sum of three terms, 
${\hat T}={\hat T}_{\rm rot} +{\hat T}_{\rm int} +{\hat T}_{\rm C}.$
Naturally the term ${\hat T}_{\rm rot}$ is associated with 
the rotational variables $(\theta_1, \theta_2, \theta_3)$, 
${\hat T}_{\rm int}$ with the internal variables 
$(R, \alpha, \beta_1, \beta_2)$ and ${\hat T}_{\rm C}$ with both. 
According to the derivation, ${\hat T}_{\rm rot}$ is expressed 
by the partial differential operators of $\theta_i$. 
We combine those differential operators into angular momentum operators 
${\hat J}'_i$ referred to the molecular axes, as usual, i.e., 
\begin{equation}
  {\hat T}_{\rm rot}
            = {\hbar^2 \over 2} 
             \sum_{\scriptstyle1\le i\le3\atop\scriptstyle1\le j\le3} 
                 \mu_{ij}   {\hat J}'_i  {\hat J}'_j , 
\label{eq:10}
\end{equation}
where the matrix $\mu$ is the submatrix given later, and ${\hat J}'_i$'s 
are the angular momentum operators 
in terms of the Euler angles of the molecular frame, 
\begin{eqnarray}
  {\hat J}'_1  & =& -i\biggl(
                  -{\cos \theta_3 \over \sin \theta_2} 
                        {\partial \over \partial \theta_1}
                  +{\sin \theta_3 } 
                        {\partial \over \partial \theta_2}
                   +{\cot \theta_2  \cos \theta_3 } 
                        {\partial \over \partial \theta_3} 
                    \biggr),
\nonumber \\
  {\hat J}'_2  & =& -i\biggl(
                    {\sin  \theta_3 \over \sin \theta_2} 
                        {\partial \over \partial \theta_1}
                  +{ \cos \theta_3} 
                        {\partial \over \partial \theta_2} 
                  -{\cot \theta_2  \sin \theta_3 } 
                       {\partial \over \partial \theta_3}
                    \biggr),
\label{eq:11}
\\
  {\hat J}'_3  & =&  -i {\partial \over \partial \theta_3} .
\nonumber 
\end{eqnarray}
Here Eq.~(\ref{eq:10}) has a form just expected 
from the classical expression Eq.~(\ref{eq:5}), 
but it should be noted that the submatrix $\mu$ is not exactly equal 
to the inverse of the inertia tensor $\boldsymbol{I}_{\rm s}$, 
due to the Coriolis coupling. 
The coefficients $\mu_{ij}$ are given as follows, in terms of the 
internal variables $(R, \alpha, \beta_1, \beta_2)$,
\begin{eqnarray}
    \mu_{11} & =&  \mu_{22} = { 1 \over \mu R^2 } ,
\nonumber \\
    \mu_{12} & =& 0  ,
\nonumber \\
    \mu_{13} & =&  {1 \over 2 \mu R^2} 
                   \cos \alpha (\cot \beta_1 + \cot \beta_2 ) ,
\label{eq:12}
\\
    \mu_{23} & =&  {1 \over 2 \mu R^2} 
                   \sin \alpha (\cot \beta_1 - \cot \beta_2 ) ,
\nonumber \\
    \mu_{33} & =& {1 \over 4} 
    \Big[ \Big( {1 \over I_{\rm A} } + {1 \over \mu R^2 } \Big) 
                             {1 \over \sin^2 \beta_1} 
      +\Big( {1 \over I_{\rm B} } + {1 \over \mu R^2 } \Big)
                             {1 \over \sin^2 \beta_2} \Big] 
        - {1 \over 2 \mu R^2} 
\nonumber \\
   & \qquad \qquad &
    +  {1 \over 2 \mu R^2} \cos2\alpha  \cot\beta_1 \cot\beta_2   ,
\nonumber 
\end{eqnarray}
where $I_{\rm A}$ and $I_{\rm B}$ are the diagonal elements of the inertia 
tensors $\boldsymbol{I}_1$ and $\boldsymbol{I}_2$, respectively. 
In the definition  Eq.~(\ref{eq:11}) of the total angular momentum operators  
${\hat J}'_i$ in the body-fixed frame, we write them in terms of 
the Euler angles $\theta_i$, which appear at the same time in the coordinates 
of the relative vector $(R, \theta, \varphi)$ between the two constituent 
nuclei, as $\theta=\theta_2$ and $\varphi=\theta_1$.  
Therefore the definition may be misleading as not to be the total angular 
momentum but to be the orbital angular momentum $\boldsymbol{L}$. 
In Appendix A, we take up simple examples of the quantization 
both in the laboratory frame and in the molecular frame, 
in order to see the relations between the coordinate sets 
and the definitions for the corresponding angular momentum operators.
There, the role of the Coriolis coupling term is also clarified. 

The internal kinetic energy operator is associated with  
the variables $(R, \alpha, \beta_1, \beta_2)$, as already mentioned.  
As usual, we introduce a volume element 
$ dV = dR d\alpha d\beta_1 d\beta_2$  instead of the original 
$ dV = D dR d\alpha d\beta_1 d\beta_2$ 
with 
$ D =  \mu^{3/2} R^2 I_{\rm A} \sin\beta_1 I_{\rm B} \sin\beta_2$, 
which means that 
the wave functions are defined with the additional factor $\sqrt D$. 
Accordingly we obtain 
\begin{eqnarray}
   {\hat T}_{\rm int} & =& {\hat O}_{\rm int} + V_{\rm add}  ,  
\label{eq:13}
\\ 
%
   {\hat O}_{\rm int}  &=&   -{\hbar^2 \over 2}
    \bigg[ {1 \over \mu} {\partial^2 \over \partial  R^2} 
          + \Big( {1 \over I_{\rm A} } + {1 \over \mu R^2 } \Big) 
                {\partial^2 \over \partial \beta_1^2} 
          + \Big( {1 \over I_{\rm B} } + {1 \over \mu R^2 } \Big) 
                {\partial^2 \over \partial \beta_2^2} 
          + {2\cos 2\alpha  \over \mu R^2 } 
                {\partial^2 \over {\partial \beta_1 \partial \beta_2}} 
\nonumber \\ 
& &  + {1 \over 4}   
    \bigg\{ \Big( {1 \over I_{\rm A} } + {1 \over \mu R^2 } \Big)
                                        {1 \over \sin^2 \beta_1} 
           +\Big( {1 \over I_{\rm B} } + {1 \over \mu R^2 } \Big)
                                        {1 \over \sin^2 \beta_2}  
           - {2 \over \mu R^2}  \bigg\}
                {\partial^2 \over \partial \alpha^2} 
\nonumber \\
 &  & - {\partial \over \partial \alpha} 
       {\cos2\alpha  \over 2 \mu R^2} \cot\beta_1  \cot\beta_2   
                {\partial \over \partial \alpha} 
\nonumber \\ 
  & & 
    -{1 \over 2 \mu R^2} 
          \Big(  \cot\beta_2 {\partial \over \partial \beta_1 } 
                + \cot\beta_1 {\partial \over \partial \beta_2 } \Big)
          \Big( \sin 2\alpha  {\partial \over \partial \alpha} 
               + {\partial \over \partial \alpha} \sin 2\alpha \Big)
\,\,\bigg], 
\label{eq:14} \\ 
   V_{\rm add} &=& -{\hbar^2 \over 8} \bigg[  
                 \Big({1 \over I_{\rm A} } + {1 \over \mu R^2 } \Big) 
                    \Big( {1  \over \sin^2 \beta_1} +1 \Big) 
                +\Big({1 \over I_{\rm B} } + {1 \over \mu R^2 } \Big) 
                    \Big( {1  \over \sin^2 \beta_2} +1 \Big) 
\nonumber \\ 
 & & 
    +{2\cos 2\alpha \over \mu R^2} \cot\beta_1 \cot\beta_2 \bigg], 
\label{eq:15}
\end{eqnarray}
where $V_{\rm add}$ is the term  so-called additional potential 
due to the new volume element.

The Coriolis coupling operator ${\hat T}_{\rm C}$ 
consists of coupling operators between the variables 
$(\theta_1, \theta_2, \theta_3)$ and $(R, \alpha, \beta_1, \beta_2)$, 
i.e., 
\begin{eqnarray}
  {\hat T}_{\rm C}  & =&   { \hbar^2 \over \mu R^2 }
     \bigg[ \  i \sin\alpha \Big( - {\partial \over \partial \beta_1 }
                  + {\partial \over \partial \beta_2 } \Big) {\hat J}'_1   
           + \cos\alpha \Big(  {\partial \over \partial \beta_1 }
               + {\partial \over \partial \beta_2 } \Big) \ i{\hat J}'_2 
\nonumber \\
 &  & \qquad\qquad 
 + {i \over 2} \sin 2\alpha 
          ( - \cot\beta_2 {\partial \over \partial \beta_1 } 
          + \cot\beta_1 {\partial \over \partial \beta_2 } )
                      {\hat J}'_3    
\nonumber  \\
  & & \qquad\qquad 
     -{i \over 4}  (\cot \beta_1 - \cot \beta_2 ) 
         \bigg( {\partial \over \partial \alpha }  \cos \alpha {\hat J}'_1   
       + {\hat J}'_1  \cos \alpha  {\partial \over \partial \alpha} \bigg) 
\nonumber  \\
  & & \qquad\qquad 
      -{i \over 4}  (\cot \beta_1 + \cot \beta_2 ) 
         \bigg( {\partial \over \partial \alpha }  \sin \alpha {\hat J}'_2   
       + {\hat J}'_2  \sin \alpha  {\partial \over \partial \alpha} \bigg) 
      \bigg]
\nonumber  \\
    & \qquad &   + {\hbar^2 \over 4} 
    \bigg[ \Big( {1 \over I_{\rm A} } + {1 \over \mu R^2 } \Big) 
                                       {1 \over \sin^2 \beta_1} 
      - \Big( {1 \over I_{\rm B} } + {1 \over \mu R^2 } \Big) 
                                       {1 \over \sin^2 \beta_2} \bigg] 
          \Big(-i {\partial \over \partial \alpha} \Big)  {\hat J}'_3 , 
\label{eq:16}
\end{eqnarray}
where the derivative operators of $\theta_i$ are again rewritten 
with the angular momentum operators ${\hat J}'_i$.

For details of some relations and explicit expressions,
see Appendices of Ref.~\citen{Ue93}, for example, 
for the angular velocities in the molecular frame, 
the classical kinetic energy in terms of time derivatives 
of the Euler angles, their quantization and symmetries of the system.

In order to make the problem to be tractable, 
we start with good $K$-quantum numbers firstly, 
which is expected to be appropriate for the system of small axial asymmetry.
For the $^{24}\rm Mg +{}^{24}Mg$ system (prolate-prolate one), for example,
it is rather simple to intuitively understand, 
because stable configurations at high spins are dominantly elongated
pole-pole ones which keep axial symmetry.
However the $^{28}\rm Si+{}^{28}Si$ system (oblate-oblate one) 
favors equator-equator configurations, which do not have the axial 
symmetry intrinsically. So secondly, the effect of $K$-mixing is 
investigated later in \S4.  

At this stage, we therefore regroup the kinetic energy operator 
as follows, 
\begin{eqnarray}
  {\hat T} & ={\hat T}' + {\hat T}'_{\rm C} ,        
\label{eq:17}
\\ 
  {\hat T}' & = {\hat T}'_{\rm rot} + {\hat T}_{\rm int} ,  
\label{eq:18}
\end{eqnarray}
where ${\hat T}'_{\rm C}$ includes the Coriolis coupling 
${\hat T}_{\rm C}$ and the $K$-mixing terms in ${\hat T}_{\rm rot}$. 
Accordingly the new rotational operator ${\hat T}'_{\rm rot}$ 
has good $K$-quantum numbers.

Let's restrict our discussion to 
{\it the rotation and vibration operator} ${\hat T}'$, 
together with the interaction potential given later. 
As the kinetic energy operator ${\hat T}'$ keeps a good $K$-quantum 
number, eigenstates of the system are of a rotation-vibration type,  
\begin{equation}
   \Psi_\lambda \sim D_{MK}^J (\theta_i) 
                              \chi_K(R, \alpha,\beta_1,\beta_2). 
\label{eq:19}
\end{equation}
%
%
Now the problem to be solved is of internal motions, i.e., motions 
associated with the internal variables $(R, \alpha, \beta_1, \beta_2)$ 
which couple with each other through the kinetic energy operator 
$\hat T'$ and the interaction potential. 
For the later use in \S3,  
we define the centrifugal potential given by ${\hat T}'_{\rm rot}$ 
with specified $J$ and $K$,
\begin{eqnarray}
 T'_{\rm rot}(J,K) &= &  {\hbar^2\over2} \bigg[
     \,\,    {1\over\mu R^2} \Big\{    J(J+1)-{3\over2}K^2 
       + {1 \over 2} \cos 2\alpha \cot \beta_1 \cot \beta_2 (K^2 -1) \Big\} 
\nonumber \\ 
 &  &  
        + \Big( {1\over I } +{1\over\mu R^2} \Big) 
         \bigg(   {K^2 -1 \over 4\sin^2 \beta_1} 
                 + {K^2 -1 \over 4\sin^2 \beta_2}  -{1 \over 2} \bigg)
  \, \bigg], 
\label{eq:20}
\end{eqnarray}
where $I$ denotes the moment of inertia of the constituent nuclei, i.e.,
$I=I_{\rm A}=I_{\rm B}$, since we are interested in the system of the 
identical constituent nuclei. 
In the expression of $T'_{\rm rot}(J,K)$, 
we use the eigenvalue $K$ instead of $\hat J'_3$. 
Note that the additional potential $V_{\rm add}$ in Eq.~(\ref{eq:15}) 
is moved into $T'_{\rm rot}(J,K)$ for convenience, and similar 
terms of ${\hat T}'_{\rm rot}$ and $V_{\rm add}$ are amalgamated. 
In the numerical calculations, the value of $I$ is estimated from the 
excitation energy of the $2^+_1$ state of the $^{28}\rm Si$ nucleus.

\subsection{Nucleus-nucleus interaction potential}

For the interaction potential, we want to have an expression  
that depends on geometrical configurations of interacting nuclei, 
i.e., a potential as a function of the Euler angles of the nuclei 
in addition to the radial distance between them. 
Proximity potential appears to be one of the most suitable 
potentials,\cite{Blocki} 
but it is rather laborious to calculate it for various configurations, 
i.e., one has to find out the shortest distance 
between two curved surfaces of arbitrarily-oriented deformed nuclei 
and to calculate curvatures etc. at the point. 
Instead, we employ a folding method. Since, in the double folding model, 
nuclear densities corresponding to geometrical molecular configurations 
are directly folded with effective nucleon-nucleon interactions, 
the model easily provides an interaction potential for the present purpose, 
i.e., as a function of the collective variables. 
As for the nucleon-nucleon interaction, we employ one that is  
called {\it density dependent} M3Y(DDM3Y),\cite{DDM3YFarid} 
\begin{equation}
 v(E,\rho,r) = f(E,\rho)g(E,r),  
\label{eq:21}
\end{equation}
where $f(E,\rho)$ gives nucleon-density dependence by
\begin{equation}
  f(E,\rho) = C(E)[1+\alpha(E)e^{-\beta(E)\rho}],  
\label{eq:22}
\end{equation}
$\rho$ denoting density of nuclear matter in which
the interacting nucleons are embedded,
and $g(E,r)$ describes the original nucleon-nucleon interaction,
\begin{equation}
  g(E,r) = \Bigg[ 7999 {e^{-4r} \over 4r} 
                 -2134 {e^{-2.5r} \over 2.5r} \Bigg]   
                  + {\hat J}(E) \delta(\boldsymbol{r}) . 
%
\label{eq:23}
\end{equation}
The first term of $g(E,r)$ is M3Y potential without OPEP 
and the second term represents that from single-nucleon 
exchange, suggested by Satchler and Love.\cite{M3YSatchler} 
$E$ is the bombarding energy per nucleon, which is chosen to be as suitable
for the resonance energies ($E=3.75$MeV corresponding to 
$E_{\rm lab}=105$ MeV for ${}^{28}\rm {Si} +{}^{28}\rm {Si}$).
At a short distance of the folding potential, i.e., with highly overlapping 
densities, DDM3Y gives weakly attractive potential.
At the normal density, for example, the density-dependent factor $f(E,\rho)$ 
reduces the interaction strength by a factor about 3/4, compared with the 
original $g(E,r)$, while it is enhanced by a factor 1.2 
at the half density, i.e., at the contact region.

The folding-model potential, however, is considered to be accurate only 
in the tail region of the nucleus-nucleus interaction. 
In the region where nuclear-density overlap goes beyond the normal density, 
it is considered to be not accurate enough.
Hence, in addition to the folding potential with the nucleon-nucleon 
interaction, we introduce a phenomenological repulsive potential, 
which would originate from the effects of the Pauli principle 
among nucleons belonging to the interacting nuclei respectively, 
or from compression effects due to the overlapping density. 
We estimate strength of the repulsive potential due to the compression 
of nuclear density, from the equation of state of nuclear matter, 
i.e., from the binding energy as a function of nuclear density. 
One may think that the picture of the density overlap is doubtful 
in low energy, but the folding model does not take into account 
density redistribution, so it is consistent to account higher densities 
in the overlapping region. Anyhow, what we are interested in is the dynamics 
of two interacting nuclei in high spins where strong centrifugal forces 
dominate. Therefore, the long-range part of interactions is crucially 
important, but not the short-range part, 
which is treated more or less in a phenomenological way.

The folding potential is defined as usual, 
\begin{eqnarray}
  U(\boldsymbol{R})  & =& \int d\boldsymbol{r}_1  \,
                        \int d\boldsymbol{r}_2   \,
                         {\rho}_1 (\boldsymbol{r}_1)  
                         {\rho}_2 (\boldsymbol{r}_2)  
                          v (\boldsymbol{r}_{12})    , 
\nonumber \\
   \boldsymbol{r}_{12} & =& 
            \boldsymbol{R} +\boldsymbol{r}_2 -\boldsymbol{r}_1 , 
\label{eq:24}
\end{eqnarray}
where $\boldsymbol{R}$ is the relative vector between the interacting 
nuclei and $\boldsymbol{r}_i$ are referred to the centers of the nuclei,  
respectively. 
The long-range attractive part of the interaction potential in the 
molecular frame $V_{\rm attr}$ is obtained from $U(\boldsymbol{R})$ 
by taking the vector $\boldsymbol{R}$ to be parallel to the $z'$-axis 
and by taking orientations of the density distributions of the 
constituent nuclei with respect to the molecular frame. 
By using Fourier transformation, 
\begin{eqnarray}
   V_{\rm attr}    &= & {1 \over 2\pi^2} \sum_{lm} i^{-l} 
                                           Y_{lm}( \hat{\hbox{\mbf R}})
                    \int dk k^2 j_l (kR) 
                     \int d{\hat k}  Y^*_{lm}( \hat{\hbox{\mbf k}})
                        \widetilde  v(\hbox{\mbf k}) 
                        \widetilde \rho_{1}(\hbox{\mbf k}) 
                        \widetilde \rho_{2}(- \hbox{\mbf k}) ,
\label{eq:25}
\\ 
      \widetilde  v(\hbox{\mbf k}) 
   &= &    \int d\hbox{\mbf r} e^{i \hbox{\mbf k} \hbox{\mbf r}} v(r) 
                    =   4\pi \int dr r^2 j_0 (kr) v(r)   ,    
\label{eq:26}
\\ 
      \widetilde \rho_{i}(\hbox{\mbf k}) 
    &= &  \int d\hbox{\mbf r}' e^{i \hbox{\mbf k} \hbox{\mbf r}'} 
                                          \rho_{i}(\hbox{\mbf r}') .
\label{eq:27}
\end{eqnarray}
The density distribution $\rho_{i}(\hbox{\mbf r}')$ 
in the molecular frame is related to that in the body-fixed frame, 
i.e., to that in the principal axes of the constituent nucleus; 
by Euler rotations, 
$ \rho_i (\hbox{\mbf r}')
           = \rho_i^B (\hbox{\mbf r}''_i)
           = \hat{\hbox{\calg R}} (\alpha_i \beta_i \gamma_i) 
                                      \rho_i^B (\hbox{\mbf r}') $, 
where $\rho_i^B (\hbox{\mbf r}''_i)$ is the density distribution 
in the principal axes and therefore 
\begin{equation}
 \rho_i^B (\hbox{\mbf r}''_i) = 
     \sum_{l=even} \rho_l(r''_i) Y_{l0}(\hat{\hbox{\mbf r}}''_i)
\label{eq:28}
\end{equation}
with the assumed axial symmetry of each constituent nucleus. 
So the Fourier transform $\widetilde \rho_{i}(\hbox{\mbf k})$ 
is given with the Euler angles included as parameters, 
\begin{eqnarray}
      \widetilde \rho_{i}(\hbox{\mbf k}) 
    &= &  \sum_l i^l \,  \widetilde \rho_{l}(k) 
           \sum_{m'} D^{l*}_{m' 0}(\alpha_i \beta_i \gamma_i) 
                                Y_{lm'}(\hat{\hbox{\mbf k}}) ,
\label{eq:29}
\\ 
     \widetilde \rho_l(k) 
    &= &  4\pi \int  dr r^2 j_l (kr) \rho_l (r) .
\label{eq:30}
\end{eqnarray}
Inserting Eq.~(\ref{eq:29}) with $i=1$ and $2$ into Eq.~(\ref{eq:25}), 
we obtain the final form of the interaction potential 
as a function of the internal variables 
$(R, \alpha, \beta_1, \beta_2)$ in the following, 
\begin{eqnarray}
     V_{\rm attr}(R ,\alpha,\beta_1 ,\beta_2) 
      &=& \sum_{l' l'' l}  {(2 \pi)}^{-3} i^{l'-l''-l} 
                        {\hat l'} {\hat l''} (l' l'' 0 0 \mid l 0) 
\nonumber \\
        &  & \qquad
             \times     F_{l' l'' l} (R) 
                        G_{l' l'' l} (\alpha,{\beta_1},{\beta_2}), 
\nonumber \\
   F_{l' l'' l}(R) &=& \int dk k^2 j_l (kR) 
                      \widetilde  v(k)  \widetilde \rho_{l'}(k) 
                                        \widetilde \rho_{l''}(k) ,  
\label{eq:31}
\\ 
    G_{l' l'' l}(\alpha,{\beta_1},{\beta_2}) 
&=& \sum_{m \ge 0} (-1)^m (2 - \delta_{m 0}) (l' l'' m -m  \mid l 0)
\nonumber \\ 
           &  & \qquad \times  \cos (2m\alpha)  
                     d^{l'}_{m 0}({\beta_1}) 
                     d^{l''}_{m 0}({\beta_2}) .                  
\nonumber 
\end{eqnarray}
It should be mentioned here that $\gamma_i$ does not appear in the 
final expression due to the $D$-function with one magnetic quantum 
number being zero which originates from the axially-symmetric density 
distribution in Eq.~(\ref{eq:28}), 
and that $\alpha_1$ and $\alpha_2$ are combined 
into $2\alpha = \alpha_1 - \alpha_2$ due to the fact 
that the vector $\boldsymbol{R}$ is parallel to $z'$-axis, i.e., 
the magnetic quantum number associated with $\boldsymbol{R}$ 
is zero. The Coulomb interaction is also folded, together with 
nuclear interaction $v(E, \rho,r)$ of Eq.~(\ref{eq:21}).

We assume the density profile of 
$\rho_i^B(\hbox{\mbf r}''_i)$ to be the Fermi distribution with 
$\rho_i^B(\hbox{\mbf r}''_i)  = 
 \rho_0 / [1+\exp \{(r''-R_{\rm N}(\hbox{\mbf r}''_i))/a_{\rm N} \}]$, 
$R_{\rm N}(\hbox{\mbf r}''_i)$ denoting the radius of 
the deformed nucleus. 
As for the deformation of the constituent $^{28}\rm Si$ nuclei,
the existence of the hexadecapole deformation 
($\beta_4=0.18 \pm 0.02$) is suggested from coupled-channel analyses 
for the elastic and inelastic neutron scattering.\cite{Howell} 
Therefore, we take the radius of each nucleus as 
$R_{\rm N}(\hbox{\mbf r}''_i) =  r_{0} A_i^{1/3} 
[1 + \beta_{\rm Q} Y_{20}(\hat{\hbox{\mbf r}}''_i) 
   + \beta_{\rm H} Y_{40}(\hat{\hbox{\mbf r}}''_i)  ] $ 
including two parameters $\beta_{\rm Q}$ and $\beta_{\rm H}$ 
for the deformations, the values of which are determined 
to be $-0.46$ and $0.22$, respectively, according to the suggested 
value for the ratio $\beta_{\rm Q} / \beta_{\rm H}$      
and their magnitudes adjusted with the $B(E2)$ value 
of the ground-rotational band of ${}^{28}\rm {Si}$.\cite{Endt} 
The value of $r_{0}$ is taken to be $1.03$fm       
from the textbook of Bohr-Mottelson,\cite{BohrTEXT1} and $a_{\rm N}$ 
to be $0.48$fm to reproduce the RMS radius of the ground state.

Next, we proceed to {\em the effect of density overlap} in the inner region,
where the folding potential is not expected to be adequate. 
An overlapping of the densities brings about a higher nuclear density 
than the normal one, which gives rise to a binding energy loss of the 
interacting system in addition to the attractive folding potential.
We take into account the effect as a repulsive potential to be
added to the folding one given in Eq.~(\ref{eq:24}).
The volume with higher density depends on the configurations of the 
constituent nuclei, especially on their relative distance.
Actually, the overlapping of two nuclei produces nuclear density from zero
to twice of the normal density.
An accurate calculation of the effect, therefore, is rather laborious.
We propose a simple approximate way.
If we assume the density profile to be of sharp cut-off or with a very small 
diffuseness, an overlapping volume has always twice of the normal density.
So the short-range repulsive effect is expected to be proportional 
to the overlapping volume, and it would be simulated by a potential 
\begin{equation}
V_{\rm rep}(R,\alpha,\beta_1,\beta_2) = 
           V_{\rm P} \int \delta (\hbox{\mbf r}_{12})
\rho'_1(\hbox{\mbf r}_1)\rho'_2(\hbox{\mbf r}_2)
d \hbox{\mbf r}_1 d\hbox{\mbf r}_2,
\label{eq:32}
\end{equation}
where the primes on the densities indicate Fermi distributions 
with a small diffuseness $a_{\rm P}$. 
The strength $V_{\rm P}$ of $V_{\rm rep}$ is chosen in the following, 
referring to the Equation of State (EOS) of nuclear matter. 
Thus the total interaction potential is given by 
\begin{equation}
V_{\rm int} = V_{\rm attr} + V_{\rm rep} , 
\label{eq:33}
\end{equation} 
where $V_{\rm attr}$ denotes the usual folding potential 
defined in Eq.~(\ref{eq:24}). 
The repulsive potential looks like a folding potential of the zero-range 
interaction, but has the primed densities instead of the normal density 
distributions. Of course, we can utilize a merit of the form 
of Eq.~(\ref{eq:32}) in the actual calculations.

To determine the strength $V_{\rm P}$, we use EOS of the nuclear matter, 
i.e., a binding energy loss per nucleon $\Delta \varepsilon$ for twice 
of the normal density which is calculated under

\begin{wrapfigure}{r}{6.6cm}
\centerline{\includegraphics[height=5.5 cm]   
                             {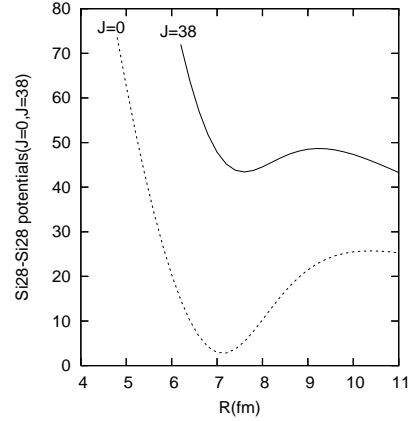}}
\caption{ 
The radial forms of the effective potentials 
in the parallel equator-equator configuration of 
$^{28}\rm Si+{}^{28}Si$,  
$V_{JK}(R)= V_{\rm int}(R,\pi/2, \pi/2,\pi/2) + T'_{\rm rot}(J,K)$
for spins $J=0$ and $J=38$ with $K=0$ are shown.
}
                  \label{fig:3}
\end{wrapfigure}

\noindent
the condition of complete overlap at the $R=0$ limit. 
Without Coulomb energy the value of $\Delta \varepsilon$ can be 
taken to be $7 \sim 11$MeV\cite{Takatsuka} 
from the values of the nuclear compression modulus 
$K_{\infty}= 180 \sim 240$MeV,\cite{Blaizot}  
which is suggested by the experiments on giant monopole resonances.
Hence the values $a_{\rm P} =0.25$fm and $V_{\rm P}=330{\rm MeV fm}^3$
are obtained to reproduce $\Delta \varepsilon = 9$MeV 
in the ${}^{28}\rm Si +{}^{28}Si$ system.
Radial forms of the folding potential are shown in Fig.~3, for the stable 
geometrical configurations (parallel equator-equator ones, see the next 
section), where the effective potentials for $J=0$ and $J=38$ are displayed. 
Details of the folding potentials about their dependences on $a_{\rm P}$ 
and $V_{\rm P}$ are already discussed in Ref.~\citen{UeSuppl}, 
where the effects of hexadecapole deformation in $^{28}\rm Si$ nuclei 
are also investigated. 

\section{Dinuclear structures of the $^{28}$Si$+^{28}$Si system}

\subsection{Stable configuration of the oblate-oblate system with high spins}

\begin{figure}[b]
\centerline{\includegraphics[height=5.0 cm]
                             {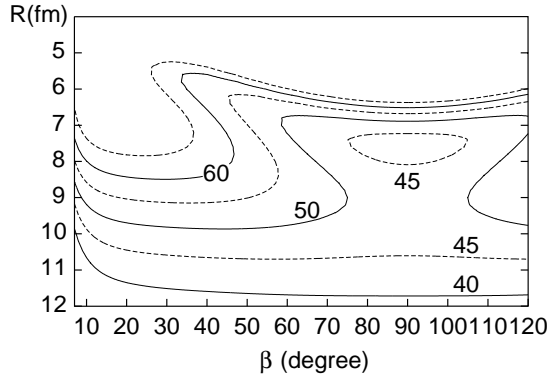}}
\caption{
%
The effective potential energy $V_{JK}$ for the $^{28}\rm Si+{}^{28}Si$ 
system with $J=38$ and $K=0$ is displayed,  
for the $R-\beta(\beta_1=\beta_2)$ degrees at $\alpha=\pi/2$. 
A local energy minimum exists at $R=7.6$fm. 
Contours are in MeV. 
The figure is essentially the same as the energy contour map 
in Refs.~\citen{Ue94} and \citen{UeSuppl}. 
}
                    \label{fig:4}
\end{figure}

In order to know dynamical aspects of multi-dimensional internal 
motion, we calculate the effective potential with specified 
spin $J$ and $K$, defined as follows:
\begin{equation}
 V_{JK}(R, \alpha, \beta_1, \beta_2)  =  
         V_{\rm int}(R, \alpha, \beta_1, \beta_2)  
        + T'_{\rm rot}(J,K) .
\label{eq:3.1}
\end{equation}
In Fig.~4, an $R-\beta(\beta_1=\beta_2)$ energy surface, i.e.,
$V_{JK}(R,\pi/2, \beta ,\beta)$ is displayed for $J=38$ and $K=0$.  
We find a local minimum point 
at $\beta_1=\beta_2=\pi/2$ and $R=7.6$fm, 
namely, at the equator-equator(E-E) configuration, 
with a rather deep potential well around the equilibrium. 
We mention that the some expressions, 
numerical results and figures in this section 
are already published in Refs.~\citen{Ue94} and \citen{UeSuppl}, 
but we show them for explanation.

\begin{wrapfigure}{r}{6.6cm}
\centerline{\includegraphics[height=5.5 cm]   
                             {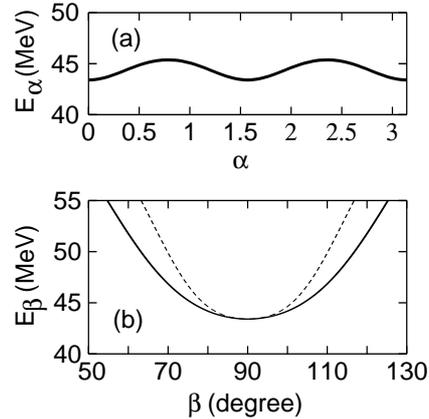}}
\caption{
(a) the $\alpha$-dependence of the effective potential $V_{JK}$ 
with $J=38$ and $K=0$, 
for the E-E configuration at $R=R_{\rm e}=7.6$fm. 
(b) $V_{JK}$ versus $\beta=\beta_1=\beta_2$ 
at $\alpha=0$ and at $\alpha=\pi/2$, 
which are displayed by dashed and solid lines, respectively.
The figure is the same as published in Refs.~\citen{Ue94} 
and \citen{UeSuppl}. 
}
                     \label{fig:5}
\end{wrapfigure}

%
In Fig.~5(a), the $\alpha$-dependence of $V_{JK}$ 
in the E-E configuration at the equilibrium distance is shown. 
(Note that our definition for the domain of the variables is 
$0 \le \alpha < \pi$ and $0 \le \beta_1, \beta_2 \le \pi$.)  
We find that the $\alpha$-dependence is extremely weak.  Another
point is that we have two local minima at $\alpha=0$ and $\pi/2$. 
Those two configurations are, however, exactly the same, 
namely, {\em parallel} E-E configuration 
($z''$-axes of the constituent nuclei are parallel).
Therefore it is necessary to impose symmetry on the wave functions. 
In Fig.~5(b), $\beta$-dependences of $V_{JK}$ with $\beta_1=\beta_2$
are compared between at $\alpha=0$ and at $\alpha=\pi/2$, 
where solid line is for $\alpha=\pi/2$ 
(the cross section of Fig.~4 at $R=R_{\rm e}=7.6$fm) 
and dashed line for $\alpha=0$.  (Note that configurations
with $\beta_1=\beta_2 \ne \pi/2$ at $\alpha=0$ are not the same as
those with the same $\beta_i$-values at $\alpha=\pi/2$, but are the
same as those with $\beta_1=\pi -\beta_2$ at $\alpha=\pi/2$.) 
The $\beta$-well at $\alpha=\pi/2$ is seen to be rather shallow, 
compared with that at $\alpha=0$. Hence, despite the weak
$\alpha$-dependence of $V_{JK}$ in the E-E configuration, 
we have significantly $\alpha$-dependent restoring force 
for $\beta$-motions around the E-E configuration.


\vskip 8 true mm

\subsection{Harmonic approximation and normal modes with a specified $K$}

In order to solve normal modes for four variables 
$(R,\alpha,\beta_1,\beta_2)$, we expand $V_{JK}$ into a quadratic 
form for $R$, $\beta _{1}$ and $\beta _{2}$, 
at the equilibrium E-E configuration, 
while for $\alpha$ we keep its dependence exactly in terms of 
$\cos (2m\alpha)$ series, such as those given in the interaction potential 
of Eq.~(\ref{eq:31}).   
Then the effective potential is expressed as 
\begin{eqnarray}
 V_{JK}(R, \alpha, \beta_1, \beta_2)  =  
  &   V_{JK}(R_{\rm e}, \alpha,{\pi \over 2},{\pi \over 2}) 
      + {k_R\over2}(R-R_{\rm e})^2  
\nonumber \\ 
 & +{1\over2}k_{\beta}^{11}(\alpha) \Delta \beta^2_1  
   +{1\over2}k_{\beta}^{22}(\alpha) \Delta \beta^2_2 
\nonumber \\
&       + k_\beta^{12}(\alpha) \Delta \beta_1  \Delta \beta_2 
         + (higher \,\, order), 
\label{eq:3.2}
\end{eqnarray}
where $\Delta \beta_i$ denotes $\beta_i - \pi/2$.
$k_{\beta}^{ij}(\alpha)$ denotes the second derivative 
$ {\partial^2 V_{JK}/  \partial \beta_i \partial \beta_j} $,
$k_{\beta}^{11}(\alpha)$ being equal to $k_{\beta}^{22}(\alpha)$. 
Although $k_{\beta}^{ij}(\alpha)$ is 
a coefficient of  $\Delta \beta_i  \Delta \beta_j$  in the expansion, 
it is a function of $\alpha$, i.e., 
we take into account $\alpha$-dependence of the coefficient, 
in addition to the $\alpha$-dependence of 
$V_{JK}(R_{\rm e}, \alpha,{\pi \over 2},{\pi \over 2})$. 
As $k_{\beta}^{11}(\alpha)$ consists of 
$\cos (2m\alpha)$ series with $m=even$ including zero,
the major part is a constant $k_0$ from $m=0$. 
We write 
$k_{\beta}^{11}(\alpha) = k_{\beta}^{22}(\alpha) = k_0 + k_2(\alpha)$, 
$k_2(\alpha)$ being a sum of contributions 
from terms with $m=even >0$. 

We introduce new coordinates in order to eliminate cross products 
of $\beta_1$ and $\beta_2$ 
both in ${\hat T}_{\rm int}$ and in the quadratic expansion of $V_{JK}$. 
The new variables describe {\it butterfly}  and {\it anti-butterfly} 
modes as follows:
\begin{eqnarray}
\beta_+ =& ( \Delta \beta_1 + \Delta \beta_2 )/  \sqrt 2  
                     = (\beta_1 + \beta_2 -\pi )/ \sqrt 2,
\nonumber \\
 \beta_- =& 
 (\Delta \beta_1 - \Delta \beta_2) /  \sqrt 2 
                     = (\beta_1 - \beta_2 ) / \sqrt 2 .
\label{eq:3.3}
\end{eqnarray}
Furthermore the inertia masses of three variables 
$\alpha, \beta_+$ and $\beta_-$ 
are approximated by the values given at the E-E configuration. 
Combining the kinetic energy operator and the expanded effective 
potential, the total hamiltonian is given as follows:
\begin{eqnarray}
 & H  =  H_0  + T'_{\rm C}+ (higher \,\, order),   
\label{eq:3.4}
\\ 
\nonumber\\
 & H_0  =  H_R + H_{\rm angl}(\beta_+,  \beta_-, \alpha),            
\label{eq:3.5}
\\ 
\nonumber\\
   & \qquad 
     H_R = -{\hbar^2\over 2\mu}{\partial^2\over\partial R^2}
                + {k_R\over2} (R-R_{\rm e})^2 ,
\label{eq:3.6}
\\ 
\nonumber\\
  &  \qquad 
     H_{\rm angl}(\beta_+,\beta_-,\alpha) = 
             H_+(\beta_+, \alpha) + H_-(\beta_-, \alpha)
\nonumber \\
  &\qquad\qquad\qquad\qquad\qquad
           -{\hbar^2 \over 4I} {\partial^2\over\partial\alpha^2}  
           + V_{JK}(R_{\rm e}, \alpha,{\pi \over 2},{\pi \over 2}),
\label{eq:3.7}
\\ 
 & \qquad \qquad  H_\pm(\beta_\pm, \alpha)  =  -{\hbar^2\over2} 
   \Big( {1\over I} +{1\pm \cos2\alpha \over\mu R_{\rm e}^2} \Big) 
                   {\partial^2\over\partial\beta_\pm^2}   
                        + {k_\pm(\alpha) \over2}  \beta^2_\pm , 
\label{eq:3.8}
\end{eqnarray}
where $+$ or $-$ sign of $\pm$ in Eq.~(\ref{eq:3.8}) corresponds to 
the $\beta_+$ and $\beta_-$ degrees of freedom, respectively, 
with 
$ k_+(\alpha) = k_0+k_2(\alpha) + k_\beta^{12}(\alpha) $
and 
$ k_-(\alpha) = k_0+k_2(\alpha) - k_\beta^{12}(\alpha) $.

Now we solve the Schr\"odinger equation with the hamiltonian $H_0$ 
for the internal four degrees of freedom, which is separated into 
two parts. One is the hamiltonian $H_R$ for the radial motion 
and nothing but that of a simple one dimensional harmonic oscillator. 
Another is $H_{\rm angl}$ for the angle variables 
$\alpha, \beta_+$ and $\beta_-$, which is also {\it almost separable} 
into $H_+$ of $\beta_+$, $H_-$ of $\beta_-$ 
and the remaining hamiltonian for $\alpha$. 
$H_+$ and $H_-$ again represent harmonic oscillators, 
although the masses and the restoring forces depend on $\alpha$. 
Hence we analytically obtain wave functions for $H_\pm$ 
and their energy quanta $\hbar \omega_\pm$ with the frequencies  
\begin{equation}
\omega_\pm = \sqrt {k_\pm(\alpha) 
           \bigg( {1\over I} +{1\pm \cos2\alpha \over\mu R_{\rm e}^2} 
                \bigg) } \,\,\, .
\label{eq:3.9}
\end{equation}
Taking into account those vibrational energies from the $\beta$-degrees 
of freedom, we introduce a reduced potential for the $\alpha$-motion, 
and obtain the Schr\"odinger equation for the $\alpha$-motion 
as follows: 
\begin{equation}
      \bigg[ -{\hbar^2 \over 4I} {\partial^2\over\partial\alpha^2}  
        +V_{JK}(R_{\rm e},\alpha, {\pi \over 2}, {\pi \over 2}) 
        +E^\beta_{n_+,n_-}(\alpha) 
 \bigg] \phi(\alpha) 
  = E_{\rm angl} \phi(\alpha) ,  
\label{eq:3.10}
\end{equation}
where $E^\beta_{n_+,n_-}(\alpha)$ denotes vibrational energy  
$(n_+ +1/2)\hbar\omega_+ + (n_- +1/2)\hbar\omega_-$  from $H_+ + H_-$, 
added as a part of the reduced potential. 
Note that, in order to obtain analytic form of $\hbar\omega_\pm$ 
in $\cos (2m\alpha)$ series, 
we expand square root in Eq.~(\ref{eq:3.9}) supposing  
$\omega_0 = \sqrt {k_0 ({1 / I} + {1 / \mu R_{\rm e}^2}) }$ 
to be the leading term.  Accordingly, we consider a solution 
$\phi(\alpha)$ of Eq.~(\ref{eq:3.10}) to be described by cosine and 
sine functions of $\alpha$, i.e., Fourier series, as the reduced 
potential 
$ V_{JK}(R_{\rm e},\alpha, {\pi \over 2}, {\pi \over 2}) 
+E^\beta_{n_+,n_-}(\alpha)$ 
is described by a sum of $\cos (2m\alpha)$.
Then Eq.~(\ref{eq:3.10}) is reduced to a secular equation,  
which is easily solved. Thus the eigenenergy of the system is given 
as follows, specified by the quantum numbers 
$(n, n_+, n_-, K, (\nu, \pi_\alpha))$, 
\begin{eqnarray}
 E^J(n,n_+,n_-,K,(\nu, \pi_\alpha)) = & E_0(R_{\rm e})
        +{\hbar^2 \over 2} \bigg[{J(J+1)-K^2 -1 \over \mu R_{\rm e}^2} 
                                    + {K^2 -2 \over 2I} \bigg]  
\nonumber \\ 
     & + \Big( n+{1\over2} \Big) \hbar \omega_R 
\nonumber \\
   &  + \big( n_+  +n_-  + 1 \big)  \hbar \omega_0 
        + E_\nu^\alpha(\pi_\alpha) ,          
\label{eq:3.11}
\end{eqnarray}
where $\nu$ denotes a dominant frequency of the $\alpha$-motion 
with $\pi_\alpha$ for the parity concerning the reflection at the 
equilibrium of $\alpha= \pi /2$. The first and second terms in the 
r.h.s. of Eq.~(\ref{eq:3.11}) are constant energies from the 
interaction potential and the centrifugal energy included in $V_{JK}$ 
at the equilibrium, respectively. 
$( n_+   +n_- +  1 )  \hbar \omega_0$ 
and $ E_\nu^\alpha(\pi_\alpha)$ are the vibrational energies 
for the $\beta$-motions without the $\alpha$-dependence 
and the energy for the $\alpha$-motion, respectively.

%
\begin{wraptable}{r}{\halftext}
  \caption{Molecular states allowed by the selection rule, 
           specified by the $K$-quantum number, the $\beta$-vibrational 
           quanta $(n_+, n_-)$ and $\nu$ for the $\alpha$-motion. }
        \label{table:1}
        \begin{center}
          \begin{tabular}{ccc} \hline \hline
           $K$       & $(n_+, n_-)$           & $\nu$ \\ \hline
            0        &  (0,0), (2,2)          &  0, 4, 8,...  \\
            0        &  (2,0), (4,0), (4,2)   &  0,2,4,6,... \\ \hline
            2        &  (0,0), (1,1), (2,2)   &  2, 6,10,...   \\
            4        &  (0,0), (1,1), (2,2)   &  0, 4, 8,...    \\
            2, 4     &  (2,0), (4,0), (4,2)   &  0,2,4,6, ... \\ \hline
            1, 3     &  (1,0), (2,1)          &  1, 3, 5,...  \\ \hline
          \end{tabular}
        \end{center}
\end{wraptable}

There is a selection rule $K \pm \nu = even$ for the $\alpha$-motion. 
Because of the parity and boson symmetries, 
$n_+$ can be taken to be larger than or equal to $n_-$.  
For the $\beta$-vibrational modes, 
we have a rule 
$(-1)^{n_+ + n_-} = (-1)^K$ due to the symmetry of each constituent 
nucleus under the space inversion. 
Details of the symmetries of the molecular system, the wave functions 
and the selection rule are given in Appendix B. 
The resultant states are summarized in Table~I. 
Note that the eigenfunction of the $\alpha$-motion is not 
necessarily the internal rotation specified with a single $\nu$-value, 
and that mixing over allowed $\nu$-states is expected.

\begin{figure}[htb]
 \hfill
 \parbox[b]{100mm}{
   \begin{center}
     \includegraphics[scale=1.0]{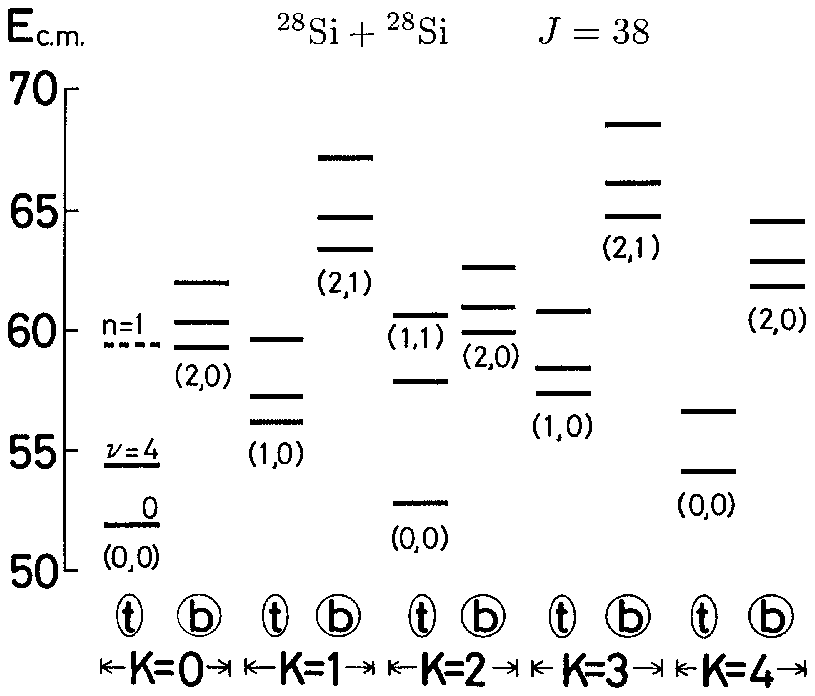} 
   \end{center}
 }
 \hfill
 \parbox[b]{35mm}{
   \begin{center}
     \includegraphics[scale=0.2]{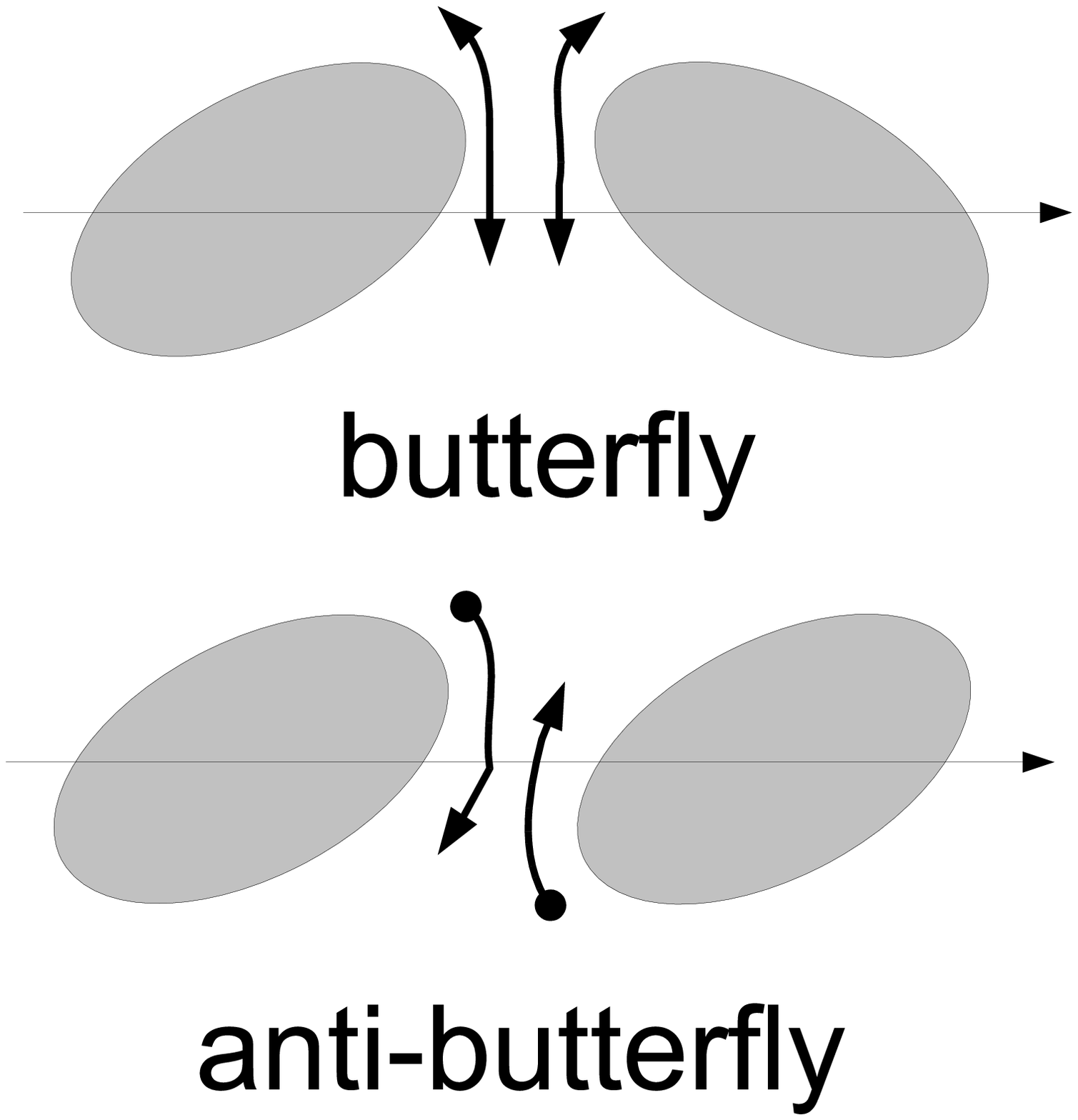}
   \end{center}
 }
 \hfill
%
\caption{  
Molecular normal modes for the $^{28}\rm Si+{}^{28}Si$ system 
for $J=38$. The quantum states are specified by 
$(n, n_+, n_-, K, (\nu,\pi_\alpha))$, 
where $n=0$ is given except for one level($n=1, \nu \sim 0$) 
displayed with dashed line. The quanta $(n_+, n_-)$ of the 
$\beta$-motions are given below the levels, and $K$ at the bottom. 
$(t)$ and $(b)$ marks assigned in the lower part of the figure 
indicate the twisting rotational mode and the butterfly modes, 
respectively.
Also given above some levels with $K=0$ are dominant values 
of the quantum number $\nu$ for the $\alpha$-motion. 
On the right-hand side, butterfly and anti-butterfly motions 
are illustrated.
The figure on the l.h.s. is the same as published 
in Refs.~\citen{Ue94} and \citen{UeSuppl}. 
}
                    \label{fig:6}
\end{figure}

In Fig.~6, molecular normal modes of  $^{28}\rm Si+{}^{28}Si$ 
with spin 38 are displayed, classified with the $K$-quantum numbers. 
The twisting-mode excitations associated with the $\alpha$-degree 
are obtained, and indicated by marked ({\bf t}) at the bottom. 
Also given above each level with $K=0$ and ({\bf t}) 
is the dominant quantum number of $\nu$ for the $\alpha$-motion, 
which means the $\alpha$-motion is approximately described 
by a single term $\cos \nu \alpha$.  
The butterfly and anti-butterfly vibrational modes are indicated 
by marked ({\bf b}). 
A pair of quanta $(n_+, n_-)$ is given below the levels. 
All those are due to the internal degrees of freedom, 
i.e., intrinsic excitations. 
Apparently the $K$-excitation and the twisting rotational mode 
appear to be lower than the $\beta$-vibrational modes. 
The excitation energy for $K=2$ is very small, smaller than 1MeV, 
and even those for $K=4$ or $\nu=4$ are smaller than 3MeV. 

In Fig.~7(a), a few examples of wave functions for the 
$\alpha$-motion are exhibited, where the $\beta$-modes are 
in the zero-point oscillation (dashed line) 
or the 2-quanta excitation of butterfly (solid line). 
We see that, with zero quanta for the $\beta$-modes, 
the amplitude is wriggling around the value of the unit,
the equilibria $\alpha=0$ and $\pi/2$ being slightly favored. 
(With exact $\nu=0$ we have a constant behavior. 
Weak $\nu =4$ mixing exists.)
With 2 quanta for the butterfly mode, however, we find surprisingly 
strong concentration around the equilibrium of $\alpha=\pi/2$. 
In Fig.~7(b), we inspect the reduced $\alpha$-potential for 
quanta $(2,0)$. 
$\,$ Compared with the potential for $(0,0)$,
we find that the minimum at $\alpha=0$ disappears, 
and the potential well at $\,$ $\alpha=\pi/2$ 
is extended to wider region, 
which sustains the localization of 
the amplitude. 
 $\,$ One

\begin{wrapfigure}{r}{6.6cm}
\centerline{\includegraphics[height=7.8 cm]   
                             {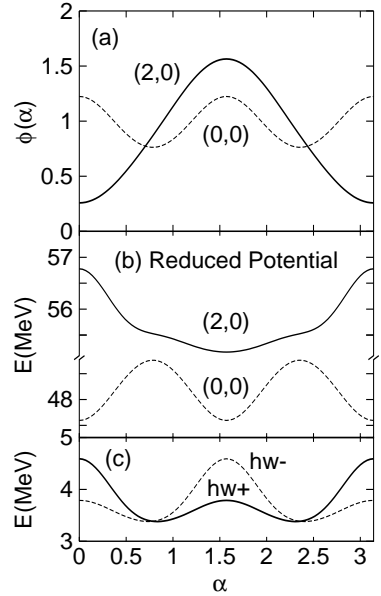}}
\caption{ (a) Wave functions 
for the $\alpha$-motion for  $J=38$ and $K=0$.
Those with the zero-point oscillation $(0,0)$ and the butterfly 
excitation $(2,0)$ for the $\beta$-degrees of freedom 
are displayed, respectively.
(b)  The reduced $\alpha$-potential 
$V_{JK}(R_{\rm e},\alpha, \pi /2, \pi /2) 
        +E^\beta_{n_+,n_-}(\alpha) $.
(c) $\alpha$-dependences of the $\beta$-energy quanta 
$\hbar \omega_\pm$.
The figure is the same as published in Refs.~\citen{Ue94} 
and \citen{UeSuppl}. 
}
                    \label{fig:7}
\end{wrapfigure}

\noindent
may wonder why the difference between $\alpha=0$ and $\pi/2$ exists. 
The reason is as follows: 
at $\alpha=0$, due to the definition of $\beta_\pm$, 
$\beta$-motion with $(n_+, n_-)=(2,0)$ does not imply 
butterfly excitation but anti-butterfly one with 2 quanta. 
Such a characteristic of the $\beta_\pm$ coordinates gives larger 
excitation energy for $(2,0)$ at $\alpha=0$ than at $\alpha=\pi/2$. 
In Fig.~7(c), the energy quanta $\hbar\omega_\pm$ versus $\alpha$ 
are shown, where we are able to confirm the point. 
Returning back to the dinuclear configuration, 
for a configuration with $\beta_1=\beta_2 < \pi/2$, 
for example, we obtain a butterfly one at $\alpha=\pi/2$, 
such as displayed in Fig.~2, 
while at $\alpha=0$ we obtain an anti-butterfly one 
with the same values of $\beta_i$. 
Hence the localization around $\alpha=\pi/2$, seen in Fig.~7(a), 
indicates nothing but a realization of a physical butterfly excitation. 
Thus, we are able to classify the levels in Fig.~6 into two groups, 
i.e., the twisting mode and the butterfly (or anti-butterfly) mode, 
respectively. Some examples of the wave functions for the normal 
modes are explicitly given in Appendix C.

\section{Rotational motion at extremely high spins 
            with triaxial deformation}

%
\begin{wrapfigure}{r}{6.6cm}
\centerline{\includegraphics[height=4 cm]  
                             {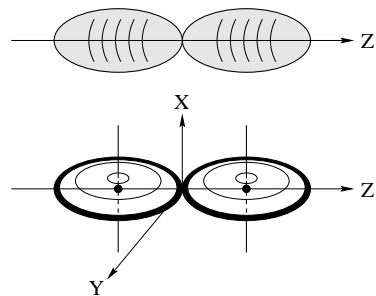}}
\caption{
Equilibrium configurations of two di-nuclear systems. 
$\,$ The upper portion 
is for $^{24}\rm Mg-^{24}Mg$ 
$\,$ and $\,$ the lower one for $^{28}\rm Si-^{28}Si$.
}
                         \label{fig:8}
\end{wrapfigure}

One of the characteristic features of the spectrum 
obtained theoretically is a series of low-energy 
$K$-rotational excitation due to axial asymmetry around molecular 
z-axis, which is in contrast with the $^{24}\rm Mg+{}^{24}Mg$ 
case.\cite{Ue89,Ue93} 
One can understand the reason immediately from Fig.~8, 
where the upper configuration($^{24}\rm Mg+{}^{24}Mg$) has 
axial symmetry as a total system, but the lower one for 
$^{28}\rm {Si} +{}^{28}\rm {Si}$ has axial asymmetry. 
Thus $K$ is not a good quantum number, namely, 
we expect the eigenstates are $K$-mixed.

A triaxial system preferentially rotates around the axis 
with the largest moment of inertia.  
By the definition of the axes in the lower panel of Fig.~8, 
we have the moments of inertia as $I_X > I_Y >> I_Z$, 
due to the nuclear shape.
Thus the system, which is seen as two pancake-like 
objects($^{28}\rm Si$'s) touching side-by-side, 
rotates around $X$-axis normal to the reaction plane.  
Such a motion is called as wobbling.

We extend our molecular model so as to include couplings between states 
with different $K$-quantum numbers.  As a result, we will obtain 
new low-lying states due to the triaxial shape of the equilibrium 
configuration. The Coriolis terms in the molecular hamiltonian 
bring those couplings. 
However, in practice, we do not treat the Coriolis terms explicitly, 
but we diagonalize the hamiltonian of the asymmetric rotator 
to obtain the rotational spectrum.

The Coriolis terms in the molecular hamiltonian 
in Eq.~(\ref{eq:16})   
gives an impression that those are quite different from the 
asymmetric rotator. So the effect of the Coriolis coupling terms 
will be examined later in \S4.2,  
to show that the molecular hamiltonian reduces to the asymmetric 
rotator hamiltonian in the sticking limit.

\subsection{Analyses by asymmetric rotator}

We describe the rotational motions of 
two pancake-like objects($^{28}\rm Si$'s) touching side-by-side
by means of the asymmetric rotator. 
Generally its hamiltonian is written as follows, 
with  the moments of inertia about the intrinsic axes
$I_x$, $I_y$, and  $I_z$, respectively; %
\begin{eqnarray}
 \hat T_{\rm rot} &=& {\hbar^2 \over 2} \left(
{ {\hat J}_x^2 \over I_x}
+{ {\hat J}_y^2 \over I_y}
+{ {\hat J}_z^2 \over I_z} \right)  
\label{eq:4.1}
\\  
&=&  {\hbar^2 \over 2} \left\{
{ {\hat J}^2 \over I_{\rm av}}
+ {1 \over \Delta} ( -{\hat J}_x^2 + {\hat J}_y^2 )
+ {1 \over I_{K}} {\hat J}_z^2 \right\} ,
\label{eq:4.2}
\end{eqnarray}
where ${\hat J}_x$, ${\hat J}_y$ and ${\hat J}_z$ denote the 
components of the angular momentum operator along the intrinsic 
axes of the body-fixed frame (the same operators as defined 
in Eq.~(\ref{eq:11})).   
$I_{\rm av}$, $\Delta$ and $I_K$ in Eq.~(\ref{eq:4.2}) are 
related to 
$I_x$, $I_y$ and $I_z$ by 
\begin{eqnarray}
{ 1 \over I_{\rm av}} &=& 
        {1 \over 2} \left( {1 \over I_x} +{1 \over I_y} \right), 
\label{eq:4.3} \\
{ 1 \over \Delta} &=& - {1 \over I_x} + {1 \over I_{\rm av}}
                   = {1 \over I_y} - {1 \over I_{\rm av}}, 
\label{eq:4.4} \\
{ 1 \over I_K} &=&  {1 \over I_z} - {1 \over I_{\rm av}}. 
\label{eq:4.5}
\end{eqnarray}
By using lowering and raising operators, 
${\hat J}_+$ and ${\hat J}_-$ of the angular momentum 
in the body-fixed frame, we obtain 
\begin{equation}
 \hat T_{\rm rot} = {\hbar^2 \over 2} \left\{
{ {\hat J}^2 \over I_{\rm av}}
+ { {\hat J}_z^2 \over I_K}
- {1 \over 2\Delta} \left( {\hat J}_+^2 + {\hat J}_-^2 \right)
                                                      \right\},
\label{eq:4.6}
\end{equation}
where ${\hat J}_{\pm} \equiv  {\hat J}_x \pm  i{\hat J}_y$, 
respectively, which give rise to couplings between different $K$'s.
The coupling strength is given by the coefficient $1/\Delta$,
which is proportional to the difference 
between $1/I_y$ and $1/I_x$.
In an intuitive understanding, 
the rotation around $x$-axis is lower in energy 
than the rotation around $y$-axis due to $I_x > I_y$. 
When the energy difference between the rotations around the molecular
$x$- and $y$-axes is larger than $K$-excitation energies, 
the $K$-mixing is expected to be rather large. 
In other words, 
an energetically favored motion,
i.e., rotation around $x$-axis would be realized by the $K$-mixing. 
%

\begin{figure}[b]
\centerline{\includegraphics[height=7 cm]
                             {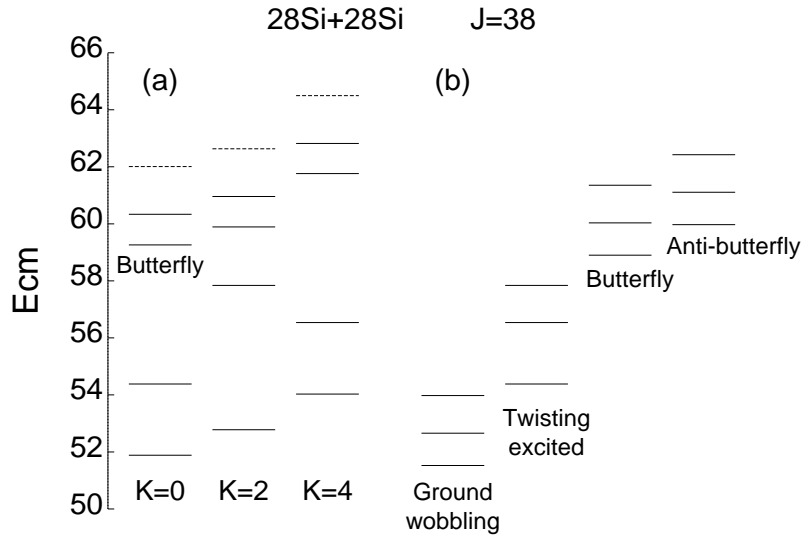}}
\caption{
Energy spectra of the $^{28}\rm Si+{}^{28}Si$ system for $J=38$. 
(a) Molecular normal modes without $K$-mixing. 
(b) After $K$-mixing, with indications of the modes under the levels. 
}
                        \label{fig:9}
\end{figure}

In order to obtain an accurate description of this triaxial rotator,
as it is well known for polyatomic molecules,
we diagonalize the hamiltonian with an inertia tensor of the axial 
asymmetry, which gives rise to mixings of $K$-projections of 
the total spin $J$.\cite{BohrTEXT2}  
The resultant motion should be called as "wobbling mode".\cite{Wobb}  
The energy spectrum is displayed in Fig.~9(b), compared with 
the spectrum without $K$-mixing in Fig.~9(a).
Now the states of low lying $K$-series are not the eigenstates 
by themselves, but are recomposed into new states.  
It is very interesting that we again obtain several states including 
the $K=0$ component as a result of $K$-mixing,
which should show up themselves in the scattering.
Those states are closely located in energy and so in good agreement 
with several fine peaks observed in the experiment.
It should be noted here that 
due to the lack of the enough information 
about the deformations of the total system, 
we assumed the same parameter for the coupling strength $\Delta$ 
for the molecular ground-band states and for the butterfly states, 
although the extent of asymmetry is generally different in each band.  
As for the magnitudes of the moments of inertia $(I_x, I_y, I_z)$, 
we estimated them as follows. 
We assumed a constant value for the relative distance, 
i.e., for $\mu R^2$, and adopted $R=R_{\rm e}=7.6$fm. 
For the contributions from the moments of inertia of 
the constituent $^{28}\rm Si$ nuclei, we estimated them 
about the $y$- and $z$-axes from the excitation energy of 
their $2^+_1$ state. As for the contribution about $x$-axis,
we assumed a factor $4/3$ larger than those about the other axes, 
due to the distribution of the nuclear density of $^{28}\rm Si$. 
For calculations about the moments of inertia of dinuclear systems, 
details are given in Appendix D.

As an analytical prescription,
in the high spin limit ($K/J  \sim 0$), the diagonalization 
in the $K$-space is found to be equivalent to solving 
a differential equation of the harmonic oscillator with 
parameters given by the moments of inertia.
Thereby, the solution is a gaussian, 
or a gaussian multiplied by an Hermite polynomial, 
\begin{equation}
  f_n(K) = H_n \left({ K \over b } \right)
 \exp \biggl[ -{1 \over 2} \biggl( { K \over b } \biggr)^2  \biggr] ,
\label{eq:4.7}
\end{equation} 
where the width $b$ is given by 
\begin{equation}
b= (2J^2  {I_K / \Delta } )^{1/4}.
\label{eq:4.8}
\end{equation} 
The eigenenergy $E_n$  is approximately given by 
\begin{equation}
 E_n = { J(J+1) \hbar^2 \over 2I_{\rm av} }
                   -{ J^2 \hbar^2 \over 2\Delta }    
 +{ \sqrt{ 2 \over \Delta \cdot I_K} J\hbar^2 } 
                   \left(n +{1 \over 2} \right) ,
\label{eq:4.9}
\end{equation}
where the second term on the r.h.s. is due to 
the coupling energy between the states with $\Delta K=2$, 
which is approximated by $K/J =0$. The third term is due to the energy 
of the harmonic oscillator, with the energy quantum, 
\begin{equation}
  \hbar \omega ={\sqrt{ 2 \over \Delta \cdot I_K} J \hbar^2} .
\label{eq:4.10}
\end{equation}
Its excitation should be with $n=even$ 
due to the symmetry between the $|K>$ and $|-K>$ components. 
Note that by the approximation $J(J+1) \sim J^2$ for the second term 
of the r.h.s. of Eq.~(\ref{eq:4.9}), the first and second terms 
can be amalgamated into $J(J+1) \hbar^2 / 2I_x$, 
which reminds that the moment of inertia of the rotation is $I_x$. 
Considering $I_x \sim I_y$, i.e., $I_K^{-1}\sim (I_z^{-1} -I_x^{-1})$, 
the energy quantum $\hbar \omega$ of Eq.~(\ref{eq:4.10}) is equivalent 
to that of the wobbling formula given in Ref.~\citen{BohrTEXT2}.  

In order to calculate angular correlations we use those analytic forms 
in Eq.~(\ref{eq:4.7}), which is simple and intuitive way to understand 
the extent of $K$-mixing.
Of course we can utilize numerical values obtained 
in the diagonalization procedure, but the values are almost the same 
as those given by the analytic form.
For the lowest state $f_0 (K)$ of Eq.~(\ref{eq:4.7}), 
we have the wave function for the wobbling ground state as 
\begin{equation}
\Psi^{JM}_\lambda \sim \sum_K  \exp (-K^2 / 2b^2)  D_{MK}^J (\theta_i) 
                      \chi_{K}(R, \alpha,\beta_1,\beta_2), 
\label{eq:4.11}
\end{equation} 
where in general, $\chi_K$ can be any molecular mode of triaxial 
deformations, such as the ground-state configuration (parallel 
equator-equator one), the butterfly mode and the anti-butterfly mode. 
The magnitude of $b$ estimated by Eq.~(\ref{eq:4.8}) is $1.85$, 
for example, for the values of the moments of inertia used in the 
calculations for the energy spectrum in Fig.~9. 
This is the largest value expected, because we assumed a static 
configuration there, in which the zero-point motions of the twisting 
and butterfly modes are neglected. 
A note for the wobbling wave functions is given 
in the last part of Appendix C.

\begin{figure}[b]
\centerline{\includegraphics[height=7.4 cm]   
                             {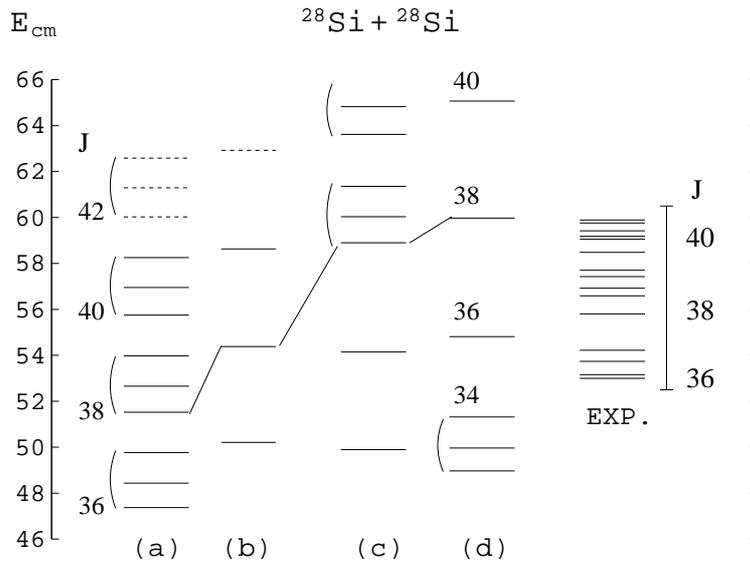}}
\caption{
Dinuclear spectrum of the $^{28}\rm Si+{}^{28}Si$ system 
in the resonance energy region is displayed. 
Levels with $J=38$ are connected by thin lines for eye-guide. 
From the left, the resonance levels theoretically obtained, 
where 
(a) the members of the molecular ground band and 
of the wobbling excited bands, 
(b) an excited band due to the twisting motion with $K=0$, 
(c) the butterfly mode with wobbling 
and (d) the anti-butterfly mode with wobbling. 
On the right-hand side, the experimental data are displayed, 
where $J=36 - 40$ indicate the spin assignments for the 
broad bumps.\cite{BettsPL} 
We selected resonance levels from the narrow peaks 
in the elastic and inelastic excitation functions, 
according to a statistical analysis on their correlation, 
with an indicative bar for the available energy region 
of the data.\cite{SainiBetts} 
}
                        \label{fig:10}
\end{figure}

In Fig.~10, theoretical energy levels of the $^{28}\rm Si+{}^{28}Si$ 
system are compared with the experimental data in the 
resonance energy region.\cite{BettsPRL2, SainiBetts} 
From the left, (a) shows the molecular ground band followed with 
the wobbling excited states, 
(b) an excited band due to the twisting motion with $K=0$, 
(c) the butterfly mode with wobbling 
and (d) the anti-butterfly mode with wobbling. 
Molecular configurations are well stable by the barrier up to $J=40$, 
while with $J=42$, an existence of the molecular resonance 
state is unlikely, as the zero-point energy of the radial motion is 
over the barrier top. 
Levels with $J=38$ are connected by thin lines for eye-guide. 
On the right-hand side, the experimental data are displayed, 
where $J=36 \sim 40$ indicate the spin assignments for the 
broad bumps.\cite{BettsPL} 
We see the density of the resonance states in the data is well 
reproduced by the calculated eigenstates, which are due to the wobbling 
motion and the excitations of the internal modes, such as butterfly etc. 
Note that the existence of the excited states of the wobbling motion 
as resonance states depends upon the stability of the triaxial structure, 
and hence the numbers of those excited states taken up in Fig.~10 are 
not definitive. 

From $J=36$ up to $J=40$, the anti-butterfly mode appears higher than 
the butterfly one as is discussed in \S3.2.   
For those anti-butterfly states we do not display the excited states 
of the wobbling motion, because each eigenstate of the anti-butterfly 
mode appears as the excited state of the butterfly state 
in the $\alpha$-motion, and thus the configuration of the 
anti-butterfly mode is not enough triaxial. 
On the other hand, with $J=34$ the anti-butterfly mode is lower 
in energy than the butterfly one, and thus we display the excited 
states of the wobbling motion for this mode. 
The reason of the lower excitation of the anti-butterfly mode is 
as follows. With relatively-low angular momentum, the constrain 
by the energy well around the equator-equator configuration 
becomes rather weak, and the stability of this configuration is 
not well guaranteed. 
We found that the equator-equator configuration is not at the local 
energy minimum below with $J=32$.  For example, with $J=30$, 
the stable configuration for the molecular ground state is 
an antibutterfly-like one of $\beta \sim 60^\circ$, the equilibrium 
distance $R_{\rm e}$ of which is much smaller than that with  $J=34$. 
Such softening of the energy surface occurs with $J=34$, 
which gives rise to lowering of the anti-butterfly mode.

%
\subsection{%
    Comparison between the molecular-model hamiltonian and the 
    asymmetric \\
    rotator's
}

The wobbling motion associated with "$K$-mixing" is an important aspect 
of the rotation of the asymmetrically deformed nucleus in high spins. 
In order to investigate such an aspect, we have introduced the asymmetric
rotator in addition to the molecular model, because the rotator model is 
simple to understand the essential feature of rotational motions. 
As the rotator model is based on more or less rigid intrinsic structure, 
it is interesting to know how the simple rotator model is related 
to the molecular model.

A triaxial system preferentially rotates around the axis with the largest 
moment of inertia. By the definition of the axes in the lower panel of 
Fig.~8, we have the moments of inertia of the total system 
as $I_X > I_Y >> I_Z$ due to the configuration. Thus the total system, 
which is seen as two pancake-like objects touching side-by-side, rotates 
around the $X$-axis which is normal to the reaction plane.  
In this context, the magnitudes of the moments of inertia is crucially 
important; the large contributions to $I_X$ from the third moments of 
inertia $I_3$ of the two constituent nuclei are expected.  
So here we study two examples of the molecular model, 
one of which is with $I_3 =0$, and the other is with $I_3 \ne 0$.

\begin{figure}[t]
\centerline{\includegraphics[
                             height=6.5 cm]  
                             {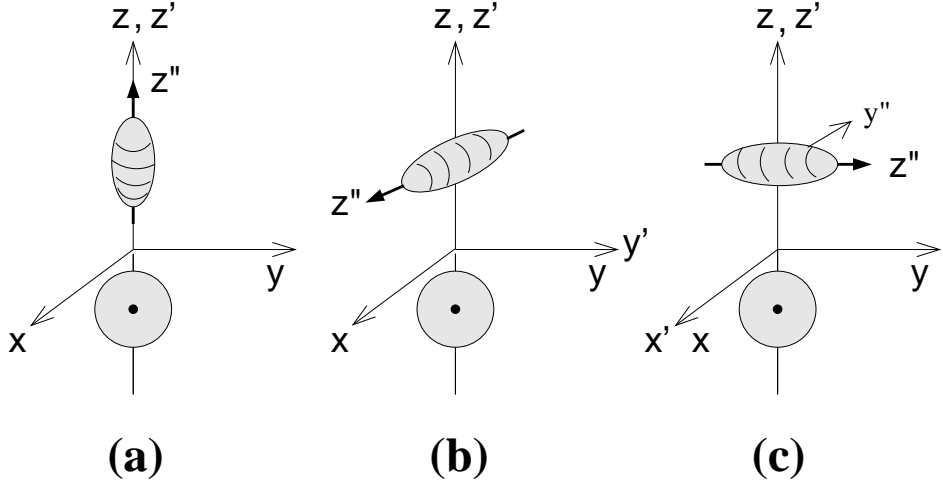}}
\caption{ 
Some examples for the geometrical configurations of the system 
consisting of a spherical nucleus and an axially-symmetric 
deformed nucleus. 
The molecular $z'$-axis is set to be parallel to the $z$-axis, 
as the whole rotation by $\Omega(\theta_1,\theta_2,\theta_3)$ 
will start with those configurations. 
In (a), the symmetry axis of the deformed nucleus ($z''$-axis) 
is parallel to the $z'$-axis, due to the Euler rotations with 
$(\alpha,\beta)=(0,0)$. In (b), the $z''$-axis is parallel 
to the $x'$-axis, due to $(\alpha,\beta)=(0,\pi/2)$. 
In (c), the $z''$-axis is parallel to the $y'$-axis, 
due to $(\alpha,\beta)=(\pi/2,\pi/2)$. 
}
                 \label{fig:11}
\end{figure}
%

We take up a resonant system consisting of 
"a spherical nucleus and a deformed nucleus". 
In order to see only the rotational motion, we assume 
that the two nuclei are bound and stay at a constant relative 
distance $R$, which reduces the degrees of freedom of the system. 
For the first example (case 1), an axially-symmetric deformation 
is assumed for the constituent nucleus, in which  $I_3$ is taken 
to be zero. Thus the coordinates are taken as 
$(q_i)= (\theta_1, \theta_2, \theta_3, \beta)$, 
where $\theta_i$ denote the Euler angles for the rotation of the 
molecular axes. The internal degree of freedom is described with 
$\beta$. Those conditions are taken to be corresponding with 
the degrees of freedom in \S2.   
On the other hand, in the second example (case 2), an 
axially-asymmetric deformation is assumed, i.e., $I_3$ is not 
zero, which gives rise to a degree of freedom $\gamma$. 
And then the coordinates are 
$(q_i)= (\theta_1, \theta_2, \theta_3, \beta, \gamma)$. 
Of course, the introduction of the $\gamma$-degree of 
freedom is an extension from the description in \S2, 
which is expected from the nuclear density distribution 
illustrated in Fig.~8.

Typical configurations are illustrated in Fig.~11, 
where in (a), (b) and (c) the configurations are set with 
$(\alpha,\beta)=(0,0)$, $(0,\pi/2)$ and $(\pi/2,\pi/2)$, respectively. 
The configuration (a) is axially symmetric for case 1, while those of 
(b) and (c) are axially asymmetric (the same shape). 
%
Note that the shape of the constituent deformed nucleus is prolate, 
but the figure is useful both for the prolate nucleus 
and for the oblate nucleus, of course. 
We describe the orientations of the principal axes of the constituent 
deformed nucleus with the Euler angles $(\alpha, \beta)$, 
in which the degree of freedom associated with $\alpha$ is essentially 
the same as that associated with $\theta_3$ of the total system, 
and thus $\alpha$ does not appear in $(q_i)$. 
Actually, the rotations of the whole system and the constituent 
nucleus are described by the relation, 
\begin{equation}
  \Omega_n ({\tilde \alpha}, {\tilde \beta}, {\tilde \gamma})
        =   \Omega'_n (\alpha, \beta, \gamma) 
             \Omega_M (\theta_1, \theta_2, \theta_3) ,
\label{eq:4.12}
\end{equation}
where $\Omega_n$ and $\Omega'_n$ denote Euler rotations 
for the constituent deformed nucleus with respective angles, and 
$\Omega_M$ denotes rotations of the molecular axes.
On the r.h.s. of Eq.~(\ref{eq:4.12}), $\Omega'_n$ denote the successive 
rotation after $\Omega_M$;
firstly the axes of the constituent deformed nucleus rotate 
up to the directions of the molecular axes by $\Omega_M$, 
and secondly they rotate referring to the molecular axes by $\Omega'_n$.
Since the successive rotation is decomposed into 
$\Omega'_n(\gamma) \Omega'_n(\beta) \Omega'_n (\alpha) $, we have 
the rotations with angles $\theta_3$ and $\alpha$ around the same 
axes obtained after $\Omega_M (\theta_1, \theta_2)$. 
Thus we can take the angle of the third rotation of the whole system 
simply to be $\theta_3 + \alpha$, which involve the freedom $\alpha$.  
Here we regard the value of $\alpha$ as the initial condition for the 
starting configurations before rotation, which are displayed in Fig.~11.

We write the classical kinetic energy with the angular velocities, 
and then we replace these angular velocities with time derivatives 
of those coordinates $(q_i)$. 
The classical kinetic energy is expressed in the form 
$T={1 \over 2} \sum g_{ij}{\dot q}_i{\dot q}_j$  
and we quantize it by using the general formula 
for the curve-linear coordinate system. The quantum mechanical 
expression for the kinetic energy is given by  
\begin{equation}
  {\hat T} = -{\hbar^2 \over 2}
              \sum_{ij} {1 \over \sqrt g}{\partial \over \partial q_i} 
               \sqrt g(g^{-1})_{ij}{\partial \over \partial q_j}, 
\label{eq:4.13}
\end{equation}
where $g$ and $g^{-1}$ denote the determinant 
and the inverse matrix of $(g_{ij})$, respectively. 
The metric tensor $(g_{ij})$ is composed with the submatrices $g_{rot}$ 
for the whole rotational degrees and $g_{int}$ for the internal degrees 
of freedom as 
\begin{equation}
(g_{ij})= \left( 
          \begin{array}{cc}
          g_{\rm rot}  & g_{\rm C} \\
          {}^tg_{\rm C}         & g_{\rm int}
          \end{array}
                                       \right), 
\label{eq:4.14}
\end{equation}
where $g_{\rm C}$ denotes the nondiagonal part which corresponds to 
the Coriolis coupling, ${}^tg_{\rm C}$ being the transpose of 
$g_{\rm C}$. The way to obtain those components of $g_{ij}$ is 
described in detail in Ref.~\citen{Ue93}. 
Here we briefly see their definitions: 
\begin{eqnarray} 
g_{\rm rot}&=& {}^tV(\theta_2, \theta_3) 
         (\mbox{\boldmath $I$}_\mu + {}^tR(\alpha, \beta, \gamma) 
          \mbox{\boldmath $I$}_n R(\alpha, \beta, \gamma))
                 V(\theta_2, \theta_3), 
\label{eq:4.15} \\
g_{\rm int}&=& {}^tV(\beta, \gamma) 
                \mbox{\boldmath $I$}_n  V(\beta, \gamma),  
\label{eq:4.16} \\
g_{\rm C}&=& {}^tV(\theta_2, \theta_3) 
             {}^tR(\alpha, \beta, \gamma) 
             \mbox{\boldmath $I$}_n V(\beta, \gamma),  
\label{eq:4.17}
\end{eqnarray}
where $V(\theta_2, \theta_3)$ and $V(\beta, \gamma)$ denote 
the transformation matrices between the derivatives of 
the Euler angles and the angular velocities of the molecular 
axes and of the constituent deformed nucleus, respectively, 
$R(\alpha, \beta, \gamma)$ being the rotation matrix. 
$\mbox{\boldmath $I$}_\mu$ denotes the inertia tensor for the 
two constituent nuclei as point-masses, i.e., the diagonal 
moments for $x$- and $y$-axes being $\mu R^2$, 
while $\mbox{\boldmath $I$}_n$ denotes the inertia tensor for the 
constituent deformed nucleus in its principal axes, respectively. 
The moments of inertia of the constituent deformed nucleus 
(the diagonal elements of $\mbox{\boldmath $I$}_n$) are taken 
as follows: 
for the axially symmetric nucleus (case 1), $I_1=I_2=I_0$ and 
$I_3=0$, and for the axially asymmetric nucleus (case 2), 
$I_1=I_2=I_0$ and the value of $I_3$ being not zero. 
For the latter case, in general, $I_1 \ne I_2$ may be used 
for the static asymmetric deformation, as was tried in 
the theory of the asymmetric rotator,\cite{Davydov}
but for simplicity we avoid this tedious calculations. 
Our consideration is focused on the appearance of the $\gamma$-degree 
of freedom associated with $I_3 \ne 0$, and for this purpose 
the assumption $I_1=I_2$ brings no problem.  
Note that $\gamma$ is spurious for the deformed nucleus 
with the axial symmetry, i.e., for case 1 with $I_3=0$. 
With the aid of mathematical software, we can easily obtain 
the elements $g$ and $(g^{-1})_{ij}$, for example,
$g=(\mu R^2)^2 I_0^2 \sin^2 \beta \sin^2 \theta_2$ for case 1, 
and $g=(\mu R^2)^2 I_0^2 I_3 \sin^2 \beta \sin^2 \theta_2$ 
for case 2, respectively.

As the classical kinetic energy consists of three parts, 
i.e., the rotation of the whole system, the internal motions 
and their couplings, the quantum mechanical operator for the 
kinetic energy ${\hat T}$ is also given as a sum of three terms, 
\begin{equation}
{\hat T}={\hat T}_{\rm rot} + {\hat T}_{\rm int} + {\hat T}_{\rm C} .
\label{eq:4.18}
\end{equation}
Naturally the term ${\hat T}_{\rm rot}$ is associated with 
the rotational variables $(\theta_1, \theta_2, \theta_3)$, 
${\hat T}_{\rm int}$ with the internal variables $(\beta, \gamma)$ 
and ${\hat T}_{\rm C}$ with both. 
According to the derivation, ${\hat T}_{\rm rot}$ is expressed 
by the partial differential operators of $\theta_i$. 
We combine those differential operators into angular momentum 
operators ${\hat J}'_i$. 
Thus we obtain the following kinetic energy terms, 
\begin{equation}
  {\hat T}_{\rm rot}
            = {\hbar^2 \over 2} 
             \sum_{\scriptstyle1\le i\le3\atop\scriptstyle1\le j\le3} 
                 \mu_{ij}   {\hat J}'_i  {\hat J}'_j  ,
\label{eq:4.19}
\end{equation}
where the matrix $\mu$ is the submatrix given later, 
and ${\hat J}'_i$ are the angular momentum operators 
in terms of the Euler angles of the molecular frame, as usual, 
which is already given in Eq.~(\ref{eq:11}).   
The coefficients $\mu_{ij}$ are determined due to 
the moments of inertia corresponding to the geometrical 
configuration, and are given in terms of the parameter 
$\alpha$ and the internal variable $\beta$ as follows; 
\begin{eqnarray}
    \mu_{11} & =&  \mu_{22} = { 1 \over \mu R^2 } ,
\nonumber \\
    \mu_{12} & =& 0  ,
\nonumber \\
  \mu_{13} & =&{1 \over \mu R^2} \cos \alpha \cot \beta ,
\label{eq:4.20} \\
  \mu_{23} & =&{1 \over \mu R^2} \sin \alpha \cot \beta ,
\nonumber \\
  \mu_{33}&=& \Big( {1 \over I_0 } + {1 \over \mu R^2 } \Big) 
            {1 \over \sin^2 \beta} 
                    - {1 \over \mu R^2} .
\nonumber
\end{eqnarray}
Note that the above expressions of $\mu_{ij}$ are the same 
between two examples, which are obtained under the assumption 
$I_1=I_2=I_0$.

The internal kinetic energy operator is associated with  
the variable $\beta$ for case 1. 
We obtain 
\begin{equation}
 {\hat T}_{\rm int}(\beta) = -{\hbar^2 \over 2}
       \bigg( {1 \over I_0 } + {1 \over \mu R^2 } \bigg) 
                {\partial^2 \over \partial \beta^2} 
    -{\hbar^2 \over 8}
            \bigg({1 \over I_0 } + {1 \over \mu R^2 } \bigg) 
                 \bigg( {1  \over \sin^2 \beta} +1 \bigg) ,     
\label{eq:4.21}
\end{equation}
where the second term on the r.h.s. is the additional potential 
due to the new volume element $ dV =d\beta$ instead of the original 
$ dV =  \mu R^2 I \sin\beta d\beta$. 
For the asymmetrically deformed constituent nucleus with 
$I_3 \ne 0$ (case 2), we have an additional term associated with 
the $\gamma$-degree of freedom, i.e., 
\begin{equation}
 {\hat T}_{\rm int}(\beta, \gamma) = {\hat T}_{\rm int}(\beta) 
   -{\hbar^2 \over 2} \bigg\{ {1\over \sin^2 \beta}
        \Big({1\over I_0}\cos^2 \beta +{1\over \mu R^2} \Big) 
      + {1\over I_3} \bigg\}  {\partial^2 \over \partial \gamma^2} , 
\label{eq:4.22} 
\end{equation}  
where indications $(\beta)$ and $(\beta, \gamma)$ on 
${\hat T}_{\rm int}$ are for distinction between cases 1 and 2.

The Coriolis coupling operator ${\hat T}_{\rm C}$ 
consists of coupling operators between the variables 
$(\theta_1, \theta_2, \theta_3)$ and $\beta$ for case 1, i.e., 
\begin{equation}
 {\hat T}_{\rm C}(\theta_i; \beta)=  -{ \hbar^2 \over \mu R^2 }
           \bigg\{ - \sin\alpha 
   \Big( -i {\partial \over \partial \beta}\Big) {\hat J}'_1   
                         + \cos\alpha 
   \Big( -i {\partial \over \partial \beta}\Big) {\hat J}'_2 
 \bigg\},
\label{eq:4.23} 
\end{equation}
where the derivative operators of $\theta_i$ are rewritten 
with the angular momentum operators ${\hat J}'_i$. 
As for case 2, the couplings are between the variables 
$(\theta_1, \theta_2, \theta_3)$ and $(\beta, \gamma)$, 
for which we obtain  
\begin{eqnarray}
  {\hat T}_{\rm C}(\theta_i; \beta, \gamma)  
                  = {\hat T}_{\rm C}(\theta_i; \beta) 
        &  - { \hbar^2 \over \mu R^2  \sin \beta}
           \bigg\{ \cos\alpha 
           \Big( -i {\partial \over \partial \gamma}\Big) 
               {\hat J}'_1   
                         + \sin\alpha 
           \Big( -i {\partial \over \partial \gamma}\Big)
               {\hat J}'_2 \bigg\}
\nonumber \\
        & - { \hbar^2 \over  \sin \beta} \cot \beta
               \Big({1 \over I_0 }+{1 \over \mu R^2 } \Big) 
            \Big( -i {\partial \over \partial \gamma}\Big)     
                        {\hat J}'_3  .
\label{eq:4.24} 
\end{eqnarray}

Now we investigate those kinetic energy operators obtained. 
Firstly we proceed with case 1. Generally the hamiltonian of 
the system is composed of the kinetic energy and the interaction 
between the constituent nuclei $V(\beta)$, i.e., 
\begin{equation}
 \hat{H}= 
  \hat{T}_{\rm rot} + \{ {\hat T}_{\rm int}+V(\beta) \} 
              + \hat{T}_{\rm C}. 
\label{eq:4.25}
\end{equation}
Without the coupling $\hat{T}_{\rm C}$, the internal motion 
is determined by the eigenvalue equation, 
\begin{equation}
 \hat{H}_{\rm int} \, \chi(\beta)= 
  \{ {\hat T}_{\rm int}+V(\beta) \} \chi(\beta)= E \chi(\beta) .
\label{eq:4.26}   
\end{equation}
For the case of a very strong confinement by the interaction 
$V(\beta)$, the motion associated with the $\beta$-degree is 
expected to approximately follow Eq.~(\ref{eq:4.26}). 
And we write the rotational motion of the system, 
including $\hat{T}_{\rm C}$, as 
\begin{equation}
 {\hat H}_{\rm rot} = {\hat T}_{\rm rot} + {\hat T}_{\rm C}.
\label{eq:4.27}
\end{equation}
To analyze the rotational motion given by ${\hat H}_{\rm rot}$, 
it is important to assume a dominant configuration such as 
the equator-equator one of the $^{28}\rm Si+{}^{28}Si$ system, 
which appears in low energy due to the interaction between 
the constituent nuclei. 
For we have all the possible rotational motions generally 
to appear in the energy spectrum of ${\hat H}_{\rm rot}$,
which make the problem extremely complicated.
The dominant configuration assumed in this investigation 
is an axially asymmetric one with the strong confinement
which is seen in Fig.~11(b) and/or Fig.~11(c). 
Thus we put $\beta \sim \pi/2$ for $\mu_{ij}$ of 
${\hat T}_{\rm rot}$ in Eq.~(\ref{eq:4.20}), 
i.e., $\cot \beta$ to be zero. 
As for $\alpha$, its value is not essential, and we can 
choose any value for $\alpha$.  
The Coriolis coupling ${\hat T}_{\rm C}(\theta_i; \beta)$ 
of Eq.~(\ref{eq:4.23}) reduces conveniently into one term, 
for the configuration with $\alpha=0$ illustrated in Fig.~11(b), 
and we obtain 
\begin{equation}
 {\hat H}_{\rm rot} =  { \hbar^2 \over 2}
       \bigg\{  {J(J+1) \over \mu R^2}
       + \Big({1 \over I_0 } - {1 \over \mu R^2 } \Big) 
        \Big( - {\partial^2 \over \partial \theta_3^2} \Big) 
   -{2 \over \mu R^2}\Big( -i{\partial \over \partial \beta} \Big)
            {\hat J}'_2     \bigg\},         
\label{eq:4.28} 
\end{equation}
where the first term of the r.h.s. is the angular momentum 
${\hat J}^2$ replaced by $J(J+1)$, and the last term gives 
the Coriolis coupling. Note that for another configuration with 
$\alpha=\pi/2$ illustrated in Fig.~11(c), $-{\hat J}'_1$ appears 
instead of ${\hat J}'_2$. 
In general, the Coriolis coupling induces $K$-mixing associated with 
the vibrational excitations, which is exemplified by the third term
of the r.h.s. of Eq.~(\ref{eq:4.28}); 
by using the creation and annihilation operators, $a^*$ and $a$ 
of the vibrational motion obtained by Eq.~(\ref{eq:4.26}), 
the term is rewritten as 
$\sim (a^* - a)({\hat J}_- - {\hat J}_+)$, 
where ${\hat J}'_{\pm}={\hat J}'_1 \pm {\hat J}'_2$ are 
the lowering and raising operators of the $K$-quantum numbers. 
This description corresponds to that of the molecular hamiltonian 
given in \S2.    

On the other hand, there is a picture in which the internal degree 
$\beta$ is frozen due to strong adhesion between the constituent 
nuclei, i.e., {\em the sticking limit}.\cite{Bass} 
In order to obtain relation to the rotator hamiltonian, 
we approach from such a picture, in which the sharing of the 
angular momenta is determined classically, i.e., 
%
\begin{equation}
\mbox{\boldmath $J$}=(\mu R^2 +I_0) \mbox{\boldmath $\omega$}, 
\qquad
\mbox{\boldmath $L$}=\mu R^2 \mbox{\boldmath $\omega$}, \qquad 
\mbox{\boldmath $S$}=I_0 \mbox{\boldmath $\omega$}, 
\label{eq:4.29}
\end{equation}
where $\mbox{\boldmath $\omega$}$ denotes the angular velocity of the 
whole system. Corresponding to the above relations, quantum mechanical 
ones are derived in Eq.~(\ref{eq:A.18}) of Appendix A as follows, 
\begin{equation}
  \hat{\mbox{\boldmath $S$}}/I_0 
  =\hat{\mbox{\boldmath $L$}}/\mu R^2 
  =\hat{\mbox{\boldmath $J$}}/(\mu R^2 + I_0) ,
\label{eq:4.30}
\end{equation}
where $\hat{\mbox{\boldmath $S$}}$, $\hat{\mbox{\boldmath $L$}}$ 
and $\hat{\mbox{\boldmath $J$}}$ denote the spin of the constituent 
deformed nucleus, the orbital and the total angular momenta, 
which are defined as the rotations on the same plane. 
Although the relations in Eq.~(\ref{eq:4.30}) are limited among the 
rotations around the same axis, they are approximately applicable to 
the present analysis, because we are considering the rotation of the 
whole system around the largest moment of inertia; for example, 
the axis of the rotation is $y'$ for the configuration in Fig.~11(b). 
Therefore $\hat{\mbox{\boldmath $S$}}$ is associated with the 
$\beta$-degree, and $\hat{\mbox{\boldmath $J$}}$ corresponds to 
${\hat J}'_2$ with respect to the configuration in Fig.~11(b).  
Following Eq.~(\ref{eq:4.30}), we use a relation, 
\begin{equation}
  \Big(-i {\partial \over \partial \beta} \Big)\ /I_0 =
               {\hat J}'_2/(\mu R^2 + I_0) ,
\label{eq:4.31}
\end{equation}
and rewrite the total kinetic energy 
${\hat T}={\hat T}_{\rm rot} + {\hat T}_{\rm int} + {\hat T}_{\rm C}$. 
Then the result turns out to be the hamiltonian of 
the asymmetric rotator, i.e., 
\begin{equation}
  \hat{T}= {\hbar^2 \over 2} \bigg\{
              { ({\hat J}'_1)^2 \over \mu R^2 }  
              +{({\hat J}'_2)^2 \over \mu R^2 + I_0} 
              +{({\hat J}'_3)^2 \over I_0 } \bigg\} .
\label{eq:4.32}
\end{equation}
Note that due to the configuration in Fig.~11(b), 
we put $\alpha=0$ and $\beta =\pi/2$ 
for $\mu_{ij}$ of ${\hat T}_{\rm rot}$ in Eq.~(\ref{eq:4.20}), 
and ${\hat T}_{\rm C}(\theta_i; \beta)$ of Eq.~(\ref{eq:4.23}), 
respectively. 
Note also that we drop the additional potential 
in ${\hat T}_{\rm int}(\beta)$ 
(the second term of the r.h.s. of Eq.~(\ref{eq:4.21})), 
because the new volume element is not applied 
for the spin degree of freedom $\beta$. 
Thus we successfully obtained the asymmetric rotator hamiltonian. 
However, ${\hat T}$ in Eq.~(\ref{eq:4.32}) is too simple, 
because the $\beta$-degree is frozen and disappears. 
So this way does not exactly correspond to what is shown 
in \S4.1.    
There we have the hamiltonian for the activated $\beta$-degrees 
in addition to the rotator one. Although the above $\hat T$ 
in Eq.~(\ref{eq:4.32}) is well correspondent to the kinetic energy 
operator given in \S2.2 and gives a rotator hamiltonian, 
eventually, the results of the diagonalization of 
${\hat T}_{\rm C}$ in \S2.2  
may not give rise to the energy spectrum given in Fig.~9.

Next we examine the case 2, in which two internal degrees 
of freedom $(\beta, \gamma)$ are treated. 
In Fig.~12, the configuration with the oblate deformed 
constituent nucleus is illustrated corresponding to Fig.~11(b), 
where a larger value of $I_3$ than that of $I_0$ is assumed. 
So the axis with the largest moment of inertia is $x'$ in this case. 
According to the configuration, we again put $\alpha=0$ 
and $\beta =\pi/2$ into  ${\hat T}_{\rm rot}$, 
${\hat T}_{\rm int}(\beta, \gamma)$
and ${\hat T}_{\rm C}(\theta_i; \beta, \gamma)$, 
respectively, which gives 
\begin{equation}
 {\hat T}_{\rm int}(\beta, \gamma) = 
  -{\hbar^2 \over 2} \bigg\{ 
       \Big( {1 \over I_0 } + {1 \over \mu R^2 } \Big) 
        \Big( {\partial^2 \over \partial \beta^2} 
                                    +{1 \over 2} \Big)     
     + \Big({1\over I_3} + {1\over \mu R^2} \Big) 
          {\partial^2 \over \partial \gamma^2} \bigg\} , 
\label{eq:4.33}
\end{equation} 
and 
\begin{equation}
  {\hat T}_{\rm C}(\theta_i; \beta, \gamma) = 
                  - { \hbar^2 \over \mu R^2 }
          \bigg\{  
  \Big(-i{\partial \over \partial \beta}\Big){\hat J}'_2 
 +\Big(-i{\partial \over \partial \gamma}\Big) {\hat J}'_1   
                         \bigg\}.
\label{eq:4.34}
\end{equation}
We again proceed in the picture of the sticking limit. 
Relations similar to Eq.~(\ref{eq:4.29}) are assumed as follows: 
%
%
\begin{equation} 
\mbox{\boldmath $J$}=(\mu R^2 + I_3) 
       \mbox{\boldmath $\omega$}, \qquad
\mbox{\boldmath $L$}=\mu R^2 
        \mbox{\boldmath $\omega$}, \qquad 
\mbox{\boldmath $S$}=I_3 \mbox{\boldmath $\omega$} , 
\label{eq:4.35}
\end{equation}
where $\mbox{\boldmath $\omega$}$ denotes the angular velocity 
of the whole system again, the rotation being around the $x'$-axis. 
Due to the same relation as Eq.~(\ref{eq:4.30}) except $I_0$ being 
replaced with $I_3$, we use a relation, 
\begin{equation}
  \Big(-i {\partial \over \partial \gamma} \Big)\ /I_3 =
               {\hat J}'_1/(\mu R^2 + I_3) .
\label{eq:4.36}
\end{equation}
By Eq.~(\ref{eq:4.36}), only the $\gamma$-degree is frozen out, 
and as a result we obtain the total kinetic energy operator, 
\begin{eqnarray}
  {\hat T}&=& {\hbar^2 \over 2}  \bigg\{
              { ({\hat J}'_1)^2 \over \mu R^2 + I_3} 
              +{({\hat J}'_2)^2  \over \mu R^2 } 
              +{({\hat J}'_3)^2 \over I_0 }  \bigg\}
\nonumber \\
    && -{\hbar^2 \over 2}
       \Big( {1 \over I_0 } + {1 \over \mu R^2 } \Big) 
        \Big( {\partial^2 \over \partial \beta^2} 
                                    +{1 \over 2} \Big)     
    - { \hbar^2 \over \mu R^2 }
  \Big(-i{\partial \over \partial \beta}\Big){\hat J}'_2 .
\label{eq:4.37}
\end{eqnarray}
%

\begin{wrapfigure}{r}{6.6cm}
\centerline{\includegraphics[
                             height=5.0 cm]
                             {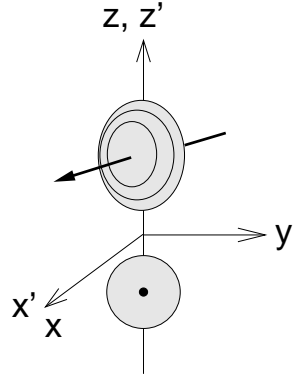}}
\caption{ An example for the geometrical configurations 
of the system consisting of a spherical nucleus and a nucleus 
with oblate deformation, for intuitive understanding. 
The molecular $z'$-axis is set to be parallel to the $z$-axis, 
and the symmetry axis of the deformed nucleus ($z''$-axis) 
is parallel to the $x'$-axis, due to $(\alpha,\beta)=(0,\pi/2)$. 
}
                 \label{fig:12}
\end{wrapfigure}
%

\noindent
Thus we finally obtain the asymmetric rotator hamiltonian 
accompanied by the vibrational mode $\beta$. 
It is noted that the equilibrium of the $\beta$-vibration 
is assumed to be $\, \beta = \pi/2 \,$ in the derivation. 
In this model, the pole orientation of the constituent 
deformed nucleus is fluctuating around the direction of $x'$-axis 
seen in Fig.~12, while the whole system with the moment of 
inertia $\mu R^2 + I_3$ rotates approximately around 
the $x'$-axis with rather confined configuration. 
This is just the same picture adopted in \S4.1,  
where the two constituent nuclei keep in touch and are 
sticking in the rotational motions along the same axis 
as the whole rotating system. 
As for the moments of inertia of the system, 
strong confinements due to the nucleus-nucleus 
interaction are supposed to give rise to induced 
deformations and/or a neck formation of the constituent nuclei, 
which bring nonzero moments $I_3$ of the constituent nuclei 
in addition to the original moments $I_0$ of the deformed 
nuclei with the axial symmetry. 
Thus, we expect that the above analysis is physically meaningful.

\section{Summary}  

The interaction between two nuclei is described 
with the internal collective variables, 
i.e., the orientations of the poles of the constituent nuclei 
in the rotating molecular frame.
In the dinuclear system with oblate-deformed constituent nuclei, 
an equator-equator touching configuration with the parallel principal 
axes is found to be the equilibrium at high spins.   
In the $^{28}\rm Si +{}^{28}Si$ system, 
the relative distance between the two $^{28}\rm Si$ nuclei is 
$7 - 8$fm, indicating a nuclear compound system 
with hyperdeformation. The barrier position is $9 - 10$fm, 
greatly outside from that of usual optical potentials.
Molecular configurations are well stable by the barrier 
up to $J=40$, while with $J=42$, an existence of the molecular resonance 
state with narrow widths is unlikely, as the zero-point energy of the 
radial motion is over the barrier top. This theoretical maximum spin 
is in accord with the bumps observed in grazing angular momenta.

Couplings among various molecular configurations are taken into account 
by the method of normal mode around the equilibrium configuration, 
which gives rise to the molecular modes of excitation, such as the radial 
vibration, the butterfly motion, the anti-butterfly motion and so on. 
The twisting mode ($\nu =4$) is found to be the lowest excitation. 
Vibrational energy quanta for the butterfly and the anti-butterfly 
modes are about $4$MeV, but the excitation energies of those modes 
have to be twice, $8$MeV, since states of $K=even$ with one vibrational 
quantum are not allowed due to the boson symmetry. Thus, the energies 
are close to those for the radial excitation. Although the excited 
state of the radial mode is not bound in the present calculations, 
the possibility of the radial-mode resonance is not completely excluded, 
because it is likely that the interaction between two $^{28}\rm Si$ 
would be more attractive than the present folding potential with 
the frozen density.

A triaxial system preferentially rotates around the axis with 
the largest moment of inertia.  By the definition of the axes 
in the lower panel of Fig.~8, we have 
the moments of inertia of the total system as $I_X > I_Y >> I_Z$. 
Thus the total system, which is seen as two pancake-like objects 
touching side-by-side, rotates around the $X$-axis which is normal 
to the reaction plane.  As the axial symmetry is slightly 
broken, wobbling motion appears in that way.

We extend our molecular model so as to include couplings between states 
with different $K$-quantum numbers.  Usually, the Coriolis coupling terms 
are diagonalized, but we do not treat them explicitly.   
In practice, we use the asymmetric rotator as an intuitive model. 
By the diagonalization of the rotator hamiltonian in the $K$-space, 
we obtain new low-lying states due to a triaxial shape of 
the equilibrium configuration. 
In the high spin limit ($K/J  \sim 0$), the diagonalization is found 
to be equivalent to solving a differential equation of the harmonic 
oscillator with spring constants given by the moments of inertia.
Thereby, the analytic solution is obtained to be a gaussian, 
or a gaussian multiplied by an Hermite polynomial, which is a useful tool 
for the analyses of the molecular states with the triaxial configuration. 

Since the Coriolis terms in the molecular hamiltonian appear to be 
quite different from the asymmetric rotator, it is necessary and 
meaningful to study the relations between the molecular hamiltonian 
and the asymmetric rotator's. 
The analysis turns out that the hamiltonian of the molecular model 
with the $\gamma$-degree of freedom reduces to that of 
an asymmetric rotator in the sticking limit.\cite{Bass} 
Thus the intuitive use of the asymmetric rotator is warranted, 
and it provides a very simple understanding with easy calculations 
of the effects of $K$-mixing on the energy spectrum.

Finally it should be mentioned that an extension of the molecular 
model is possible so as to include the $\gamma_i$-degrees of freedom. 
For example, possible $\gamma$-vibrations of the constituent nuclei 
could be taken into account. But we do not include those surface 
vibrations and the corresponding $\gamma_i$-degrees of freedom, 
considering that the dominances of the members of the ground-state 
band in the decays are reported for the $^{28}\rm Si + {}^{28}Si$ 
system\cite{Beck2000} and for the $^{24}\rm Mg + {}^{24}Mg$ 
system,\cite{Salsac} respectively. 
A molecular model with two asymmetric rotators 
of the constituent nuclei is not pursued for the moment, 
which does not appear rewarding for elaboration. 
Furthermore, the constituent nuclei are expected to be strongly 
confined to form the whole deformed system, in which the 
$\gamma_i$-degrees of freedom are approximately frozen. 
Hence, we adopt the asymmetric rotator for the whole system 
as a sticking limit of the $\gamma_i$-degrees of freedom.

We have intuitively expected that the moment of inertia 
$I_X$ is the largest for the configuration in Fig.~8.  
Namely, we implicitly assume $I_1 \ne I_2$ and  $I_3 \ne 0$ 
for the moments of inertia of the constituent nuclei in their 
principal axes, due to additional deformations likely induced by 
the interactions at the contact configurations of the two nuclei,  
while in \S2,   
without the induced deformations, we have assumed 
the axial symmetry of $^{28}\rm Si$ and the intrinsic 
moment $I_3=0$. (The latter gives  $I_Y > I_X >> I_Z$.) 
The dynamical process of the transition between those two states 
of the constituent nuclei with $I_3 =0$ and $I_3 \ne 0$ 
is an interesting problem, which should be clarified in future. 
The nuclear structure with large induced deformations may be 
close to that obtained by the Nilsson-Strutinsky calculations for 
$^{56}$Ni.\cite{Bengtsson} 
However the experiments exhibit the nature of the dinuclear complex 
in resonances, i.e., the dominance of binary decays,\cite{Beck2000} 
which suggests the contact of the two $^{28}\rm Si$ nuclei is not 
violent enough for rearrangements of the nuclear structure of 
the constituent nuclei in the molecular model.

The molecular states obtained in the present paper are expected 
to be the origin of a large number of resonances observed, 
and hence theoretical analyses have been made for the angular 
distributions and the angular correlations. 
The results have been compared with the recent experiment 
performed in Strasbourg\cite{Beck2000, Nouicer99} 
to give good agreements with the data, 
which will be given in the succeeding paper.\cite{UeNewII}

\section*{Acknowledgements}
The authors thank Drs. C. Beck, R. Freeman and F. Haas 
for stimulating discussions in their collaborations.
The authors are grateful for the discussion and 
for the hospitality of Dr. B. Giraud in the visits at Saclay.

This work was supported in part 
by the Grant-in-Aid for Scientific Research from the Japanese 
Ministry of Education, 
Culture, Sports, Science and Technology (12640250).

\appendix
\section{
Relation between the Coordinate Systems 
    and the Angular Momentum Operators in the Molecular Model}

\begin{wrapfigure}{r}{6.6cm}
\centerline{\includegraphics[
                             height=5.0 cm]
                             {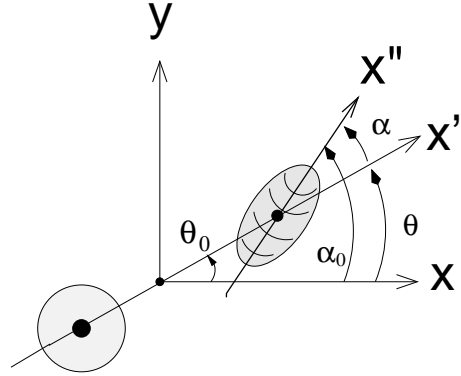}}
\caption{ The Coordinates of the system which consists of 
a spherical nucleus and a deformed nucleus with axial symmetry. 
Both the relative vector and the symmetry axis of the deformed 
nucleus are assumed to be parallel to the $xy$-plane.
}
                 \label{fig:13}   
\end{wrapfigure}

In the present appendix, the description of the angular momenta 
for the total kinetic energy operator in the molecular model 
is ascertained with respect to the coordinate system. 
For this purpose, we take up a two-body problem in which one body 
is deformed and has own rotational degrees of freedom. 
In the laboratory frame the description of the kinetic energy of 
the system is rather simple 
as we consider the energies from the rotational motion of the two-body 
relative vector (orbital motion) and the spin degrees separately. 
However in the body-fixed frame, i.e., in the molecular frame, 
the description is not so easy, because the total system is not 
simple as a rigid rotator. 
Firstly, the coordinates for the molecular frame and those 
of the internal degrees of freedom associated with the frame 
should be chosen appropriately. 
Secondly, although we can obtain the kinetic energy operator $\hat T$ 
by using the formula for the curve-linear coordinates 
which is given later in Eq.~(\ref{eq:A.3}), 
we have three parts of $\hat T$, i.e., the rotational energy, 
the energy associated with the internal degrees of freedom and the 
couplings between them, the roles of which have to be clarified. 
As is shown below, the property of the operator, for example, 
to be the total angular momentum or to be the orbital 
angular momentum, is determined not only by the coordinates for the 
rotational motions themselves but also by the moments of inertia 
associated with. This is natural as the angular momentum 
in the classical description. 
Two examples are given; one is for the present molecular frame, 
and the other is for a new molecular frame, in which the couplings 
between the whole rotation and the internal motions disappear.

Consider a resonant system consisting of 
"a spherical nucleus and an axially-symmetric deformed nucleus". 
In order to see the rotational motion, we assume 
that the two nuclei are bound and stay at a constant relative distance 
$R$, which reduces the degrees of freedom of the system. 
To limit the degrees of freedom to be "two dimensional", 
we again assume that the symmetry axis of the deformed nucleus 
is in the reaction plane. 
As is illustrated in Fig.~13, referring to the space-fixed axes, 
the orientation of the relative vector of the two constituent nuclei 
is described by the angle $\theta_0$, and the angle of 
the orientation of the symmetry axis is denoted as $\alpha_0$. 
The classical kinetic energy is given with two angular velocities 
$\omega=\dot \theta_0$ and $\omega_n=\dot \alpha_0$, 
as $T=1/2 \cdot (\mu R^2 \omega^2 + I_n \omega_n^2)$, 
where $\mu$ and $I_n$ denote the reduced mass of the two nuclei 
and the moment of inertia of the deformed nucleus, respectively. 
The corresponding kinetic energy operator is given by 
\begin{equation}
  {\hat T} = {\hbar^2 \over 2\mu R^2} \hbox{\mbf L}^2 
              + {\hbar^2 \over 2I_n} \hbox{\mbf S}^2,
\label{eq:A.1}
\end{equation}
where $\hbox{\mbf L}$ and $\hbox{\mbf S}$ denote the orbital 
angular momentum and the spin of the deformed nucleus, as usual. 
Note that by definition, the angular momenta are those for 
the one dimensional rotations, i.e., they are given by 
$\hbox{\mbf L}=-i\partial/\partial \theta_0$ and 
$\hbox{\mbf S}=-i\partial/\partial \alpha_0$, respectively.

In the description by the molecular model, we take the molecular 
$x'$-axis which is parallel to the relative vector of the two 
nuclei, as illustrated in Fig.~13, 
and the coordinate is denoted as $\theta$ (the angle $\theta$ 
is the same as $\theta_0$, i.e., $\theta = \theta_0$). 
And the angle of the orientation of the symmetry axis is described 
referring to the molecular $x'$-axis as $\alpha = \alpha_0-\theta_0$. 
Then due to $\omega'_n=\omega_n - \omega$ 
with  $\omega'_n=\dot \alpha$ and $\omega=\dot \theta$, 
the classical kinetic energy is written as 
\begin{equation}
T={1\over 2} (I_{\rm total}\ \omega^2
    +2I_n\omega \omega'_n + I_n{\omega'_n}^2) ,
\label{eq:A.2}
\end{equation}
where $I_{\rm total}$ denotes the moment of inertia 
of the whole system given by $I_{\rm total}=\mu R^2+I_n$. 
Replacing these angular velocities with time derivatives 
of the coordinates $(q_i)=(\theta , \alpha)$, 
we write a classical kinetic energy in the form 
${1 \over 2} \sum g_{ij}{\dot q}_i{\dot q}_j$. 
And then we quantize it by using the general formula 
for the curve-linear coordinate system, 
\begin{equation}
  {\hat T} = -{\hbar^2 \over 2}
           \sum_{ij} {1 \over \sqrt g}{\partial \over \partial q_i} 
           \sqrt g(g^{-1})_{ij}{\partial \over \partial q_j}, 
\label{eq:A.3}
\end{equation}
where $g$ and $g^{-1}$ denote the determinant and the inverse matrix of 
$(g_{ij})$, respectively. In this case, the metric tensor is given by 
\begin{equation}
(g_{ij})=  \left(  \begin{array}{cc}
          \mu R^2+I_n  & I_n \\
          I_n          & I_n
          \end{array}                \right), 
\label{eq:A.4}
\end{equation}
and hence the inverse matrix is obtained as 
\begin{equation}
(g_{ij})^{-1} =  \left( \begin{array}{cc}
          1/\mu R^2    &  -1/\mu R^2    \\
          -1/\mu R^2  & 1/\mu R^2 + 1/I_n
          \end{array}                    \right). 
\label{eq:A.5}
\end{equation}
As the classical kinetic energy consists of the three parts, i.e.,
the rotation of the whole system, the internal motions and their 
couplings, the quantum mechanical operator for the kinetic energy 
${\hat T}$ is also given as a sum of three terms, 
which appears to be  
\begin{equation}
  {\hat T} =  -{\hbar^2 \over 2}
\biggl\{ 
   {1 \over \mu R^2} {\partial^2 \over \partial \theta^2} 
 -{2 \over \mu R^2}{\partial^2 \over \partial \theta \partial \alpha}
  +\Big({1 \over \mu R^2}+ {1 \over I_n}\Big)
                                {\partial^2 \over \partial \alpha^2} 
\biggr\}.
\label{eq:A.6}
\end{equation}
We define the total angular momentum  $\hbox{\mbf J}_\theta$ 
and the operator for the internal (rotational or vibrational) 
motion of the deformed nucleus $\hbox{\mbf S}_\alpha$, 
by $\hbox{\mbf J}_\theta=-i\partial/\partial \theta$ 
and $\hbox{\mbf S}_\alpha=-i\partial/\partial \alpha$, 
respectively, 
and rewrite Eq.~(\ref{eq:A.6}) with those operators as follows, 
\begin{eqnarray}
 {\hat T} &=&   {\hbar^2 \over 2} 
                \bigl\{ \hbox{\mbf J}_\theta^2/\mu R^2 
          -2\hbox{\mbf J}_\theta  \hbox{\mbf S}_\alpha/\mu R^2
               +(1/\mu R^2 + 1/I_n )\hbox{\mbf S}_\alpha^2 \bigr\}    
\label{eq:A.7}
  \\
%
%
   &=&  {\hbar^2 \over 2\mu R^2}
        \bigl(\hbox{\mbf J}_\theta-\hbox{\mbf S}_\alpha \bigr)^2
        +{\hbar^2 \over 2I_n}\hbox{\mbf S}_\alpha^2 ,
\label{eq:A.8}
\end{eqnarray}
which  corresponds to Eq.~(\ref{eq:A.1}) and exhibits the role of 
the Coriolis terms in the molecular model. 
Note that the coordinate $\theta$ for the molecular $x'$-axis is 
the same angle as $\theta_0$, 
but the role of $\theta$ is different from  $\theta_0$. 
The molecular axis $x'(\theta)$ represents the motion of the whole 
system with the moment of inertia  $I_{\rm total}$, 
while $\theta_0$ is the angle of the relative vector between 
the two constituent nuclei, which represents the orbital motion. 
The process by using the formula Eq.~(\ref{eq:A.3}) of the general 
quantization for the curve-linear coordinates 
gives a simple example for the kinetic energy operator 
described in \S2,   
which clarifies the property of the molecular coordinate $\theta$.

We can introduce the angular momentum operators 
$\hbox{\mbf J}_\theta$ and $\hbox{\mbf S}_\alpha$, of course, 
by the direct transformation for the differential operators.
We obtain, due to the relations between the arguments 
$(\theta_0=\theta, \alpha_0=\theta+\alpha)$,  
\begin{eqnarray}
{\partial \over \partial \theta}
&=&
{\partial \theta_0 \over \partial \theta}{\partial \over \partial \theta_0}
 +
{\partial \alpha_0 \over \partial \theta}{\partial \over \partial \alpha_0}
=
{\partial \over \partial \theta_0} +{\partial \over \partial \alpha_0}, 
\label{eq:A.9} \\
{\partial \over \partial \alpha}
&=&
{\partial \theta_0 \over \partial \alpha}{\partial \over \partial \theta_0}
  +
{\partial \alpha_0 \over \partial \alpha}{\partial \over \partial \alpha_0}
=
{\partial \over \partial \alpha_0}, 
\label{eq:A.10}
\end{eqnarray}
and accordingly we can confirm the relations, 
$\hbox{\mbf J}_\theta=\hbox{\mbf L}+\hbox{\mbf S}$ 
and 
$\hbox{\mbf S}_\alpha =\hbox{\mbf S}$. 
As for the wave functions, let us start those with the eigenvalues 
$M$ and $m$ for $\hbox{\mbf L}$ and  $\hbox{\mbf S}$, respectively. 
The total wave function $\Psi$  is defined by 
$ \Psi (\theta_0, \alpha_0) = N e^{iM\theta_0} e^{im\alpha_0}$, 
the arguments of which could be replaced by the coordinate 
transformation as $(\theta_0=\theta, \alpha_0=\theta + \alpha)$, 
and accordingly we obtain 
$ \Psi (\theta, \alpha) = N e^{iM\theta} e^{im(\theta + \alpha)}
  =N e^{i(M+m)\theta} e^{im\alpha}$. 
Note that the function $e^{im\alpha}$ is the same one  
obtained from the operation of the unitary transformation of 
the whole rotation 
${\hat R_{n}(\theta)}=
     e^{-i\theta ( \hbox{\mbf n}\cdot \hbox{\mbf S})}$ 
on the spin function in the laboratory system, i.e., 
${\hat R_{ n}(\theta)} e^{im\alpha_0}=e^{im\alpha}$,
where $\hbox{\mbf n}$ denotes the unit vector normal to the plane.  
(For the general rotations in three dimensional space, 
the transformations are described by $D$-functions.) 
The resultant part 
$e^{i(M+m)\theta}$ properly corresponds to the wave function 
for the degree of freedom of "the whole rotation of the system", 
and thus we again confirm that the eigenvalue $J$ of 
$\hbox{\mbf J}_\theta$ satisfies the usual rule $J=M+m$.

In the molecular model, firstly we consider the whole rotating 
system with a stable (equilibrium) configuration expected, 
and secondly we investigate the internal degrees of freedom 
associated with it. 
With the strong nucleus-nucleus interaction, the motions of 
the constituent nuclei may be perfectly confined, and 
hence we sometimes consider the internal degrees of freedom 
to be frozen, i.e., {\em the sticking limit}.\cite{Bass} 
Thus it is worth while looking the kinetic energy 
of the molecular model in the sticking limit. 
In the classical kinetic energy Eq.~(\ref{eq:A.2}), we put 
$\omega'_n =0$ ($\omega = \omega_n $) and accordingly we obtain 
the energy of the rotator 
$T= I_{\rm total}\omega^2 /2$. 
However, in the quantum mechanical expression in Eq.~(\ref{eq:A.6}) 
and/or (\ref{eq:A.8}), it is clear above that the molecular model 
does not directly correspond to the sticking model, 
because the rotational energy is not given by 
$\hbox{\mbf J}^2\hbar^2/2I_{\rm total}$.
To make a model corresponding to the sticking model in quantum 
mechanics, we define a new coordinate $\Theta$, 
"the angle for the center of the moments of inertia" 
in stead of the Euler angle $\theta$ for the molecular $z'$-axis, 
as 
$\Theta = (\mu R^2 \cdot \theta_0 + I_n\cdot \alpha_0)/(\mu R^2+I_n)$. 
The other coordinate is again "the internal angle" 
$\alpha'=\alpha_0 - \theta_0$, which is the same as in the 
molecular model, and the corresponding moment of inertia is given by 
$I_{\rm internal}=\mu R^2 \cdot I_n/(\mu R^2+I_n)$. 
The set of the coordinates $(\Theta, \alpha')$ gives the classical 
kinetic energy expression without the Coriolis coupling term, as
\begin{equation}
T={1\over 2} (I_{\rm total}\Omega^2 + I_{\rm internal}{\omega'_n}^2),
\label{eq:A.11}
\end{equation}
where $\Omega$ denotes the angular velocity $\dot \Theta$. 
Thus we obtain the kinetic energy operator 
\begin{equation}
{\hat T} = {\hbar^2 \over 2I_{\rm total}} \hbox{\mbf J}_\Theta^2 
          + {\hbar^2 \over 2I_{\rm internal}} \hbox{\mbf S}_\alpha'^2 ,
\label{eq:A.12} 
\end{equation}
where by putting $\hbox{\mbf S}_\alpha'=0$ we reach the kinetic energy 
of the rigid-rotator type. Note that the choice of those coordinates 
follows the usage of the center of mass coordinate and the relative 
vector for two-body problem. Unfortunately this set of the coordinates 
would be limited on the rotations in a plain, 
because it is not easy to define "the center of the moments of inertia" 
for the multi-dimensional internal rotations.
Now, by the direct transformation for the differential operators, 
we again calculate operators 
$\hbox{\mbf J}_\Theta$ and $\hbox{\mbf S}_\alpha'$ 
due to the relations 
\begin{eqnarray}
\theta_0 &=& \Theta -I_n/(\mu R^2+I_n) \cdot \alpha', 
\label{eq:A.13}
\\  
    \alpha_0 &=& \Theta +\mu R^2/(\mu R^2+I_n) \cdot \alpha',
\label{eq:A.14}
\end{eqnarray}
which appear as 
\begin{eqnarray}
{\partial \over \partial \Theta}
  &=&
{\partial \over \partial \theta_0}+{\partial \over \partial \alpha_0}, 
\label{eq:A.15} \\
%
{\partial \over \partial \alpha'}
    &=&
-{I_n \over \mu R^2+I_n} {\partial \over \partial \theta_0}
  +{\mu R^2 \over \mu R^2+I_n} {\partial \over \partial \alpha_0}. 
%
\label{eq:A.16}
\end{eqnarray}
Thus again we have a usual description for the total 
angular momentum associated with $\Theta$ as a sum of the orbital 
angular momentum and the spin, 
i.e., $\hbox{\mbf J}_\Theta=\hbox{\mbf L}+\hbox{\mbf S}$. 
As for the spin for the internal rotation $\hbox{\mbf S}_\alpha'$, 
the definition turns out to be 
\begin{equation}
\hbox{\mbf S}_\alpha'=  -i \partial/\partial \alpha'
  = (\mu R^2  \cdot \hbox{\mbf S} -I_n \cdot \hbox{\mbf L})
      /(\mu R^2+I_n) .
\label{eq:A.17}
\end{equation}
For the sticking limit, we put $\hbox{\mbf S}_\alpha'=0$, 
and then we obtain 
\begin{equation}
\hbox{\mbf S}/I_n = \hbox{\mbf L}/\mu R^2  
                  = \hbox{\mbf J}/(\mu R^2 + I_n) ,
\label{eq:A.18}
\end{equation}
which is known as the rule of angular momentum sharing 
in the sticking model.\cite{Bass}   
Thus the coordinate system taken up here properly gives
the whole rotation and the internal motion without the coupling 
as a quantum mechanical description of the sticking model.

\section{
Symmetries of the System 
      and Construction of the Wave Functions}

Here we deal with symmetry properties of the system and their 
associated restrictions on the wave function, from which selection 
rules for quantum numbers are deduced.
Following those results, practical expressions for wave functions 
are given in the next Appendix C.

Firstly we note the coordinate transformation rules for boson and 
parity operations in the molecular frame. Here we do not describe 
how to obtain the rules. One could refer the derivations given 
in the Appendix D of Ref.~\citen{Ue93}. 

\vskip 4 true mm
\noindent
{\it Boson symmetry}  
 
We have the exchange operator $\hbox{\calg P}_{12}$, 
which acts on both the molecular coordinates and the internal variables, 
and transforms them as follows; 
\begin{equation}
   \hbox{\calg P}_{12}: 
( \theta_1 , \theta_2 , \theta_3 , \alpha , R, \beta_1 , \beta_2 ) 
             {\hbox to 15pt{\rightarrowfill}} 
            ( \pi + \theta_1 , \pi - \theta_2 ,  - \theta_3 , 
                            \alpha , R, \pi - \beta_2 , \pi - \beta_1 ). 
\label{eq:B.1}
\end{equation}

\noindent
{\it Inversion symmetry} ({\it parity})  
 
The inversion operator $\hbox{\calg P}$ acts as follows;
\begin{equation}
   \hbox{\calg P} : 
( \theta_1 , \theta_2 , \theta_3 , \alpha , R, \beta_1 , \beta_2 ) 
             {\hbox to 15pt{\rightarrowfill}} 
            ( \pi + \theta_1 , \pi - \theta_2 , \pi - \theta_3 , 
                           -\alpha , R, \beta_1 , \beta_2 ). 
\label{eq:B.2}
\end{equation}
\vskip 4 true mm

\noindent
{\it  Wave functions of the system with good symmetries }  
 
Since the axial symmetry of constituent nuclei is assumed, 
the variables $\gamma_i$ are not necessary.  
Each nucleus has positive parity, and thus its density 
profile is invariant under space inversion. 
Accordingly, the basis wave function  
$D_{MK}^J (\theta_i) \chi_{K}(R,\alpha, \beta_1,\beta_2) $  
should be invariant under the inversion operation upon a constituent 
nucleus, 
\begin{equation}
\hbox{\calg I}_i: (\alpha_i , \beta_i) {\hbox to 15pt{\rightarrowfill}} 
                           (\alpha_i +\pi, \pi -\beta_i ).    
\label{eq:B.3}
\end{equation}
Before we examine the symmetry for $\hbox{\calg I}_i$,
it should be noted that the transformations $\hbox{\calg I}_i$ 
affect the rotational variable $\theta_3$ 
as well as the internal variables $\alpha, \beta_1$ and $\beta_2$, 
because orientation of the molecular $x'$-axis changes 
according to a change of the orientation of a constituent nucleus. 
For example, we take up 
$\hbox{\calg I}_1^2 :  (\alpha_1 , \beta_1) 
       {\hbox to 15pt{\rightarrowfill}} (\alpha_1 + 2\pi, \beta_1 )$, 
which should be equal to unity, because it gives just $2\pi$ rotation 
of one constituent nucleus around the molecular $z'$-axis.    
By operating $\hbox{\calg I}_1^2$ on $\theta_3$ and $\alpha$ of 
$D^J_{MK}(\theta_i)\chi_{K}(R,\alpha, \beta_1,\beta_2)$,
according to $\theta_3=(\alpha_1 +\alpha_2)/2$
and $\alpha=(\alpha_1 -\alpha_2)/2$, we obtain a cyclic condition 
including the factor from the transformation on $\theta_3$,
\begin{equation}
  D^J_{MK}(\theta_i)\chi_{K}(R,\alpha, \beta_1,\beta_2)
   = (-1)^K  D^J_{MK}(\theta_i)\chi_{K}(R,\alpha +\pi, \beta_1,\beta_2).
\label{eq:B.4}
\end{equation}

Now in order to examine symmetries about the inversion operations 
$\hbox{\calg I}_i$, we set trial wave functions concretely.
By introducing harmonic approximation with the variables
$\beta_+ = (\Delta \beta_1 + \Delta \beta_2)/\sqrt 2$
and
$\beta_- = (\Delta \beta_1 - \Delta \beta_2)/\sqrt 2$
with $\Delta \beta_i \equiv  \beta_i -\pi/2$,
the internal motions are described with
\begin{equation}
\chi_{K}(R,\alpha, \beta_1,\beta_2) 
   = f_n(R)\phi_K(\alpha)\varphi_{n_+}^+(\beta_+ ,\alpha) 
                       \varphi_{n_-}^-(\beta_- ,\alpha) , 
\label{eq:B.5}
\end{equation}
where  \hfill $n$,  \hfill $n_+$  \hfill 
and  \hfill $n_-$  \hfill
denote  \hfill oscillator  \hfill quantum  \hfill numbers, 
 \hfill respectively. \hfill
For 
$\chi_{K}(R,\alpha, \beta_1,\beta_2)$ of Eq.~(\ref{eq:B.5}), 
Eq.~(\ref{eq:B.4}) means as 
\begin{equation}
 \phi_K(\alpha)=(-1)^K \phi_K(\alpha+\pi). 
\label{eq:B.6}
\end{equation}
Note that $\alpha$-dependences in $\varphi_{n_+}^+$ and 
$\varphi_{n_-}^-$ originate from those in the oscillator hamiltonians 
$H_\pm(\beta_\pm, \alpha)$ for the $\beta_+$ and  $\beta_-$ degrees of 
freedom, which are almost separable from the $\alpha$-degree but not 
completely, as is seen in Eqs. (\ref{eq:3.5}) $-$ (\ref{eq:3.8}) 
in \S3.       
Naturally, the periodical property of the $\alpha$-degree of freedom 
is of period $\pi$.  Furthermore due to the geometrical identification
of the configurations specified with $\alpha=\pi/2$ and $\alpha=0$,
we have a relation between butterfly function 
$\varphi_{n_+}^+ (\beta_+, \alpha)$ and
anti-butterfly one $\varphi_{n_-}^- (\beta_-, \alpha)$.
According to Eq.~(\ref{eq:3.9}), we have those oscillator energies 
$\hbar\omega_+
=\hbar\sqrt{k_+(\alpha)\{1/I + (1+\cos2\alpha)/\mu R_{\rm e}^2 \}}$ 
and
$\hbar\omega_-
=\hbar\sqrt{k_-(\alpha)\{1/I + (1-\cos2\alpha)/\mu R_{\rm e}^2 \}}$, 
respectively, where $k_+(\alpha)$ and 
$k_-(\alpha)$ denote spring moduli for respective modes. The moduli 
are defined by the coefficients of $\Delta\beta_i \Delta\beta_j$
in the harmonic expansion, and have been written in the text as 
$k_+(\alpha)=k_0+k_2(\alpha)+k^{12}_\beta(\alpha)$ and 
$k_-(\alpha)=k_0+k_2(\alpha)-k^{12}_\beta(\alpha)$, respectively,
where $k_0$ is a constant, and $k_2(\alpha)$ and $k^{12}_\beta(\alpha)$ 
consist of $\cos(2m\alpha)$ series with $m=even$ and $m=odd$, 
respectively. Since $\cos2(\alpha+\pi/2)=-\cos2\alpha$ 
and this is also the case in $k^{12}_\beta(\alpha)$, 
i.e., $k^{12}_\beta(\alpha+\pi/2)=-k^{12}_\beta(\alpha)$, 
we have relations 
$\hbar\omega_+(\alpha+\pi/2)=\hbar\omega_-(\alpha)$ and 
$\hbar\omega_-(\alpha+\pi/2)=\hbar\omega_+(\alpha)$. 
Thus the oscillator functions 
 $\varphi_{n_+}^+ (\beta_+, \alpha)$ and
 $\varphi_{n_-}^- (\beta_-, \alpha)$ also follow the same relations,
such as
\begin{equation}
 \varphi_{n'}^+ (\beta, \alpha) 
             = \varphi_{n'}^-(\beta, \alpha+\pi/2)
             = \varphi_{n'}^+(\beta, \alpha+\pi),
\label{eq:B.7}
\end{equation}
where the variable $\beta$ denotes $\beta_+$ or $\beta_-$. 
As for the transformation on $\Delta \beta_i$,
the operator $\hbox{\calg I}_i$ is the inversion by the definition,
and so we have 
$\hbox{\calg I}_1 :  (\theta_3, \alpha, \beta_+ , \beta_-) 
                               {\hbox to 15pt{\rightarrowfill}} 
           (\theta_3 + \pi/2, \alpha + \pi/2, -\beta_-, -\beta_+)$
and
$\hbox{\calg I}_2 :  (\theta_3, \alpha, \beta_+ , \beta_-) 
                     {\hbox to 15pt{\rightarrowfill}} 
            (\theta_3 + \pi/2, \alpha - \pi/2, \beta_-, \beta_+)$.
By utilizing the above relations,
we perform symmetrization of the basis wave function about 
$\hbox{\calg I}_i$.
Starting with the wave function 
        $D_{MK}^J (\theta_i) \chi_{K}(R,\alpha, \beta_1,\beta_2)$
with Eq.~(\ref{eq:B.5}) for $\chi_{K}(R,\alpha, \beta_1,\beta_2)$,
we obtain 
\begin{eqnarray}
   \Psi_\lambda   & \equiv & 
       (1 + \hbox{\calg I}_1)(1 + \hbox{\calg I}_2 ) \cdot
        D_{MK}^J (\theta_i)  f_n(R) \phi_K(\alpha)
      \varphi_{n_+}^+(\beta_+ ,\alpha) \varphi_{n_-}^-(\beta_- ,\alpha) 
\nonumber \\
  & =& D_{MK}^J (\theta_i)  f_n(R)   \{ 1+ (-1)^{n_+  + n_-  -K} \}
       h_{Kn_+ n_-}(\alpha, \beta_+, \beta_-)  ,
\label{eq:B.8}
\end{eqnarray}
with 
\begin{eqnarray}
       h_{Kn_+ n_-}(\alpha, \beta_+, \beta_-) 
& \equiv &
\phi_K(\alpha)
\varphi_{n_+}^+(\beta_+ ,\alpha) \varphi_{n_-}^-(\beta_- ,\alpha)  
\nonumber \\
&\quad &  +(-i)^K \phi_K(\alpha+\pi/2)\varphi_{n_-}^+(\beta_+ ,\alpha)
                         \varphi_{n_+}^-(\beta_- ,\alpha) .
\label{eq:B.9}
\end{eqnarray}
From the phase in the braces of Eq.~(\ref{eq:B.8}),
we obtain the selection rule for $\beta$-mode quanta as
\begin{equation}
                (-1)^{n_+ + n_-} = (-1)^K.
\label{eq:B.10}
\end{equation}
Note that the second term of the r.h.s. of Eq.~(\ref{eq:B.9}) 
originates from $\hbox{\calg I}_1$ and $\hbox{\calg I}_2$, 
with $n_+$ and $n_-$ exchanged.

Subsequently, we perform boson symmetrization and parity-projection 
by operating  ${1 \over 2} (1+ \hbox{\calg P}_{12})$ 
and ${1 \over 2} \{1 + (-1)^p \hbox{\calg P}\}$ to $ \Psi_\lambda $. 
We obtain the symmetrized wave functions as follows: 
\begin{eqnarray}
   \Psi_\lambda  &  \sim &
D_{MK}^J(\theta_i) f_n(R) \Big[ h_{Kn_+ n_-}(\alpha, \beta_+, \beta_-)
+(-1)^{p+K} 
       h_{Kn_+ n_-}(-\alpha, -\beta_+, \beta_-)   \Big]
\nonumber \\
&\qquad &   +(-1)^{p+J-K}
D_{M,-K}^J (\theta_i) f_n(R) 
\nonumber \\
&\qquad& \qquad \times 
      \Big[ h^*_{-Kn_+ n_-}(-\alpha, \beta_+, \beta_-)
                    +(-1)^{p+K} 
                h^*_{-Kn_+ n_-}(\alpha, -\beta_+, \beta_-)  \Big]. 
\label{eq:B.11}
\end{eqnarray}
Note that as the $\hbox{\calg P}_{12}$- (or $\hbox{\calg P}$-) operation 
gives $-K$ for the total rotation, we set $-K$ in Eq.~(\ref{eq:B.11}) for 
$h_{Kn_+ n_-}(\alpha, \beta_+, \beta_-)$-functions consistently, 
as follows. Since Eq.~(\ref{eq:B.6}) indicates the sign for the cycle 
for $|K|$, we define as $\phi_{-K}(\alpha)=\phi_K(\alpha)$. 
With the real functions $\phi_K(\alpha)$, the effects 
$-K$ in $h_{Kn_+ n_-}(\alpha, \beta_+, \beta_-)$-functions appear 
only in the phase $(-i)^K$ of the second terms of the r.h.s. of 
Eq.~(\ref{eq:B.9}) for $K=odd$ as complex conjugate. 
We further reduce  $h_{Kn_+ n_-}(\alpha, \beta_+, \beta_-)$
in Eq.~(\ref{eq:B.11}) with the arguments $-\alpha$ and/or $-\beta_+$. 
Since the reduced potential in Eq.~(\ref{eq:3.10}) has 
reflection symmetries at $\alpha=0$ and $\alpha=\pi/2$,
we are able to classify $\phi_K(\alpha)$ by parities with respect to 
those points, i.e.,
\begin{eqnarray}
  \phi_K(-\alpha)&=&\pi_{\alpha=0} \cdot \phi_K(\alpha),
\nonumber \\
        \phi_K(\pi/2 -\alpha)&=&\pi_{\alpha=\pi/2} \cdot 
                 \phi_K(\pi/2 +\alpha), 
\label{eq:B.12}
\end{eqnarray}
where $\pi_{\alpha=0}$ and $\pi_{\alpha=\pi/2}$ denote
the parities with respect to the reflection points, respectively. 
Note that a relation
$\pi_{\alpha=\pi/2}=(-1)^K \pi_{\alpha=0}$
is known because of
$\phi_K(\alpha+\pi)=(-1)^K \phi_K(\alpha)$ by Eq.~(\ref{eq:B.6}).
Due to 
$\hbar\omega(-\alpha)= \hbar\omega(\alpha )$
for each $\beta_+$- or $\beta_-$-mode, we also know 
$ \varphi_{n'}^+ (\beta, -\alpha) = \varphi_{n'}^+(\beta, \alpha)$
and
$\varphi_{n'}^- (\beta, -\alpha)  = \varphi_{n'}^-(\beta, \alpha)$.
By applying Eq.~(\ref{eq:B.12}) to
$h_{Kn_+ n_-}(\alpha, \beta_+, \beta_-)$ of Eq.~(\ref{eq:B.9}), 
we rewrite the internal wave functions in Eq.~(\ref{eq:B.11}).
For the first line, for example,  we obtain
\begin{eqnarray}
&  h_{Kn_+ n_-}(\alpha, \beta_+, \beta_-)
   +(-1)^{p+K} h_{Kn_+ n_-}(-\alpha, -\beta_+, \beta_-)
\nonumber \\ 
& = 
\big\{ 1+ \pi_{\alpha=0} \cdot (-1)^{p+ n_-} \big\} 
   \Big\{ \phi_K(\alpha)
\varphi_{n_+}^+(\beta_+ ,\alpha) \varphi_{n_-}^-(\beta_- ,\alpha)  
\nonumber \\ 
& \qquad \qquad\qquad\qquad\qquad  
     +(-i)^K \phi_K(\alpha+\pi/2)\varphi_{n_-}^+(\beta_+ ,\alpha)
                         \varphi_{n_+}^-(\beta_- ,\alpha) \Big\}, 
\label{eq:B.13}
\end{eqnarray}
where the rule Eq.~(\ref{eq:B.10}), $(-1)^K = (-1)^{n_+ + n_-}$ is 
used. Hence, relations are obtained as
$ \pi_{\alpha=0} \cdot (-1)^{p+n_-} =1 $,
and for even parity states  we have
\begin{eqnarray}
  \pi_{\alpha=0} & =&  (-1)^{n_-},
\nonumber \\
            \pi_{\alpha=\pi/2} & =&  (-1)^{n_+},
\label{eq:B.14}
\end{eqnarray}
which specify parities of $\alpha$-mode in connection 
with the $\beta$-mode quanta. Under the parity selection rule 
in the $\alpha$-motion in Eq.~(\ref{eq:B.14}), 
we can rewrite the functions of Eq.~(\ref{eq:B.13}) into
$h_{Kn_+ n_-}(\alpha, \beta_+, \beta_-)$. 
Thus, the final form of the total wave function with the symmetries 
is as follows: 
\begin{eqnarray}
   \Psi_\lambda  &  \sim & 
   D_{MK}^J (\theta_i) f_n(R) h_{Kn_+ n_-}(\alpha, \beta_+, \beta_-)
\nonumber \\
&\qquad &  +(-1)^{J +n_+ } D_{M,-K}^J (\theta_i) f_n(R) 
                    h_{-K n_+ n_-}(\alpha, \beta_+, \beta_-),
\label{eq:B.15}
\end{eqnarray}
with the definition of $h_{K n_+ n_-}(\alpha, \beta_+, \beta_-)$
in Eq.~(\ref{eq:B.9}) and with the selection rules given 
in Eq.~(\ref{eq:B.10}) and Eq.~(\ref{eq:B.14}).

\vskip 4 true mm
 
\noindent
{\it Restrictions on quantum numbers} 

As a summary, we have selection rules given 
in Eqs. (\ref{eq:B.10}) and (\ref{eq:B.14}), some conditions 
for the wave functions such as given in Eq.~(\ref{eq:B.7}) 
and the total wave function given in Eq.~(\ref{eq:B.15}) 
with Eq.~(\ref{eq:B.9}). In the following, we give some practical 
selection rules reduced from those relations.

Due to the cyclic condition 
$\phi_K(\alpha +\pi)=(-1)^K \phi_K(\alpha)$,
if we expand $\phi_K(\alpha)$ with periodic functions such as
$\cos \nu\alpha$ or $\sin \nu\alpha$,  
we have a general restriction for rotational quantum numbers 
$ (K \pm \nu) = 2m $, $m$ being an integer. 
And then, with a specified $K$, one of the two parities of 
Eq.~(\ref{eq:B.14}) is enough for specifying $\phi_K(\alpha)$ 
to be cosine type or sine one, 
the other selection rule being automatically fulfilled. 
Because of the symmetry of each constituent nucleus under space 
inversion, $n_+$ can be taken to be larger than or equal to $n_-$.  
In the case that $n_+$ is equal to $n_-$, we have $K=even$ 
from $(-1)^{n_+ +n_-}=(-1)^K$ of Eq.~(\ref{eq:B.10}).
And Eq.~(\ref{eq:B.9}) turns out to be 
$\{ \phi_K(\alpha) + (-i)^K \phi_K(\alpha+\pi/2) \}
\varphi_{n'}^+(\beta_+ ,\alpha) \varphi_{n'}^-(\beta_- ,\alpha)$,
which suggests $ (K \pm \nu) =4m$.
(See also Eq.~(\ref{eq:C.2}) for $K$-, $\nu$-rules.)
As for $K=0$, we obtain  the phase rule $(-1)^{J+n_+} =1$ 
from Eq.~(\ref{eq:B.15}), i.e., 
$n_+ = even$ and $n_- = even$ for $J=even$. 
With $K=\nu=0$, 
the state has no $\alpha$-dependence and hence 
the second term of Eq.~(\ref{eq:B.11}) become to be the same as 
the first term, which gives the phase rule $(-1)^{p+J}=1$, 
i.e., $J=even$ for the positive-parity states 
and $J=odd $ for the negative-parity states, respectively.

\section{
Explicit Expressions of Wave functions for the Normal Modes}

In the present Appendix, we give some examples of explicit 
expressions for the wave functions.
As is shown in Appendix B, the internal wave functions are 
approximately a product of $f_n(R)$, $\phi_K(\alpha)$, 
$\varphi_{n_+}^+(\beta_+,\alpha)$ and $\varphi_{n_-}^-(\beta_-,\alpha)$, 
which are essentially oscillator wave functions 
except for $\phi_K(\alpha)$.
The property of $\alpha$-motion may be rotational,
or may be vibrational, depending on the strength of the reduced 
potential which confines the alpha-degree of freedom.  
Note that the reduced potential is determined by the interaction 
between two nuclear surfaces as well as by quantum states 
of the other degrees of freedom.
Especially in the molecular ground state, where the additional 
potentials from the normal-mode motions such as the butterfly mode is 
weak, the property of $\phi_K(\alpha)$ is determined by the interaction.
The confinement potential obtained from the folding model is weak
as is shown in \S3,   
but its reality is not confirmed yet.
Hence we present both types for $\phi_K(\alpha)$ functions,
i.e., cosine and sine series as well as gaussian functions.
First we take up the rotational type
and adopt cosine series for  $\phi_K(\alpha)$,
assuming  $K, \nu =even$ and $n_+, n_- =even$.
Note that $e^{i\nu \alpha}$ may be convenient for describing
  $h_{Kn_+ n_-}(\alpha, \beta_+, \beta_-)$ in Eq.~(\ref{eq:B.9}),
but it does not fulfill Eqs. (\ref{eq:B.12}) and (\ref{eq:B.14}).
Note also that for $K=odd$,
  $h_{Kn_+ n_-}(\alpha, \beta_+, \beta_-)$ 
includes both cosine and sine functions due to the shift by $\pi/2$
in $\phi_K(\alpha)$.

Now we write down the total wave function according to 
Eqs. (\ref{eq:B.9}) and (\ref{eq:B.15}).
For positive-parity states with $K=even$ and $n_+ =even$ 
($n_-=even, \pi_{\alpha=\pi/2}=\pi_{\alpha=0}=1$ ),
by applying $\cos\nu(\alpha+\pi/2) =(-1)^{\nu/2}\cos\nu\alpha$ for
$\phi_K(\alpha+\pi/2)$,
we have
\begin{equation}
                 \Psi_\lambda   
\sim 
  \Big[ D_{MK}^J (\theta_i) +(-1)^J  D_{M,-K}^J (\theta_i) \Big]
    f_n(R) h_{Kn_+ n_-}(\alpha, \beta_+, \beta_-),
\label{eq:C.1}
\end{equation}
with
\begin{eqnarray}
       h_{Kn_+ n_-}(\alpha, \beta_+, \beta_-)  &\equiv &
\sum_{\nu =even} C_\nu \cos\nu\alpha   
\Big\{\varphi_{n_+}^+(\beta_+,\alpha) \varphi_{n_-}^-(\beta_-,\alpha)  
\nonumber \\
  & & \qquad\qquad\qquad
   +(-1)^{K+\nu \over 2} \varphi_{n_-}^+(\beta_+ ,\alpha)
                         \varphi_{n_+}^-(\beta_- ,\alpha) \Big\}.
\label{eq:C.2}
\end{eqnarray}
If we take a single value of $\nu$, the $\alpha$-motion is, of course, 
rotational. The quantum state $(n,n_+,n_-,K,\nu)=(0,0,0,0,0)$ 
corresponds to the molecular ground state.
Actually the solutions obtained in \S3   
is close to that with $\nu=0$.
For $K=0$, the next rotational state is with $\nu=4$, 
namely, the {\em twisting rotational mode}.
The explicit functions for $h_{Kn_+ n_-}(\alpha, \beta_+, \beta_-)$
of those states are given as
\begin{eqnarray}
 & \nu=0: & \quad
    h_{000}(\alpha, \beta_+, \beta_-)  = {\sqrt {1 \over \pi}}
   \varphi_0^+(\beta_+ ,\alpha) \varphi_0^-(\beta_- ,\alpha)  ,
\nonumber \\ 
 & \nu=4: & \quad
 h_{000}(\alpha, \beta_+, \beta_-)  = {\sqrt {2 \over \pi}} \cos 4\alpha
   \, \varphi_0^+(\beta_+ ,\alpha) \varphi_0^-(\beta_- ,\alpha)  .
\label{eq:C.3}
\end{eqnarray}
Due to the weak $\alpha$-dependence of the folding potential, 
solutions receive small mixings over $\nu$. The molecular ground 
state has the $\nu=4$ component about 3\%    
in the probability, as well as the $\nu=0$ 
component in the $\nu=4$ excited state. 
Those coefficients $C_\nu$'s are adopted for the calculations of 
the partial decay widths and the angular correlations, 
with the approximation of the constant vibrational quanta for 
$\hbar_{\beta_+}$ and $\hbar_{\beta_-}$ 
($\hbar_{\beta_+}=\hbar_{\beta_-}$), 
the value of which is taken to be $4$MeV. 
As for the relative motion, it is completely separated 
from the $(\alpha, \beta_+, \beta_-)$-degrees of freedom, and it is 
described by oscillator wave function $f_n(R)$ with the center at the 
equilibrium distance $R_{\rm e}$. According to the experimental resonance 
energy $E_{\rm cm}=55.8$MeV for $J=38$, the molecular ground state with 
$n=0$ is a suitable assignment as the theoretical eigenenergy is $51.5$MeV.
With $n=1$, the radially-excited state appears at higher than experimental 
energy, and it may not be observed in experiments, because of the broad 
decay widths expected. On the other hand, there is a possibility that 
a stronger interaction appears due to induced deformations, which give rise 
to lowering the eigenenergy of the radially-excited state to $55.8$MeV. 
Note that characteristics of probability distributions among the decay 
channels receive no effect from the choice of the radial motion with 
given $n$, which is completely separated from the angle degrees of freedom.
Only the magnitudes of the partial decay widths commonly become 
larger as we take a higher $n$-value.

If we coherently sum up over $\nu$ in Eq.~(\ref{eq:C.2}), 
the $\alpha$-motion can have a localized property such as a zero-point 
oscillation. For example, for simplicity, we take up $n_+ = n_- =0$,
and then the selection rule for $K$ and $\nu$ is $K \pm \nu =4m$ 
with $m$ being an integer, because two terms in the braces of the r.h.s. 
of Eq.~(\ref{eq:C.2}) must have the same sign.  We can write simply as
\begin{equation}
h_{K00}(\alpha, \beta_+, \beta_-)=  \sum_{K+\nu =4m} C_\nu \cos\nu\alpha   
            \, \varphi_0^+(\beta_+ ,\alpha) \varphi_0^-(\beta_- ,\alpha) .
\label{eq:C.4}
\end{equation}
Now, if we adopt gaussian type coefficients for $C_\nu$ by 
\begin{equation}
C_\nu = {4 \over 1+\delta_{\nu 0}} 
                 \Big( { 1 \over a \pi^3} \Big)^{1/4}
                 \exp \Big( -{\nu^2 \over 2a} \Big),
\label{eq:C.5}
\end{equation}
we obtain localized $\alpha$-motion such as
\begin{eqnarray}
 gaussian: &\quad & 
    h_{K00}(\alpha, \beta_+, \beta_-)  = 
\nonumber \\
    &\quad & \Big( {a \over 4\pi} \Big)^{1/4}
  \Big[ \exp \Big( -{a \over 2} \alpha^2 \Big)
+(-1)^{K/2} \exp\Big\{-{a \over 2} \Big(\alpha -{\pi\over 2}\Big)^2\Big\}\Big]
\nonumber \\
  & \quad & 
\times   \varphi_0^+(\beta_+ ,\alpha) \varphi_0^-(\beta_- ,\alpha) ,
\label{eq:C.6}
\end{eqnarray}
where the normalization constant is given for the case with no 
substantial overlapping between two terms in the square bracket 
on the r.h.s. of Eq.~(\ref{eq:C.6}). Note that due to the application 
of the selection rule $K \pm \nu =4m$ on $\cos \nu\alpha$, the 
resultant $\alpha$-function (the second line of Eq.~(\ref{eq:C.6})) 
is periodic with period $\pi$, 
and especially for $K=4n$ the function is with period $\pi/2$.
Our expression for the variable $\alpha$ in Eq.~(\ref{eq:C.6}) is 
given for the region $-\pi/4 \le \alpha <  3\pi/4$, so that the next 
gaussian peak at $\alpha=\pi$ can be omitted.

The normal modes for the $\beta_+$- and $\beta_-$-motions are named as 
{\em butterfly} and {\em anti-butterfly modes}, respectively,
the quantum numbers of which are $(n_+,n_-)$.
Due to the selection rule $(-1)^{n_+ + n_-} = (-1)^K$,
the lowest butterfly and anti-butterfly states with $K=0$
appear with $(n_+, n_-)=(2,0)$ and $(0,2)$, respectively.
The physical butterfly motion corresponds to the configuration displayed
in Fig.~2, where the motion of the axis $z''_2$ is confined around 
downside with $\alpha_2 \sim \pi$ 
and the vibrational motions with $\Delta \beta_1 \sim \Delta \beta_2$,
for example, with the quanta $(2,0)$.
When the configuration described 
with variables $\alpha_i$ and $\Delta \beta_i$ is transformed by 
$\hbox{\calg I}_2 : (\alpha_2, \beta_2) {\hbox to 15pt{\rightarrowfill}} 
         (\alpha_2 +\pi, \pi -\beta_2 ) $
in Eq.~(\ref{eq:B.3}), $(\beta_+, \beta_-)$ is transformed into 
$(\beta_-, \beta_+)$, which means that the physical butterfly motion 
is also described with configurations  with the motion of the axis 
$z''_2$ confined around upside with $\alpha_2 \sim 0$ 
and the vibrational motions with $\Delta \beta_1 \sim -\Delta \beta_2$,
with the quanta $(0,2)$ as the example.
Those symmetric terms for $(n_+, n_-)$ exchange are described in 
Eq.~(\ref{eq:B.9}) and Eq.~(\ref{eq:C.2}), where wave functions for 
the lowest butterfly state $h_{K20}(\alpha, \beta_+, \beta_-)$  
consist with one term with $(2,0)$ of $\alpha \sim \pi/2$ and another 
term with $(0,2)$ of $\alpha \sim 0$; for example, for butterfly,
\begin{eqnarray}
 \Big( {a \over  4\pi} \Big)^{1/4}
    \left[ (-1)^{K/2} 
\right.    &
\exp \left\{ 
   -{a \over 2} \left(\alpha -{\pi\over 2}\right)^2 
     \right\}
              \varphi_2^+(\beta_+) \varphi_0^-(\beta_-) 
\nonumber \\ 
  & \left.
   + \exp \left( -{a \over 2} \alpha^2 \right)
            \varphi_0^+(\beta_+) \varphi_2^-(\beta_-) \right] , 
\label{eq:C.7}
\end{eqnarray}
with a zero-point oscillation assumed for the $\alpha$-degree,
while for anti-butterfly,
\begin{eqnarray}
       \Big( {a \over 4\pi} \Big)^{1/4} 
    \left[ (-1)^{K/2}  
\right.    &
\exp \left\{
   -{a \over 2} \left(\alpha -{\pi\over 2} \right)^2 
     \right\}
              \varphi_0^+(\beta_+) \varphi_2^-(\beta_-) 
\nonumber \\ 
  & \left.
   + \exp \left( -{a \over 2} \alpha^2 \right)
            \varphi_2^+(\beta_+) \varphi_0^-(\beta_-) \right] .
\label{eq:C.8}
\end{eqnarray}

We have solved the Schr\"odinger equation for the $\alpha$-degree
in \S3,       
where we worked with the reduced potential for the butterfly 
quanta $(2,0)$ and obtained the solutions $\phi_K(\alpha)$.
A butterfly state appears with the lowest energy 
with a localization around $\alpha=\pi/2$ as is seen in Fig.~7. 
The motion for the $\alpha$-degree of an anti-butterfly state 
which corresponds to the quanta $(2,0)$, i.e., $\phi_K(\alpha)$ for 
the second term of Eq.~(\ref{eq:C.8}) is obtained with an excitation.
In order to calculate the partial widths and the angular correlations 
for the butterfly states, we use a simple but a typical expression, 
in which $\phi_K(\alpha)$ are described with dominant two coefficients 
$C_0$ and $C_2$ in Eq.~(\ref{eq:C.2}).  
Furthermore to obtain a typical expression, we impose a symmetry between 
the $\alpha$-motions in the butterfly and anti-butterfly states. Then we 
have the expression for the {\it butterfly} state with $K=even$ as
\begin{eqnarray}
    h_{K20}(\alpha, \beta_+, \beta_-) = {1 \over 2 \sqrt \pi} 
    && \big[  (-1)^{K/2} (1 - {\sqrt 2}\cos 2\alpha ) 
            \varphi_2(\beta_+) \varphi_0(\beta_-) 
\nonumber \\ 
    && + (1 + {\sqrt 2}\cos 2\alpha ) 
              \varphi_0(\beta_+) \varphi_2(\beta_-) \big], 
\label{eq:C.9}
\end{eqnarray}
where the oscillator $\varphi_{n'}^\pm(\beta_\pm, \alpha)$ are also 
simplified to be independent upon $\alpha$ with an averaged oscillator 
energy.
The corresponding pair of the {\it anti-butterfly} state is given by 
\begin{eqnarray}
    h_{K02}(\alpha, \beta_+, \beta_-) = {1 \over 2 \sqrt \pi} 
    & \big[ & (-1)^{K/2} (1 - {\sqrt 2}\cos 2\alpha ) 
            \varphi_0(\beta_+) \varphi_2(\beta_-) 
\nonumber \\
    &+&  (1 + {\sqrt 2}\cos 2\alpha ) 
                \varphi_2(\beta_+) \varphi_0(\beta_-) \big].
\label{eq:C.10}
\end{eqnarray}
As the dynamical solutions for the $\alpha$-motions in the reduced 
potential with the quanta $(2,0)$, $\phi_K(\alpha)$ are simplified to be
$(1-{\sqrt 2}\cos 2\alpha)$ for the butterfly mode,
while they are $(1+{\sqrt 2}\cos 2\alpha)$ for the anti-butterfly mode.  
Note that the results of our dynamical calculations 
with the reduced potential are rather close to those typical expressions.

\vskip 2.5 true mm

\noindent
{\it Wobbling motion ($K$-mixed states)}

\vskip 1 true mm

Following the discussion on $K$-mixed states in \S4.1, 
we define a wave function for the wobbling motion.
The mixing weights are given by a gaussian function of $K$,
by which we superpose $D^J_{MK}(\theta_i)$ such as
\begin{equation}
\sum_{K} {\tilde C}_K \sqrt{2J+1 \over 8\pi^2} D^J_{MK}(\theta_i) =
\sqrt{2J+1 \over 4\pi} e^{iM\theta_1} d^J_{MK}(\theta_2) 
\sum_{K} {\tilde C}_K {e^{iK\theta_3} \over \sqrt{2\pi}},
\label{eq:C.11}
\end{equation}
with 
\begin{equation}
     {\tilde C}_K=  
                  \left\{ \begin{array}{ll} 
\sqrt{ 2 \over b\sqrt\pi } 
  \exp\Big[-{1\over 2}\Big({K \over b}\Big)^2\Big]   &\mbox{for $K=even$}
 \\ 
 \qquad 0    &\mbox{for $K=odd$}, 
\end{array}
\right.
\label{eq:C.12}
\end{equation}
where $|\Delta K| =2$ due to the nature of couplings due to the axial 
asymmetry. The term with $K=0$ is important for including the components 
of the elastic scattering. The coherent summation over 
${\tilde C}_K e^{iK\theta_3} / \sqrt{2\pi}$ 
again gives us a gaussian function $W (\theta_3)$ of period $\pi$ 
due to $K=even$. We obtain,  for $0\le \theta_3 <  2\pi$,
\begin{equation}
W(\theta_3) =  
\sqrt{b\over 2\sqrt\pi} 
\bigg[ \exp\Big\{-{b^2\over2}\theta_3^2\Big\}
+ \exp\Big\{-{b^2\over2}(\theta_3 -\pi)^2\Big\}
+ \exp\Big\{-{b^2\over2}(\theta_3 -2\pi)^2\Big\}\bigg].
\label{eq:C.13}
\end{equation}

The wobbling basis function of Eq.~(\ref{eq:C.11}) would be applied 
with the internal modes
$\chi_K (R, \alpha, \beta_+, \beta_-)$ of Eq.~(\ref{eq:B.5}),
which receives the transformations 
$\hbox{\calg I}_i$ given in Eq.~(\ref{eq:B.3}), 
and is symmetrized as in Eq.~(\ref{eq:B.8}).
When we sum over $D^J_{MK}(\theta_i)$ of different $K$-values,
relative phases of those functions should be chosen properly. The 
coefficients for the wobbling motion in Eq.~(\ref{eq:C.11}) is given 
for the axially asymmetric configurations, where the molecular axis 
with the largest moment of inertia is $x'$, i.e., $I_{x'} > I_{y'}$, 
such as for an equator-equator one illustrated in Fig.~8.
However the butterfly configuration such as illustrated in Fig.~2 
has $I_{x'} < I_{y'}$, because, with $\alpha_1=0$ and $\alpha_2=\pi$, 
i.e., $\theta_3=\pi/2$ and $\alpha=-\pi/2$ by definition,
the $x'$-axis moves to the direction of the initial $y'$-axis.
To recover the condition $I_{x'} > I_{y'}$, we need to reset the 
$x'$-axis on to the intial direction, which brings additional phases
$e^{-i\pi/2}=(-i)^K$ on $D^J_{MK}(\theta_i)$. 
The expressions in Eqs. (\ref{eq:C.6})$\sim$(\ref{eq:C.10}) satisfy 
this relative phase convention. Of course, we are able to adopt the 
rotational equation of motion with $I_{x'} < I_{y'}$ in \S4.1, 
to obtain the alternative phase $(-1)^{K/2}$ for the $K$-bases.

\section{
Moments of Inertia of Dinuclear Systems}

The expression of the inertia tensor of dinuclear systems is already 
given in Eq.~(\ref{eq:8}), which is defined by the configuration 
referring to the molecular axes. 
We again write it here, for convenience, i.e., 
\begin{equation}
    \boldsymbol{I}_{\rm s}  =     \boldsymbol{I}_\mu (R)  
   +{}^t R'(\alpha_1 \beta_1 \gamma_1) \, \boldsymbol{I}_1 
                                      R'(\alpha_1 \beta_1 \gamma_1)  
   +{}^t R'(\alpha_2 \beta_2 \gamma_2) \, \boldsymbol{I}_2 
                                      R'(\alpha_2 \beta_2 \gamma_2),   
\label{eq:D.1}
\end{equation}
where the first term of the r.h.s. denotes the moments of inertia 
of two-ion centers, and the second and third terms are individual 
contributions from the constituent nuclei with 
rotation matrices $R'(\alpha_i \beta_i \gamma_i)$. 
The diagonal components $I_{11}$ and $I_{22}$ of the inertia tensor 
$\boldsymbol{I}_\mu (R)$ are $\mu R^2$, the others being zero. 
The inertia tensors of the two constituent nuclei, 
$\boldsymbol{I}_1$ and $\boldsymbol{I}_2$ 
are defined in the coordinate frames of their principal axes. 
Then, they are diagonal, elements of which are determined 
by the excitation energies of the members of the ground rotational 
bands of the constituent nuclei. 
Except for the relative vector of the two-ion centers, the whole 
dinuclear configuration is determined by the orientations of the 
principal axes of the constituent nuclei, due to Euler rotations 
$\Omega'_i (\alpha_i, \beta_i, \gamma_i) \Omega_M(\theta_1, \theta_2)$. 
In them, the first rotation $\Omega_M(\theta_1, \theta_2)$ is 
concerned about the molecular axes, and the second rotation 
$\Omega'_i(\alpha_i \beta_i \gamma_i)$ is that of each constituent 
nucleus referring to the molecular axes. 
Thus the moments of inertia about the molecular axes are obtained 
with the rotation matrices $R'(\alpha_i \beta_i \gamma_i)$.

We estimate magnitudes of the moments of inertia from the shape of 
the molecular configuration displayed in Fig.~8. 
Inserting $\alpha_1 =\alpha_2 =0$ and $\beta_1 =\beta_2 = \pi/2$, 
we obtain the diagonal elements of $\boldsymbol{I}_{\rm s}$ to be 

\vfill
\eject

\begin{eqnarray}
  I_x & =& \mu R^2 + I_a + I_b,
\nonumber \\
  I_y & =& \mu R^2 + I_A + I_B,
\label{eq:D.2}
\\
  I_z & =&  I_A + I_B , 
\nonumber 
\end{eqnarray}
with the nondiagonal elements being zero. $I_A$ etc. denote the 
moments of inertia of the constituent nuclei, individually in 
their principal axes, i.e., the diagonal elements $(I_1, I_2, I_3)$ 
of $\boldsymbol{I}_1$ are written as $(I_A, I_A, I_a)$, 
and those of $\boldsymbol{I}_2$ as $(I_B, I_B, I_b)$, 
and their nondiagonal elements are zero. 
Note that for the states of the ground rotational band of the 
$^{28}\rm Si$ nucleus, due to the axial symmetry, $I_1 = I_2$ is 
assumed and further $I_a = I_b =0$ is adopted in \S2. 

The value of the  moment of inertia $\hbox{\calg I}$ for the 
$^{28}\rm Si$ ground band, is determined from the excitation 
energy  $E_{\rm x} =1.78$MeV  of the $2^+_1$ state of the 
$^{28}\rm Si$ nucleus, i.e., by the relation, 
\begin{equation}
             {\hbar^2 \over 2 \hbox{\calg I} } I(I+1) = E_{\rm x},
\label{eq:D.3}
\end{equation}
$I$ being the spin value ($I=2$), and the value of $\hbox{\calg I}$ 
is used in the numerical calculations in \S3. Note that 
$I_A = I_B = \hbox{\calg I}$ is denoted by $I$ in Eq.~(\ref{eq:20}).

On the other hand, moments of inertia of rigid bodies have been often 
investigated in the study of rotational spectra.\cite{BohrTEXT2-mom}
In \S4.1, with respect to the nuclear shape of the whole system, 
we adopt $I_a$ and $I_b$ of nonzero value, where induced deformation 
is expected. Moments of inertia can be obtained by integrating over 
nuclear volume, such as 
\begin{equation}
  I_i = \int_V \rho (\boldsymbol{r}) (r^2 -x_i^2) dV ,
\label{eq:D.4}
\end{equation}
where $x_i$ denote the coordinates in the principal axes, and 
$\rho (\boldsymbol{r})$ is a nuclear density distribution. 
The density profile of $^{28}\rm Si$ appears in the calculations of 
the folding potential with DDM3Y force in \S2.3, and its parameters 
are given there. 
With induced deformation, $I_1$ is not necessary to be equal 
to $I_2$, but we take the value of $I_1 =I_2$ tentatively, 
due to the axial symmetry of the density profile. 
Due to the oblate shape of the density profile, we obtain the vales of 
$I_1 = I_2 < I_3$, the values of which are $(164, 164, 222)$ 
in the unit of $\rm M_n fm^2$, with $\rm M_n$ being the nucleon mass. 
Compared with the value of moment of inertia, $70 \rm M_n fm^2$ 
estimated by Eq.~(\ref{eq:D.3}), the value $164 \rm M_n fm^2$ is about 
two times larger than that, which is well known for rotational spectra 
of nuclei.\cite{BohrTEXT2-mom} 
Since the moments of inertia of rigid body are too large, we renormalize 
the values of moments of inertia obtained by Eq.~(\ref{eq:D.4}), 
to be consistent with the excitation energy of the $2^+_1$ state of 
the $^{28}\rm Si$ nucleus. This means that we multiply a factor 
$0.42$ on the moments of inertia of Eq.~(\ref{eq:D.4}). 
Thus a value $I_a =I_b = 93 \rm M_n fm^2$ is adopted, and 
Eqs.~(\ref{eq:D.2}) give the values for moments of inertia of the whole 
system. Note that $I_1 =I_2$ is broken with induced deformation of 
the $^{28}\rm Si$ nucleus, but we have no information about those 
changes of the moments of inertia.  Hence we adopt the same value 
$70 \rm M_n fm^2$ for $I_A$ with the assumption $I_1 =I_2$, 
as well as for $I_B$.

As for dinuclear configurations in \S4.2, since one of the 
constituent nuclei is spherical, i.e., 
$\boldsymbol{I}_2 =0$ in Eq.~(\ref{eq:D.1}), we obtain the expressions 
of moments of inertia by simply inserting $I_B =I_b =0$ into 
Eq.~(\ref{eq:D.2}). 
Note that those expressions are tailored for the configurations 
displayed in Fig.~11(b) and Fig.~12, where $I_A$ and $I_a$ are denoted 
as $I_0$ and $I_3$, respectively, because of no $\boldsymbol{I}_2$. 
For case 1, $I_a =0$ is adopted with the axially-symmetric deformation, 
and for case 2, $I_a \ne 0$ is adopted with the axially-asymmetric one, 
in which the ratio $I_a / I_A$ can be estimated by Eq.~(\ref{eq:D.4}).

\end{document}